\RequirePackage[2023-11-01]{latexrelease}
\documentclass[11pt,showpacs,superscriptaddress,nofootinbib,preprintnumbers]{revtex4-1}
\usepackage[utf8]{inputenc}

\usepackage{amsmath,amsthm,amssymb,mathtools,physics,cancel,mhchem,tikz,ulem}
\usepackage{graphicx}
\usepackage{dcolumn}
\usepackage{multirow}
\usepackage{array}
\usepackage{soul}
\usepackage{booktabs,lmodern,array,wasysym,microtype,xcolor}
\usepackage[T1]{fontenc}
\usepackage{wasysym} 
\newcommand{\Strong}{\CIRCLE}   
\newcommand{\Medium}{\LEFTcircle} 
\newcommand{\Limited}{\Circle}  
\newcommand{\NA}{---}      
\usepackage[hidelinks]{hyperref}
\hypersetup{colorlinks=true,
     linkcolor=blue,
     filecolor=blue,
     urlcolor=blue,
     citecolor=blue,
     }

\graphicspath{{./figures/}}

\newcolumntype{L}[1]{>{\raggedright\arraybackslash}p{#1}}
\newcolumntype{C}[1]{>{\centering\arraybackslash}p{#1}}

\makeatletter
\def\selectlanguage#1{}
\let\l@english\relax
\makeatother

\begin{document}

\begin{abstract}
Since the discovery of the Higgs boson, the long-standing task at hand in particle physics is the search for new physics beyond the Standard Model, which accounts for only about 5\% of the Universe.
In light of this situation, the neutrino sector has drawn significant attention due to neutrino oscillations, which require physics beyond the Standard Model and have prompted a wide array of active and planned experimental programs.
Notably, neutrino facilities offer substantial potential to search for new physics beyond neutrino oscillations, owing to their precision measurement capabilities, diverse experimental configurations, and various neutrino sources.
This white paper summarizes the landscape of new physics that can be probed at current and future neutrino experiments, categorized into laboratory-produced and cosmogenic signals.
We discuss recent experimental results interpreted through the lens of new physics, as well as detailed plans and projected sensitivities of next-generation facilities.
This summary is based on presentations from the 4th Workshop on New Physics Opportunities in Neutrino Facilities (NPN 2024), held at IBS in Daejeon, Korea, on June 3-5, 2024.
Particular emphasis is placed on accelerator-based neutrino experiments and a range of neutrino programs in East Asia. 
We also outline key tasks necessary to realize the promising new physics opportunities ahead.

\end{abstract}

\title{New Physics Opportunities at Neutrino Facilities:\\ 
BSM Physics at Accelerator, Atmospheric, and Reactor Neutrino Experiments}

\author{Koun Choi}
\email{koun@ibs.re.kr}
\affiliation{Center for Underground Physics, Institute for Basic Science, Daejeon 34126, Korea}

\author{Doojin Kim}
\email{doojin.kim@usd.edu}
\affiliation{Department of Physics, University of South Dakota, Vermillion, SD 57069, USA}
\affiliation{Mitchell Institute for Fundamental Physics and Astronomy$,$ Department of Physics and Astronomy$,$ Texas A\&M University$,$ College Station$,$ TX 77843$,$ USA}

\author{Jong-Chul Park}
\email{jcpark@cnu.ac.kr}
\affiliation{Department of Physics and Institute for Sciences of the Universe,\\ Chungnam National University, Daejeon 34134, Republic of Korea}

\author{Seodong Shin}
\email{sshin@jbnu.ac.kr}
\affiliation{Laboratory for Symmetry and Structure of the Universe, \\
Department of Physics, Jeonbuk National University, Jeonju, Jeonbuk 54896, Korea}
\affiliation{School of Physics, Korea Institute for Advanced Study, Seoul 02455, Korea}
\affiliation{Center for Theoretical Physics of the Universe, \\
Institute for Basic Science, Daejeon 34126, Korea}

\author{Pouya Bakhti}
\affiliation{Laboratory for Symmetry and Structure of the Universe, \\
Department of Physics, Jeonbuk National University, Jeonju, Jeonbuk 54896, Korea}

\author{Ki-Young Choi}
\affiliation{Department of Physics and Institute of Basic Science, Sungkyunkwan University, 2066 Seoburo, Suwonsi, Gyeonggido, 16419, Korea}

\author{Chang Hyon Ha}
\affiliation{Department of Physics, Chung-Ang University, Seoul 06974, Korea}

\author{Kazumi Hata}
\affiliation{Research Center for Neutrino Science, Tohoku University, Sendai, 980-8578, Japan}

\author{Wooyoung Jang}
\affiliation{Department of Physics, University of Texas, Arlington, TX 76019, USA}

\author{Yu Seon Jeong}
\affiliation{Department of Physics and Institute of Basic Science, Sungkyunkwan University, 2066 Seoburo, Suwonsi, Gyeonggido, 16419, Korea}

\author{Young Ju Ko}
\affiliation{Department of Physics, Jeju National University, Jeju, 63243, Korea}

\author{Hyun Su Lee}
\affiliation{Center for Underground Physics, Institute for Basic Science, Daejeon 34126, Korea}

\author{Weijun Li}
\affiliation{Department of Physics, University of Oxford, Oxford, OX1 3PJ, United Kingdom}

\author{Yu-Feng Li}
\affiliation{Institute of High Energy Physics, Chinese Academy of Sciences, Beijing, 100049, China}
\affiliation{School of Physical Sciences, University of Chinese Academy of Sciences, Beijing 100049, China}

\author{Mehedi Masud}
\affiliation{High Energy Physics Center, Chung-Ang University, Seoul 06974, Korea}
\affiliation{Secci\'{o}n F\'{i}sica, Departamento de Ciencias, Pontificia Universidad Cat\'{o}lica del Per\'{u}, Av. Universitaria 1801, Lima 15088, Per\'{u}}

\author{Kenny C. Y. Ng}
\affiliation{Department of Physics, The Chinese University of Hong Kong,\\
Shatin, Hong Kong, China}

\author{Jungsic Park}
\affiliation{Department of Physics, Kyungpook National University, Daegu 41566, Korea}

\author{Min-Gwa Park}
\affiliation{Laboratory for Symmetry and Structure of the Universe, \\
Department of Physics, Jeonbuk National University, Jeonju, Jeonbuk 54896, Korea}
\affiliation{Department of Physics and Astronomy, Northwestern University, Evanston, IL 60208, USA}

\author{Komninos-John Plows}
\affiliation{Department of Physics, University of Liverpool, Liverpool, L69 7ZE, United Kingdom}

\author{Meshkat Rajaee}
\affiliation{Laboratory for Symmetry and Structure of the Universe, \\
Department of Physics, Jeonbuk National University, Jeonju, Jeonbuk 54896, Korea}

\author{Eunil Won}
\affiliation{Department of Physics, Korea University, Seoul 02841, Korea}

\author{Byeongsu Yang}
\affiliation{Institute for Universe and Elementary Particles,\\
Chonnam National University, Gwangju 61186, Korea}

\author{Seong Moon Yoo}
\affiliation{Department of Physics and Institute of Basic Science, Sungkyunkwan University, 2066 Seoburo, Suwonsi, Gyeonggido, 16419, Korea}

\author{Jaehoon Yu}
\affiliation{Department of Physics, University of Texas, Arlington, TX 76019, USA}

\author{Seokhoon Yun}
\affiliation{Center for Theoretical Physics of the Universe, \\
Institute for Basic Science, Daejeon 34126, Korea}
\affiliation{Department of Physics, Kyungpook National University, Daegu 41566, Korea}

\maketitle

\tableofcontents

\section{Introduction}

Neutrinos, as fundamental components of weak interactions in the Standard Model (SM), have played a pivotal role in guiding the search for physics beyond the Standard Model (BSM). The discovery of neutrino flavor oscillations~\cite{Super-Kamiokande:1998kpq,SNO:2002tuh,KamLAND:2002uet,K2K:2002icj} has motivated the development of BSM frameworks, particularly those that account for neutrino masses---an aspect not accommodated within the SM. 
Possible symmetries or symmetry breakings in the neutrino or broader lepton sector, such as charge-parity (CP) violation or extended gauge symmetries, may help resolve fundamental questions about the Universe that lie beyond the reach of the SM.
In order to explore the neutrino sector with precision, a wide range of current and upcoming experimental programs study neutrinos originating from reactors, the Sun, the atmosphere, accelerators, astrophysical sources, and the early Universe.
Importantly, the capabilities of these experiments extend beyond the study of neutrino properties alone. Their high sensitivity, diverse detection techniques, large beam intensity, and broad energy coverage make them powerful tools for probing a wider class of BSM phenomena. In particular, neutrino experiments offer unique opportunities to search for new particles and interactions, such as dark photons, dark matter (DM), axion-like particles (ALPs), dark $Z$ bosons, and signatures of grand unification.

In light of this situation, we launched a workshop series titled {\it New opportunities at the next-generation neutrino experiments}, aimed at fostering collaboration and idea exchange between theorists and experimentalists on how to fully leverage upcoming neutrino facilities. The first workshop was held at the University of Texas at Arlington on April 12-13, 2019. 
The main contents discussed in the workshop, particularly focused on the topics of dark matter and neutrino related BSM, are summarized in the white paper~\cite{Arguelles:2019xgp}.
It was followed by the second workshop at the University of Pittsburgh on February 10-12, 2022 (postponed due to the COVID-19 pandemic), in joint with the activities for the related SNOWMASS 2021 community planning studies. 
The third workshop was held at SLAC on July 11-13, 2023, with a focus on astrophysical neutrinos from supernovae and the Sun.
The fourth workshop, {\it New Physics Opportunities at Neutrino Facilities} (NPN 2024), took place at the Institute for Basic Science (IBS) in Korea from June 3-5, 2024, which was the first event outside the US.
Although various BSM search opportunities in a broad range of experiments have been extensively discussed in the previous workshops, the main focus of those events has been given to the atmospheric neutrino experiments, solar neutrino experiments, and reactor neutrino experiments, mostly based in North America or operated by the US institutions.
Therefore, the organizers of the fourth workshop decided to place particular emphasis on the capabilities and potential of accelerator-based neutrino experiments, as well as the increasing number of active neutrino programs in East Asian countries.
Due to the rapidly growing high energy communities in this area, numerous cutting-edge neutrino programs have been launched and proposed, as will be explained later.

This paper documents the new pioneering work presented at the 4th workshop in Korea.
We categorize BSM particle signals by their production origin---laboratory-produced and cosmogenic---as described in Sec.~\ref{sec:landscape}, and summarize key recent results from currently operating neutrino experiments in Sec.~\ref{sec:current}.
Section~\ref{sec:futurelab} introduces laboratory-based next-generation neutrino programs, including accelerator-based, reactor, and neutrinoless double beta decay experiments, with detailed discussions of new physics search prospects provided in each corresponding subsection.
In contrast, Sec.~\ref{sec:futurecosmo} focuses on next-generation neutrino observatories targeting cosmogenic neutrinos and other BSM signals.
The key tasks in experimental design and simulation tools to support the realization of these new physics opportunities are outlined in Sec.~\ref{sec:timeline}, and our conclusions are presented in Sec.~\ref{sec:conclusions}.
While a wide range of neutrino experiments are underway or in development globally, this workshop and the present paper place particular emphasis on the rapidly growing and increasingly influential East Asian neutrino programs, as well as the broad potential of accelerator-based experiments.

\section{Landscape on BSM Physics at Neutrino Experiments} 
\label{sec:landscape}

The search for new physics at neutrino experiments spans a wide range of possibilities, which can broadly be categorized into two primary classes: laboratory-produced signals and cosmogenic signals. Laboratory-produced signals arise from new particles or interactions generated directly at or near the neutrino/proton/electron beam source---often through exotic meson decays, beam-target interactions, or in conjunction with the production of standard neutrinos. This category also includes the signals from reactors and the transitions or decays of nuclear isotopes. The accelerator-produced signals can be tightly correlated with the beam timing and geometry, enabling precise experimental control and background mitigation, while the reactor and rare isotope transition signals benefit from huge fluxes and low backgrounds, respectively.
On the other hand, cosmogenic signals originate beyond the laboratory and encompass a broad range of new physics candidates, ranging from DM to exotic cosmic messengers produced in the early universe or high-energy astrophysical environments. Unlike laboratory signals, these events are often characterized by broader angular and timing distributions and may demand different detection strategies and background treatments. Together, these two categories define the evolving new physics landscape accessible at upcoming neutrino experiments.

\subsection{Laboratory-produced BSM signals}

Neutrino oscillations opened a host of fundamental questions---chief among them, the origin of neutrino mass, the possibility of lepton number violation, and whether neutrinos are Dirac or Majorana particles. To address these questions, a broad experimental program has been developed to precisely measure neutrino mixing parameters, search for neutrinoless double beta decay, and explore the potential existence of sterile neutrinos. 

Alongside these efforts, neutrino experiments have revealed several anomalies and deviations from standard scenarios, including the LSND anomaly~\cite{LSND:2001aii} and the MiniBooNE low-energy excess~\cite{MiniBooNE:2008yuf,MiniBooNE:2018esg,MiniBooNE:2020pnu} in short-baseline accelerator-based experiments, the 5 MeV bump observed in reactor antineutrino spectra~\cite{DayaBay:2016ssb,NEOS:2016wee,DoubleChooz:2019qbj,Serebrov:2020kmd,STEREO:2020hup,RENO:2020dxd,PROSPECT:2022wlf,Danilov:2024fwi} against the theory prediction~\cite{Mueller:2011nm,Huber:2011wv}, and the Gallium anomaly~\cite{Abdurashitov:2005tb,GALLEX:1997lja,Barinov:2021asz,Barinov:2022wfh}.
These experimental anomalies can be explained by new physics scenarios, such as the inclusion of one or more sterile neutrinos, non-standard neutrino interactions (NSIs), the introduction of damping factors caused by quantum decoherence~\cite{Farzan:2008zv, Bakhti:2015dca,Arguelles:2022bvt} or Lorentz violation~\cite{Hollenberg:2009tr}, or dark-sector solutions, including DM and new mediator particles~\cite{Batell:2022xau}. 

The presence of one or more sterile neutrinos mixing with active neutrinos, with a mass-squared difference on the order of \(1 \, \mathrm{eV}^2\), can explain the observed appearance and disappearance of neutrinos in various experiments. However, a model with only one sterile neutrino is insufficient to simultaneously resolve all short-baseline anomalies. Moreover, there is a significant tension between appearance and disappearance data, reaching the \(4\sigma\) to \(5\sigma\) level. This tension can be mitigated by considering larger uncertainties, such as allowing for free flux normalization~\cite{Dentler:2018sju}. 

The NSIs at low energies can be described using effective four-fermion dimension-six operators. Neutral-current (NC) NSIs impact flavor changes through propagation in matter as well as production and detection, while charged-current (CC) NSIs only affect the production and detection of neutrinos~\cite{Biggio:2009nt}, which can be constrained by short-baseline experiments like DUNE near detector~\cite{Bakhti:2016gic}. 
Although NSIs resolve certain discrepancies, such as the tension between solar neutrino measurements and the KamLAND reactor neutrino experiment's determination of solar neutrino oscillation parameters, the inclusion of NC NSI introduces additional degeneracies of the neutrino oscillation parameters arising from the transition symmetry of the oscillation probabilities by the replacement of the Hamiltonian $H$ with $-H^*$ ~\cite{Gonzalez-Garcia:2013usa, Bakhti:2014pva}.
The fact that such a symmetry coincides with a large value of NSI leads to two solutions for the first and second octants of $\theta_{12}$, known as the Large Mixing Angle (LMA) and LMA-Dark solutions, respectively. Consequently, this symmetry renders oscillation experiments insensitive to the mass ordering of neutrinos. 
Similarly with the standard matter effect, NC NSI has a less significant impact at low energies and low matter densities; hence, its effect is diminished in low-energy and short-baseline experiments, such as MOMENT and COHERENT~\cite{Coloma:2017ncl,Coloma:2017egw,Denton:2022nol}.
This makes them particularly effective at breaking degeneracies induced by NSIs, which enables more precise measurements of \( \delta_{CP} \)~\cite{Bakhti:2016prn}. 
Assuming the LMA solution as the true one, only JUNO achieves a sensitivity greater than \( 4\sigma \) for determining the neutrino mass ordering.
In contrast, DUNE and T2HK have a sensitivity at \( 2\sigma \)~\cite{Bakhti:2020fde}.

In contrast to NSIs with effective operators after integrating out heavy new fields, scenarios with light mediators also induce NSIs. 
The new light particles may couple exclusively to active neutrinos, known as secret neutrino interactions (SNI), leptons (leptophilic interactions), or couple to baryons as well.
SNI can have significant implications for processes such as charged meson decay, \( Z \)-boson decay, supernova cooling, Big Bang nucleosynthesis, and may even provide insights into resolving the Hubble tension~\cite{Berryman:2022hds}. The secret coupling of neutrinos to a new light vector boson enhances the three-body decay of charged mesons via the longitudinal polarization of the vector boson. This effect can be probed in both meson decay and neutrino experiments~\cite{Bakhti:2017jhm, Bakhti:2018avv, Bahraminasr:2020ssz, Bakhti:2023mvo, Feng:2022inv, Dev:2024twk}. Similarly, leptophilic interactions can affect meson, \( Z \)-boson, and \( W \)-boson decays as well as neutrino scattering experiments~\cite{Laha:2013xua}.

Quantum decoherence modifies the evolution of the neutrino density matrix, introducing a damping factor in the flavor transition probabilities~\cite{Farzan:2008zv, Bakhti:2015dca}. Similar effects can arise from mechanisms such as wave packet decoherence, Lorentz violation, gravitational effects on neutrino oscillations, and other related phenomena. These damping factors offer a potential explanation for observed neutrino anomalies~\cite{Blennow:2005yk,Arguelles:2022bvt}.

In addition to searches motivated by non-zero neutrino masses and various experimental anomalies, neutrino experiments offer a powerful platform to probe feebly interacting new physics particles, including light dark matter (LDM) and dark-sector mediators.\footnote{In this paper, we only consider beam-produced DM as LDM. Although we mostly consider DM with masses below $\mathcal O (1 - 10\,{\rm GeV})$, some experiments can produce heavier ones.} The high beam intensities and energy scales required for neutrino production also result in the prolific generation of source particles---such as photons, electrons, and neutral/charged mesons---that can give rise to new physics particles within the beam target or dump. Depending on the details of the underlying model, these particles may produce detectable signals through decay or scattering, leading to a variety of experimental signatures such as $e^\pm$, $\gamma$, $e^+e^-$, $\gamma\gamma$, or $e^\pm\gamma$. Although neutrino-induced processes can act as significant backgrounds, the advanced detector capabilities---featuring excellent timing, angular, position, and energy resolution---enable efficient background rejection. This opens the door to exploring a broad range of previously inaccessible parameter space for new physics.

The idea of introducing dark sector species, including LDM and mediators, is emerging as a promising alternative for explaining anomalies---particularly those observed in short-baseline neutrino experiments. In these experiments, charged mesons are focused by the horn magnet system and decay into neutrinos, inducing neutrino ``beams''. However, certain dark-sector models permit charged mesons to undergo exotic three-body decays into a light mediator~\cite{Barger:2011mt,Carlson:2012pc,Laha:2013xua,Bakhti:2017jhm,Krnjaic:2019rsv,Bakhti:2023mvo}, which may subsequently decay into a dark-matter pair, depending on the specifics of the underlying model. These focused new-physics signals can account for the MiniBooNE excess while remaining consistent with existing experimental constraints~\cite{Dutta:2021cip,Dutta:2025fgz}, particularly the null results observed in MiniBooNE's dump-mode measurements~\cite{MiniBooNEDM:2018cxm}.

The production of rare processes requires high-intensity sources, which is why typical neutrino experiments are either reactor-based or rely on high-intensity beams. In proton-beam-based experiments, incident protons strike a target or dump, producing a number of hadrons, including both charged and neutral mesons. Earlier studies (e.g., Snowmass report~\cite{Batell:2022xau} and references therein) have explored the use of neutral mesons, whose rare decays can involve LDM or mediators, as well as proton bremsstrahlung processes, in which dark-sector mediators are emitted as initial- or final-state radiation. In contrast, more recent studies have identified additional production channels involving secondary particles such as photons and $e^\pm$~\cite{Dutta:2020vop,Celentano:2020vtu}, which emerge from electromagnetic cascade showers, and charged mesons~\cite{Barger:2011mt,Carlson:2012pc,Laha:2013xua,Bakhti:2017jhm,Krnjaic:2019rsv,Dutta:2021cip,CCM:2023itc}, which can undergo exotic three-body decays. These developments demonstrate that neutrino experiments can probe a broader range of parameter space than previously anticipated. More precise estimates of the fluxes of such source particles are typically obtained through dedicated Monte Carlo simulations using tools such as \texttt{GEANT4}~\cite{GEANT4:2002zbu}.  

Theoretically, these dark sectors are often coupled to the SM through well-motivated portal interactions. Renormalizable portals---such as the vector portal, e.g., kinetic mixing between the photon/$Z$-boson and a new U(1) gauge boson~\cite{Holdom:1985ag,Okun:1982xi}, the Higgs portal (scalar mixing)~\cite{Patt:2006fw,Kim:2008pp}, and the neutrino portal (sterile neutrino mixing)~\cite{Minkowski:1977sc}---provide minimal extensions that introduce new light states. 
The production of portal particles, followed by their decay into DM, enables experimental searches for LDM, often probing regions of parameter space directly relevant to the observed relic abundance.

One of the extensively studied mediators at neutrino experiments is the ALP~\cite{Peccei:1977hh,Weinberg:1977ma,Wilczek:1977pj,Svrcek:2006yi}. Depending on the available energy and the specifics of the underlying model, ALPs can be produced through a variety of mechanisms, including the Primakoff process, Compton-like scattering, resonant production, $e^\pm$-induced bremsstrahlung, and associated production channels. A produced ALP then travels to the detector of interest, where it may decay or scatter off the detector material within the fiducial volume, or be converted to a photon in the presence of detector magnetic fields~\cite{Bonivento:2019sri}. 
Various studies have proposed searching for ALPs at reactors~\cite{Dent:2019ueq,AristizabalSierra:2020rom} and accelerator-based neutrino experiments including CCM~\cite{CCM:2021jmk}, DUNE~\cite{Kelly:2020dda,Brdar:2020dpr}, FASER/FASER$\nu$~\cite{Beacham:2019nyx}, IsoDAR~\cite{Waites:2022tov} and the liquid scintillator counter ($\nu$EYE)~\cite{Seo:2023xku} at Yemilab, PASSAT~\cite{Bonivento:2019sri,Dev:2021ofc}, PIP-II~\cite{Toups:2022yxs}, and SHiP~\cite{Alekhin:2015byh,Albanese:2878604}. Most recently, the NEON collaboration, originally designed to search for the Coherent Elastic $\nu$-Nucleus Scattering (CE$\nu$NS) by the reactor neutrinos,  reported new limits on ALPs coupling to SM photons and electrons using a 2.8 GW nuclear reactor~\cite{NEON:2024kwv}.  

Among the notable experimental anomalies, the reactor antineutrino anomaly (RAA)~\cite{Mention:2011rk} arises from a deficit in the observed reactor antineutrino flux at short baselines (within a few tens of meters) compared to theoretical predictions, such as those from the Huber-Mueller (HM) model\cite{Huber:2011wv,Mueller:2011nm}. 
However, more advanced models---including the Kurchatov Institute (KI) model~\cite{Kopeikin:2021ugh}, based on the latest $\beta$-spectrum data, and the Estienne-Fallot (EF) model~\cite{Estienne:2019ujo}, which employs total absorption $\gamma$-ray spectroscopy---can largely account for the RAA, potentially resolving the discrepancy within SM uncertainties. 
Nonetheless, BSM interpretations have also been proposed, such as the conversion of reactor neutrinos into eV-scale sterile neutrinos. This possibility has been actively probed by short-baseline reactor experiments, including DANSS~\cite{Danilov:2022bss}, NEOS~\cite{NEOS:2016wee}, Neutrino-4~\cite{Serebrov:2020kmd}, PROSPECT~\cite{PROSPECT:2022wlf}, and STEREO~\cite{STEREO:2020hup}.
Beyond sterile neutrinos, these experiments also serve as platforms for searching for other BSM particles, such as dark photons and ALPs, given that nuclear reactors are strong sources of photons, as mentioned above.

In contrast to the RAA, which reflects an overall deficit in the observed flux relative to the HM model, a local excess in the prompt energy range 4 to 6 MeV is reported by RENO~\cite{RENO:2015ksa}. This feature has been confirmed by other experiments including Daya Bay~\cite{DayaBay:2015lja}, Double Chooz~\cite{DoubleChooz:2014kuw}, NEOS~\cite{NEOS:2016wee}, Neutrino-4~\cite{Serebrov:2020kmd}, and DANSS (preliminary)~\cite{Danilov:2022bss} at the 4$\sigma$ level.
Furthermore, the research reactors with 100\% $^{235}$U fuel, such as STEREO~\cite{STEREO:2020hup} and PROSPECT~\cite{PROSPECT:2022wlf}, have excluded the no-bump scenario with significances of 3.5$\sigma$ and 2$\sigma$, respectively.
This experimental anomaly is dubbed the {\it 5 MeV bump}~\cite{Huber:2016xis}.
Theoretical flux models that significantly alleviate the RAA paradoxically exacerbate the bump anomaly.
While updated flux models incorporating recent nuclear physics calculations and measurements, such as the Hayen-Kostensalo-Severijns-Suhonen (HKSS) model~\cite{Hayen:2019eop}, may partially account for the excess, they do so at the cost of worsening the overall flux fit, i.e., the RAA. As an alternative, a BSM scenario involving a sterile neutrino with NSIs has been proposed~\cite{Berryman:2018jxt}.
To precisely identify the origin of the 5 MeV excess---which contrasts with the RAA deficit---it would be highly beneficial to explore alternative reactor neutrino detection channels beyond inverse beta decay (IBD). 
Motivated by this, a complementary detection channel has been proposed: observing inelastic neutrino neutral current scattering on ${}^{13}$C, which produces a sharp 3.685 MeV photon from the de-excitation of ${}^{13}\mathrm{C}^\ast$ in liquid scintillator (LS) detectors~\cite{Bakhti:2024dcv}.
Interestingly, this alternative channel has been shown to rule out a BSM scenario proposed in Ref.~\cite{Berryman:2018jxt}, based on a nominal application of current NEOS data~\cite{Siyeon:2017tsg}.

\subsection{Cosmogenic BSM signals}

Neutrinos originating from the Universe have provided valuable insights into BSM physics, owing to their high fluxes and diverse sources of production.
Solar neutrino observations have led to the discovery of neutrino oscillations, one of the strongest experimental motivations for BSM physics.
A variety of solar neutrino programs are actively being operated or planned to exactly determine the solar oscillation parameters~\cite{Bakhti:2023vzn}. 
These measurements also offer opportunities to probe potential BSM symmetries in the neutrino sector~\cite{Bakhti:2020hbz, Park:2023hsp} or to search for new particles produced in solar nuclear reactions, which could be detected via Compton-like scattering in neutrino detectors~\cite{DEramo:2023buu}.
In solar neutrino measurements, there are discrepancies such as the solar metallicity problem, Day-Night Asymmetry, and upturn deficit~\cite{Maltoni:2015kca}. 
A super-light sterile neutrino has been proposed to explain the deficit in the upturn of the solar neutrino spectrum through resonance enhancement~\cite{deHolanda:2003tx, deHolanda:2010am,Bakhti:2013ora}.

Atmospheric neutrino experiments probe a wide range of phenomena, including atmospheric oscillation parameters, CP violation, and possible BSM interactions and particles in the neutrino sector~\cite{Bakhti:2022axo,Bakhti:2023mvo,Super-Kamiokande:2022lyl}.
Notably, NSIs play a significant role in the flavor transitions of atmospheric neutrinos, owing to their high energies~\cite{Bakhti:2022axo}, and also affect solar neutrinos because of the high matter densities in the Sun~\cite{Bakhti:2020hbz}.
Neutrinos also play a critical role in both the cooling and explosion mechanisms of core-collapse supernovae. The production of neutrino-philic BSM particles in such environments can significantly alter the expected neutrino signatures, offering a potential window into new physics in future observations~\cite{Akita:2023iwq}.

Neutrino detectors play a vital role in the broader campaign to identify DM. For example, Galactic and extragalactic DM may annihilate or decay directly into neutrinos, producing a flux of mono-energetic neutrinos with energies equal to the DM mass (in the case of annihilation) or half the mass (in the case of decay). Neutrino detectors currently lead in sensitivity for probing such processes across a wide energy range---from the MeV scale to beyond the PeV scale~\cite{Arguelles:2019ouk,Miranda:2022kzs,Rott:2014kfa,IceCube:2023ies}---despite the fact that higher-order corrections can introduce accompanying electromagnetic particles along with the neutrino signal~\cite{Kachelriess:2007aj,Bell:2008ey}.

The second direction of searching for DM in neutrino experiments is observing the neutrino signals from the celestial objects such as our Sun and Earth, where DM can be captured and settle at the core after repeated elastic scattering~\cite{Press:1985ug,Krauss:1985ks}.
Eventually, a substantial population of DM particles could build up at the core, producing a neutrino flux from their annihilation.
These have enabled neutrino detectors to probe DM-nucleon scattering cross sections in a manner complementary to direct detection experiments. In particular, for spin-dependent interactions and certain annihilation channels, the most stringent constraints are currently set by neutrino experiments such as IceCube~\cite{IceCube:2017oza,IceCube:2021xzo,IceCube:2023cwx} and Super-Kamiokande (SK)~\cite{Super-Kamiokande:2004pou,Super-Kamiokande:2011wjy,Super-Kamiokande:2015xms}.

Another direction of DM probes in neutrino experiments is searching for boosted dark matter (BDM).
Recently, theoretical models involving DM with sufficient boost to attain (semi-)relativistic kinetic energies, well above the detection thresholds of various experiments, have been proposed.
Early studies of BDM focused on multi-component dark sector scenarios, in which a light, subdominant DM component is boosted via annihilation~\cite{Agashe:2014yua,Kong:2014mia,Kim:2016zjx,Giudice:2017zke} where the DM relic densities are obtained by the assisted freeze-out mechanism~\cite{Belanger:2011ww}, or decay~\cite{Bhattacharya:2014yha,Kopp:2015bfa,Heurtier:2019rkz} of a heavier, dominant DM particle with a proper non-thermal production mechanism. Subsequently, other classes of mechanisms have been proposed in which DM is boosted through scattering with high-energy SM particles. These include interactions with cosmic rays~\cite{Yin:2018yjn,Bringmann:2018cvk,Ema:2018bih,Cappiello:2018hsu}, solar electrons~\cite{An:2017ojc}, stellar neutrinos~\cite{Zhang:2020nis,Jho:2021rmn}, diffuse supernova neutrino background~\cite{Das:2021lcr}, core-collapse supernova neutrinos~\cite{Lin:2022dbl,Lin:2024vzy}, and protons in blazar jets~\cite{Wang:2021jic}.
In addition, DM produced from stellar sources such as core-collapse supernovae~\cite{DeRocco:2019jti} and primordial black holes~\cite{Calabrese:2021src} is also classified as BDM, owing to its similar phenomenological signatures in neutrino experiments.
Building on these theoretical developments, experimental efforts to search for BDM have been undertaken by SK~\cite{Super-Kamiokande:2017dch,Super-Kamiokande:2022ncz} and ICARUS at Gran Sasso~\cite{ICARUS:2024lew}, complemented by the searches in DM direct detection experiments such as COSINE-100~\cite{COSINE-100:2018ged,COSINE-100:2023tcq}, PandaX~\cite{PandaX-II:2021kai,PandaX:2024pme}, CDEX~\cite{CDEX:2022fig,CDEX:2022dda}, NEWSdm~\cite{NEWSdm:2023qyb}, LUX-ZEPLIN~\cite{LZ:2025iaw}, and DarkSide-50~\cite{DarkSide-50:2025olf}, along with phenomenological studies~\cite{Giudice:2017zke,Alhazmi:2020fju,Alhazmi:2025nvt,Fornal:2020npv,Cherry:2015oca}.
Moreover, both experimental and phenomenological efforts to search for BDM are actively being pursued in future neutrino experiments such as DUNE~\cite{Alhazmi:2016qcs,Kim:2016zjx,DUNE:2020fgq,DeRoeck:2020ntj,Dutta:2024kuj}, Hyper-Kamiokande (HK)\cite{Kong:2014mia,Alhazmi:2016qcs,Kim:2016zjx,Kim:2020ipj,Dutta:2024kuj,Choi:2024ism}, JUNO~\cite{Dutta:2024kuj}, and the Short-Baseline Neutrino Program~\cite{Kim:2018veo}, as well as in currently operating experiments such as ProtoDUNE~\cite{Chatterjee:2018mej,Kim:2018veo,Coloma:2023adi}, KamLAND~\cite{Cappiello:2019qsw}, and IceCube~\cite{Kong:2014mia,Kim:2020ipj,Guo:2020drq,Cappiello:2024acu}.

In analogy with the first discoveries of previously unknown Standard Model particles, it is conceivable that BDM and other BSM particles are likewise being produced in the Earth’s atmosphere via cosmic-ray collisions in the atmosphere~\cite{Alvey:2019zaa,Arguelles:2022fqq}.
Neutrino experiments provide sensitivity to a variety of these hypothetical particles, including Milli-Charged Particles (MCP)~\cite{Plestid:2020kdm,Wu:2024iqm}, heavy neutral leptons (HNLs)~\cite{Coloma:2019htx,Atkinson:2021rnp,BookMotzkin:2024qgd}, and magnetic monopoles~\cite{Iguro:2021xsu,IceCube:2021eye,Candela:2025gwp}. In addition, IceCube recently reported a search for quantum gravity–induced decoherence using atmospheric neutrinos~\cite{ICECUBE:2023gdv}. This approach probes possible interactions between neutrinos and a fluctuating spacetime background, which can lead to decoherence of neutrino oscillations through non-unitary time evolution.


\subsection{Executive Summary of the Landscape}

\begin{table*}[t]
  \centering
  \setlength{\tabcolsep}{5pt}
  \renewcommand{\arraystretch}{1.1}
  \small
  \resizebox{\textwidth}{!}{
  \begin{tabular}{l l l}
    \toprule
    \textbf{Experiment} &  \textbf{Channel / Special feature} & \textbf{Detector technology} \\
    \hline 
    \multicolumn{3}{c}{\textbf{Accelerator-based}} \\[2pt]
    {\color{blue} DUNE–ND/FD} (LBNF) & $\nu$ CC/NC,  $\nu$–$e$; long-baseline & ND-complex, LArTPC \\
    T2K / T2HK {\color{blue} (HK FD)} & $\nu$ CC/NC,  $\nu$–$e$; long-baseline & ND-complex, Water Cherenkov (WC) \\
    COHERENT & CE$\nu$NS; $\pi$DAR (pulsed beam) & CsI / LAr / NaI  \\
    CCM & CE$\nu$NS; $\pi$DAR (pulsed beam) & LAr scintillator \\
    JSNS$^2$, JSNS$^2$-II & CE$\nu$NS; $\pi$DAR (pulsed beam) & Gd-LS  \\    
    FASER$\nu$ / SND@LHC (/ SND@HL-LHC) & $\nu$ CC/NC, $\nu$–$e$; (HL-)LHC forward & Emulsion (/ Tracker)\\
    FPF: FASER$\nu$2/FLArE/FORMOSA & $\nu$ CC/NC, $\nu$–$e$; HL-LHC forward & Emulsion/LArTPC/PS \\  
    SHiP     & $\nu$ CC/NC, $\nu$–$e$; SPS beam dump & Emulsion/ECAL/Tracker \\ 
    {\color{blue} IsoDAR@Yemilab} & IBD, $\nu$–$e$ ; underground $p$ beam & Slow LS (Cherenkov sep.)\\ 
    {\color{blue} $e$-beam dump@Yemilab} & ALP/$A'$ $\to \gamma\gamma, e^+e^-$, $\nu$–$e$; underground $e$ beam & Slow LS (Cherenkov sep.) \\  
    DAMSA / ProtoDAMSA & ALP/$A'$ $\to \gamma\gamma, e^+e^-$; $p$ or $e$ beam & Calorimeter + tracker \\
    \\
    \midrule
    \multicolumn{3}{c}    
    {\textbf{Reactor-based}} \\[2pt]
    NEOS / RENE & IBD, {\color{red} $\nu$-${}^{13}$C} & LS \\
    NEON & CE$\nu$NS & NaI \\
    KamLAND & IBD & LS \\
    {\color{blue} JUNO}  & IBD, {\color{red} $\nu$-${}^{13}$C}; long-baseline & LS \\
    JUNO–TAO   & IBD, {\color{red} $\nu$-${}^{13}$C}; short-baseline & Gd-LS (+ high-res SiPMs)  \\ 
    {\color{blue} $\nu$EYE@Yemilab} & IBD, $\nu$–$e$, {\color{red} $\nu$-${}^{13}$C}; long-baseline & Slow LS (Cherenkov sep.) \\   \\
    \midrule
    \multicolumn{3}{c}{ \textbf{Rare decay: $0\nu\beta\beta$}} \\[2pt]
    AMoRE-II & $0\nu\beta\beta$ decay of ${}^{100}$Mo & Low temperature bolometer \\    
    KamLAND(2)–Zen & $0\nu\beta\beta$ decay of ${}^{136}$Xe & Xe gas loaded LS  \\ \\
    \midrule
    \multicolumn{3}{c}{\textbf{Observatory-scale (atmospheric / astro / solar)}} \\[2pt]
    {\color{blue} DUNE–FD} & $\nu$ CC/NC,  $\nu$–$e$, $\nu$-${}^{40}$Ar 
    & LArTPC \\
    Super-/Hyper-Kamiokande & IBD, $\nu$ CC/NC, $
    \nu$-$e$, $\nu$-${}^{16}$O & WC (or WC-Gd for SK) \\
    IceCube / Gen2      &  $\nu$ CC/NC & Ice Cherenkov array \\
    {\color{blue} JUNO}    & IBD, $\nu$ CC/NC, $\nu$-$e$, {\color{red} $\nu$-${}^{13}$C} & LS \\    
    {\color{blue} $\nu$EYE@Yemilab} & IBD, $\nu$ CC/NC, $\nu$–$e$, {\color{red} $\nu$-${}^{13}$C} & Slow LS (Cherenkov sep.) \\
    \hline
  \end{tabular}
  }
  \caption{Unified overview of major neutrino experiments discussed in this paper, which are grouped by experimental family: accelerator-based, reactor-based, rare decay, and observatory-scale experiments. 
  Each row lists the \textbf{Experiment}, its detection \textbf{Channel} of neutrinos (mostly) along with {\bf Special feature}, and the corresponding \textbf{Detector technology}. 
  Experiments highlighted in blue indicate multi-purpose experiments or those employing multi-purpose detectors, while channels highlighted in red denote recently proposed new detection channel of reactor and solar neutrinos in liquid scintillator detectors, $\nu (\bar \nu) - {}^{13}$C inducing a signal of sharp $\gamma$-ray peak around 3.685 MeV by the de-excitation ${}^{13}$C$^\ast \to {}^{13}$C + $\gamma$.}
  \label{tab:unified_exp_table}
\end{table*}

\begin{table*}[t]
\centering
\begin{ruledtabular}
\begin{tabular}{lccccccccc}
\multicolumn{1}{c}{\textbf{Facility}} & \textbf{$\nu_s$} & \textbf{NSI} & \textbf{SNI} & \textbf{LDM} & \textbf{BDM} & \textbf{ALP} & \textbf{HNL} & \textbf{MCP} & \textbf{$\gamma'$ ($\to$ SM)} \\
\hline
{\color{blue} DUNE--LBNF (ND/FD)}     & \Strong   & \Strong   & \Strong  & \Strong & \NA & \Strong & \Strong  & \Medium & \Strong \\
{\color{blue} T2K/ND280}     & \Strong & \Medium & \Limited & \Medium & \NA & \Limited   & \Strong  & \Medium & \Medium \\
{\color{blue} T2HK (incl.\ IWCD)}    & \Medium & \Medium   & \Limited & \Medium & \NA & \Limited & \Strong & \Medium  & \Medium \\
COHERENT                  & \Strong & \Strong   & \Limited  & \Strong & \Limited & \Medium  & \Medium  & \Limited  & \Limited \\
CCM                   & \Medium  & \Medium & \Limited & \Strong   & \Limited   & \Strong  & \Medium  & \Limited  & \Limited \\
JSNS$^2$/JSNS$^2$-II & \Strong   & \Medium & \Limited & \Strong & \Limited & \Limited  & \Medium  & SUBMET & \Limited \\
FASER$\nu$/SND@LHC       & \Strong & \Strong & \Strong   & \Strong & \Limited & \Strong  & \Strong & \NA & \Strong \\
FPF (HL-LHC) & \Strong   & \Strong & \Strong   & \Strong & \Limited & \Strong  & \Strong & \Strong  & \Strong \\
SHiP                  & \Strong & \Medium & \Medium & \Strong & \Limited   & \Strong  & \Strong & \Strong & \Strong \\
{\color{blue} IsoDAR@Yemilab}         & \Strong   & \Strong & \Strong  & \Limited & \NA     & \Strong  & \Limited  & \Limited     & \Limited \\
{\color{blue} $e$-beam dump@Yemilab}               & \Limited     & \Limited       & \Limited   & \Strong   & \NA     & \Strong       & \Limited & \Limited & \Strong \\
DAMSA/ProtoDAMSA         & \Limited  & \Limited  & \Limited & \Medium   & \Limited  & \Strong       & \Limited & \Limited  & \Strong \\
NEOS   & \Strong   & \Medium & \Limited  & \Limited  & \Limited       & \Limited       & \Limited      & \Limited      & \Limited \\
RENE     & \Strong   & \Medium & \Limited  & \Limited  & \Limited       & \Limited       & \Limited      & \Limited      & \Limited \\
NEON         & \Limited  & \Limited & \Strong   & \Strong   & \Limited       & \Strong       & \Limited & \Limited      & \Limited \\
JUNO--TAO                             & \Strong  & \Medium  & \Limited  & \Limited  & \Limited       & \Medium       & \Strong      & \Limited      & \Strong \\
AMoRE-II (Yemilab)                   & \Medium       & \Limited       & \Medium       & \NA       & \Limited       & \Strong       & \Limited      & \Limited      & \Limited \\
 KamLAND/KamLAND-Zen                & \Medium & \Medium & \Limited  &  $\chi^-$ TeV  &    \Strong  & \Limited   & \Medium      & \Limited      & \Limited \\ \\
\hline \hline 
{\textbf{Facility}} &
\textbf{$\nu_s$} & \textbf{NSI} & \textbf{SNI} & \textbf{DM-$\nu$} & \textbf{BDM} & \textbf{ALP} & \textbf{HNL} & \textbf{MCP} & \textbf{Monopole} \\
\hline
{\color{blue} DUNE--FD}    & \Limited & \Medium & \Medium & \Strong & \Strong  & \Medium  & \Limited   & \Medium  & \Medium \\
SK  & \Strong  & \Strong & \Medium & \Strong & \Strong  & \Medium   & \Medium & \Medium & \Strong \\
{\color{blue} HK} & \Strong & \Strong   & \Medium & \Strong & \Medium  & \Medium   & \Medium & \Medium & \Medium \\
{\color{blue} JUNO} & \Strong & \Strong & \Limited & \Strong & \Medium  & \Medium & \Strong      & \Medium       & \Limited \\
IceCube/Upgrade/Gen2              & \Limited  & \Strong & \Medium & \Strong  & \Medium  & \Limited   & \Strong & \Medium & \Strong \\
{\color{blue} $\nu$EYE@Yemilab}   & \Strong & \Strong & \Limited   & \Limited  & \Strong    & \Limited   & \Medium & \Medium & \Medium \\

\end{tabular}
\end{ruledtabular}
\caption{Capability matrices for laboratory-based experiments (top) and neutrino observatories or medium baseline reactor experiments (bottom); representative BSM targets as columns with symbols: ``\Strong'' denotes cases where both theoretical proposals and experimental efforts (or at least recognition by an experimental collaboration) exist; ``\Medium'' indicates that theoretical proposals exist but have not yet been recognized by experimental collaborations, or that no dedicated analysis exists despite well-known potential within the community; ``\Limited'' denotes the absence of dedicated phenomenological or experimental studies; and ``\NA'' indicates cases that are technically challenging or not applicable to a given category (e.g., for multi-purpose detectors).
Technically, KamLAND-Zen searched for a charged excited dark sector with a tiny mass splitting ($\mathcal O (10\,{\rm MeV})$) from the TeV scale WIMP, not LDM, which is denoted as `$\chi^-$ TeV'.
The category `DM-$\nu$' indicates the observations of DM annihilation or decay into neutrinos.
The dedicated search for sub‑millicharged particles at J‑PARC is listed under its official name, ``SUBMET''.
Note that additional topics such as neutrino dipole moments and Lorentz violation are not tabulated due to the limit of the table size, although the related discussion is in the main contents later.
}
\label{tab:capability-accelerator} 
\vspace{-0.8em}
\end{table*}

\begin{table}[t]
\centering
\small
\begin{tabular}{p{0.19\columnwidth}|p{0.43\columnwidth}|p{0.17\columnwidth}|p{0.17\columnwidth}}
\hline \hline
\textbf{Experiment} & \textbf{Representative BSM target} & \textbf{Energy scale} & \textbf{Timeline/status} \\ 
\hline 
SK, T2K, NEOS, NEON, KamLAND, IceCube & Existing constraints on sterile $\nu$, NSI, HNLs, ALPs, LDM/BDM, and cosmogenic DM & MeV--PeV & Operating / current data \\ 
\hline 
COHERENT, CCM, JSNS$^2$ & CE$\nu$NS, NSI, LDM, ALPs, HNLs, and timing-correlated dark-sector signals & MeV--GeV & Operating / upgrades \\ 
\hline 
ProtoDAMSA, DAMSA & Prompt ALP and dark-photon decays to $\gamma\gamma$ or $e^+e^-$ in short-baseline beam dumps & MeV--GeV & R\&D / staged demonstrators \\ 
\hline
AMoRE-II, KamLAND2-Zen & Majoron, sterile $\nu$, Lorentz / Pauli exclusion violation, right-handed current, $\nu - \nu$ & MeV & From 2027 \\  
\hline 
DUNE-ND, DUNE-PRISM & Sterile $\nu$, NSI/SNI, HNLs, LDM, and dark-sector scattering or decay signatures & MeV--few GeV & From 2031 / PIP-II LINAC upgrade \\ 
\hline 
T2HK near detectors / IWCD & Sterile $\nu$, LDM, MCPs, and off-axis control of neutrino backgrounds & sub-GeV--few GeV & From late 2027 / planned \\ 
\hline  
SND@LHC & 
\multirow{4}{0.43\columnwidth}{Forward LHC neutrinos, LDM, sterile $\nu$, MCPs, and neutrino-philic mediators} & 
\multirow{4}{0.17\columnwidth}{100 GeV--TeV} & Run 3 (by June 2026) \\
\cline{1-1} \cline{4-4} SND@HL-LHC & & & HL-LHC (from Run 4 after 2030) \\
\cline{1-1} \cline{4-4} FASER$\nu$ & & & Run 3 \& Run 4 (by 2033) \\
\cline{1-1} \cline{4-4} FPF & & & From 2033 \\
\hline 
SHiP & HNLs, dark photons, ALPs, scalar portals, MCPs, and $\nu_\tau$ physics & MeV--few GeV & Future SPS facility from 2031 \\ 
\hline 
IsoDAR and $e$-beam dump at Yemilab & Sterile $\nu$, ALPs, dark photons, LDM, neutrino-philic mediators, and reactor-bump-related channels & MeV--sub-GeV & Proposed / future program \\ 
\hline
RENE & \multirow{3}{0.43\columnwidth}{Reactor anomalies, sterile $\nu$, ALPs, dark photons, and precision solar/reactor BSM probes} & \multirow{3}{0.17\columnwidth}{MeV} & Detector commissioning at CNU \& waiting for the installation approval \\
\cline{1-1} \cline{4-4} JUNO & & & operating \\
\cline{1-1} \cline{4-4} JUNO-TAO & & &  From late 2026 \\ 
\hline 
DUNE-FD & \multirow{4}{0.43\columnwidth}{atmospheric/cosmogenic BSM particles, NSI, HNLs, proton decay, and DM annihilation/decay} & 
\multirow{4}{0.17\columnwidth}{MeV--PeV} & From 2029 \\
\cline{1-1} \cline{4-4} HK & & & From 2028 \\
\cline{1-1} \cline{4-4} $\nu$EYE@Yemilab & & & Proposed \\
\cline{1-1} \cline{4-4} IceCube-Upgrade & & & From 2026 \\
\hline \hline
\end{tabular}
\caption{A compact complementarity map of selected current and future neutrino facilities discussed in this paper. The ``best BSM target'' column indicates representative strengths rather than exclusive physics coverage. Many facilities are sensitive to multiple scenarios. The energy scale and timeline entries are approximate and should be interpreted as broad guideposts.}
\label{tab:complement-map}
\end{table}

For a clearer understanding of the BSM landscape at neutrino experiments, we summarize the key features discussed in the previous subsections for each category of neutrino experiments covered in this paper and present them in two tables.
Table \ref{tab:unified_exp_table} presents a unified overview of major neutrino experiments categorized by experimental family, detailing their detection channels and technologies.
Experiments highlighted in blue indicate multi-purpose experiments or those employing multi-purpose detectors, while the channel highlighted in red denotes the recently proposed new detection channels of reactor and solar neutrinos in liquid scintillator detectors.

Table~\ref{tab:capability-accelerator}, on the other hand, presents capability matrices for laboratory-based experiments and neutrino observatories with respect to various BSM targets. For reference, the level of maturity of each study—ranging from purely theoretical proposals to established experimental efforts—is indicated by distinct symbols: ``\Strong'' denotes cases where both theoretical proposals and experimental efforts (or at least recognition by an experimental collaboration) exist; ``\Medium'' indicates that theoretical proposals exist but have not yet been recognized by experimental collaborations, or that no dedicated analysis exists despite well-known potential within the community; ``\Limited'' denotes the absence of dedicated phenomenological or experimental studies; and ``\NA'' indicates cases that are technically challenging or not applicable to a given category (e.g., for multi-purpose detectors).

While Tables~\ref{tab:unified_exp_table} and~\ref{tab:capability-accelerator} summarize detector technologies and BSM target coverage, respectively, it is useful to further emphasize the complementarity among the facilities discussed in this paper. Table~\ref{tab:complement-map} provides a compact guide to representative BSM strengths, approximate energy scales, and broad timeline or status information. This table is not intended to replace model-dependent sensitivity projections, which depend on production channels, decay or scattering modes, detector geometry, exposure, and background assumptions, but rather to help readers navigate the different roles played by accelerator, reactor, underground, and astrophysical neutrino facilities.

Further details, along with appropriate references, will be discussed in the later sections.

\section{BSM Physics Results in Current Experiments}
\label{sec:current}

Current neutrino experiments are actively searching for signs of BSM physics, including sterile neutrinos, NSIs, and connections to dark sectors. These searches have already placed stringent limits on many theoretical models, while tantalizing anomalies continue to demand more investigation. Due to space limitations, this section highlights only a selection of recent BSM results from currently running neutrino experiments, focusing on the studies discussed in NPN 2024.

\subsection{Super-Kamiokande}
\label{sk}

Super-Kamiokande is a 50-kiloton water Cherenkov detector located 1,000 meters underground, housing a cylindrical tank measuring 39.3 m in diameter and 41.4 m in height. The detector is optically separated into an inner volume viewed by 11,129 20-inch photomultiplier tubes (PMTs) and an outer veto region equipped with 1,885 8-inch PMTs. The experiment has been operational since April 1996.
Using 277 kton$\cdot$yr of SK exposure to $^{8}$B solar neutrinos, the recoil electron spectrum and day-night asymmetry have been analyzed to search for NSIs~\cite{Super-Kamiokande:2022lyl}. No significant signal has been observed for couplings to up and down quarks, leaving room for improvement in future experiments.
Atmospheric neutrinos observed in SK have also been used to test NSI scenarios~\cite{Super-Kamiokande:2011dam,Taani:2020rrm}. Searches for modifications in the $\mu$–$\tau$ and $e$–$\tau$ sectors found SK data to be consistent with the absence of NSI contributions.
Using the atmospheric neutrinos, SK could also probe other effects that modify the oscillatory patterns, such as the Lorentz invariance~\cite{Super-Kamiokande:2014exs} and the existence of a sterile neutrino~\cite{Super-Kamiokande:2014ndf}.

While the detector design of the SK is optimized for neutrino oscillation and proton decay searches, it has also demonstrated exceptional performance as an indirect DM detection experiment. By observing neutrinos produced from DM decay or annihilation---either directly into neutrinos or into other SM particles that yield secondary neutrinos---in the Sun~\cite{Super-Kamiokande:2004pou,Super-Kamiokande:2011wjy,Super-Kamiokande:2015xms}, the Galactic center and halo~\cite{Mijakowski:2016cph}, or the Earth~\cite{Frankiewicz:2015zma}, it has effectively complemented other detection strategies within the thermal WIMP paradigm.
It is also worth noting that there are phenomenological studies on other BSM scenarios such as  SNI~\cite{Das:2022xsz}, HNL~\cite{Coloma:2019htx}, MCP~\cite{Plestid:2020kdm}, and ALPs from the Sun, atmosphere, dark-sector particles, and core collapse supernovae~\cite{Watanabe:2001jm,Cheung:2022umw,Cui:2022owf,Alonso-Gonzalez:2024ems}.

The SK has also pioneered in direct detection of BDM, performing the search for the related signals in the two-component BDM scenario via electron scattering~\cite{Super-Kamiokande:2017dch} and the cosmic-ray BDM scenario via proton scattering~\cite{Super-Kamiokande:2022ncz}.
Furthermore, when BDM scatters off a bound nucleon and ejects the nucleon from oxygen quasi-elastically, the residual nucleus may de-excite by emitting $\gamma$ rays~\cite{Dutta:2024kuj,Choi:2024ism}, while the knocked-out neutron can be captured by ambient nuclei emitting a couple of $\gamma$ rays. By searching for the $\gamma + n$ pair, SK can access weaker coupling scenarios and parameter spaces inaccessible to direct detection or proton scattering experiments, a capability that further extends to the future Hyper-Kamiokande experiment with a gadolinium doping scenario~\cite{Choi:2024ism}.

\subsection{T2K}

The Tokai-to-Kamioka (T2K) experiment \cite{T2K:2011qtm} is a long-baseline neutrino experiment located in Japan, using SK as its far detector. The T2K neutrino beam is produced at the Japan Proton Accelerator Research Complex (J-PARC) by colliding 30 GeV protons on a graphite target. An off-axis near detector, ND280, is located 280 metres from the proton target, which is composed of several sub-detectors.

T2K has searched for HNLs in the mass range of $140$ MeV to $493$ MeV in the Time Projection Chambers (TPCs) of the ND280~\cite{T2K:2019jwa}. Such HNLs can be produced in kaon decays and can decay into two oppositely charged leptons with or without an additional neutrino. No signals of two oppositely charged tracks have been observed. Hence, this search places a competitive constraint on the HNL mixing parameter, e.g., U$_e^2 < 10^{-9}$ at 90 $\%$ C.L. for a mass of 390 MeV.
The collaboration has also searched for light sterile neutrinos~\cite{T2K:2019efw}. Several studies have demonstrated T2K's potential sensitivity to various BSM scenarios other than HNL and sterile neutrinos.
For example, theorists have explored T2K's expected sensitivities to NSIs~\cite{Chatterjee:2020kkm,Majhi:2022wyp}, non-unitary neutrino oscillations~\cite{Miranda:2019ynh}, 
MCP~\cite{Gorbunov:2021jog}, dark photon decaying to the SM particles~\cite{Araki:2023xgb}, and ultralight DM~\cite{Lin:2023xyk}, particularly in the context of addressing the tension with NO$\nu$A results.

\subsection{IceCube}
The IceCube Neutrino Observatory is a Cherenkov radiation instrumentation designed to detect high-energy neutrinos at the South Pole~\cite{IceCube:2016zyt}. Completed in December 2010, it is the world’s largest neutrino detector, covering a cubic kilometer of the Antarctic ice. The observatory features over 5,160 photomultiplier tubes as light sensors, buried between 1,450 and 2,450 meters below the surface. These sensors are deployed on 86 strings, most of which are arranged on a hexagonal grid with 125-meter horizontal spacing. A smaller, more densely packed central array called DeepCore increases sensitivity to lower-energy events. 
The observatory classifies neutrino interactions into three types: tracks, cascades, and double pulses. Tracks are cylindrical and produced by muons, while cascades are spherical and result from electron and low-energy tau neutrino interactions, as well as all neutral-current interactions. Double pulses, which are rare, occur from high-energy tau neutrinos and involve two distinct light pulses corresponding to the production and decay of a tau particle. 

The observatory’s design and technology allow it to effectively study a wide range of neutrino interactions and BSM particles. IceCube is capable of detecting astrophysical neutrinos well beyond the TeV scale, including galactic neutrinos up to 100~TeV~\cite{IceCube:2023ame}, diffuse astrophysical neutrinos up to PeV scale~\cite{Naab:2023xcz}, and neutrinos from a nearby active galaxy such as NGC 1068 up to 10~TeV~\cite{IceCube:2022der}.
In addition, atmospheric neutrinos in the energy range of 0.5--10 TeV are used to search for decoherence of neutrino oscillations through non-unitary time evolution from quantum gravity effects~\cite{ICECUBE:2023gdv}. 

IceCube also provides the most stringent bound on DM decay or annihilation in the Galactic Center for masses in the range $10\,{\rm GeV} \lesssim m_{\rm DM} \lesssim O(10)$ TeV~\cite{IceCube:2023ies,KM3NeT:2021kbg}, in the Sun for masses above $O(10)$ GeV~\cite{IceCube:2021xzo,IceCube:2022wxw}, and in the Earth for masses above 100 GeV~\cite{Renzi:2023pkn}.
For extragalactic DM decaying into neutrinos, IceCube sets the strongest bounds for DM masses at the PeV scale~\cite{IceCube:2023gku}.

\subsection{JSNS$^2$}

J-PARC Sterile Neutrino Search at J-PARC Spallation Neutron Source (JSNS$^2$) is searching for sterile neutrinos as a direct test of the LSND anomaly, employing a gadolinium-doped LS detector at the J-PARC Materials and Life Science Facility in Japan. 
The detector performance was validated during the commissioning run~\cite{JSNS2:2021aee}, followed by a series of physics runs from 2021 to 2024. 
Initial comparisons between the observed data and the anticipated outcomes for one of the four sidebands, which differ from the signal region in either prompt or delayed energy ranges, have demonstrated a strong correlation, with additional sideband data to be released shortly.  The initial results of a direct test, based on the data collected in 2022, were first achieved in 2026, consistent with the expected background~\cite{JSNS2:2026muy}.
Meanwhile, the accidental single-event rates of inverse beta decay, including both prompt and delayed signals, have been evaluated~\cite{JSNS2:2023hxl} through a specialized calibration run that leverages beam timing within a 125-microsecond time window. 
Current efforts are focused on calibrating the detector with a Cf-252 radioactive source, improving pulse shape discrimination to effectively mitigate fast neutron backgrounds, and deepening the understanding of the neutrino beam via \ce{^{12}C(\nu_e, e^-) ^{12}N_{g.s.}} reaction.

\subsection{NEOS}

NEOS, which stands for ``Neutrino Experiment for Oscillation at Short Baseline'', is an experiment designed to investigate the presence of sterile neutrinos using reactor antineutrinos in Yeong-gwang, Korea. The NEOS detector, installed in July 2015, collected data until May 2016, securing 180~days of reactor-on data and 46~days of reactor-off data for analysis. The detector was located approximately 24~meters from the reactor core and detects approximately 2,000 IBD candidates per day. The data did not provide strong evidence for the existence of light sterile neutrinos under the 3+1 neutrino hypothesis, so an exclusion limit curve was presented. A significant portion of the parameter space allowed by RAA, including the best-fit value, was excluded~\cite{PhysRevLett.118.121802}.

Following this, NEOS-II was designed and conducted to investigate the spectral bump observed around 5\,MeV in various reactor experiments and to perform a more detailed search for sterile neutrinos. The refurbished detector was installed in the same location as Phase-I. The experiment, conducted from September 2018 to October 2020, covered a full cycle of reactor operation and included 112 days of reactor-off data collected before and after the cycle. The analysis focuses on two main aspects: one addressing the extraction of yields and spectra for each isotope related to the 5 MeV bump, and the other improving the sterile neutrino search results obtained in NEOS-I. The results, which will be announced soon, are expected to play an important role in resolving the anomalies in reactor neutrinos.

\subsection{NEON}

\begin{figure*}[t]
		\begin{center}
			\begin{tabular}{c c}
      \includegraphics[width=0.485\textwidth]{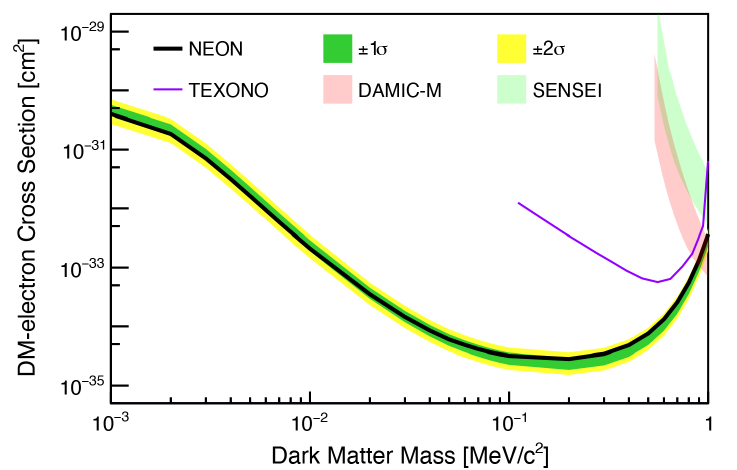} &
      \includegraphics[width=0.515\textwidth]{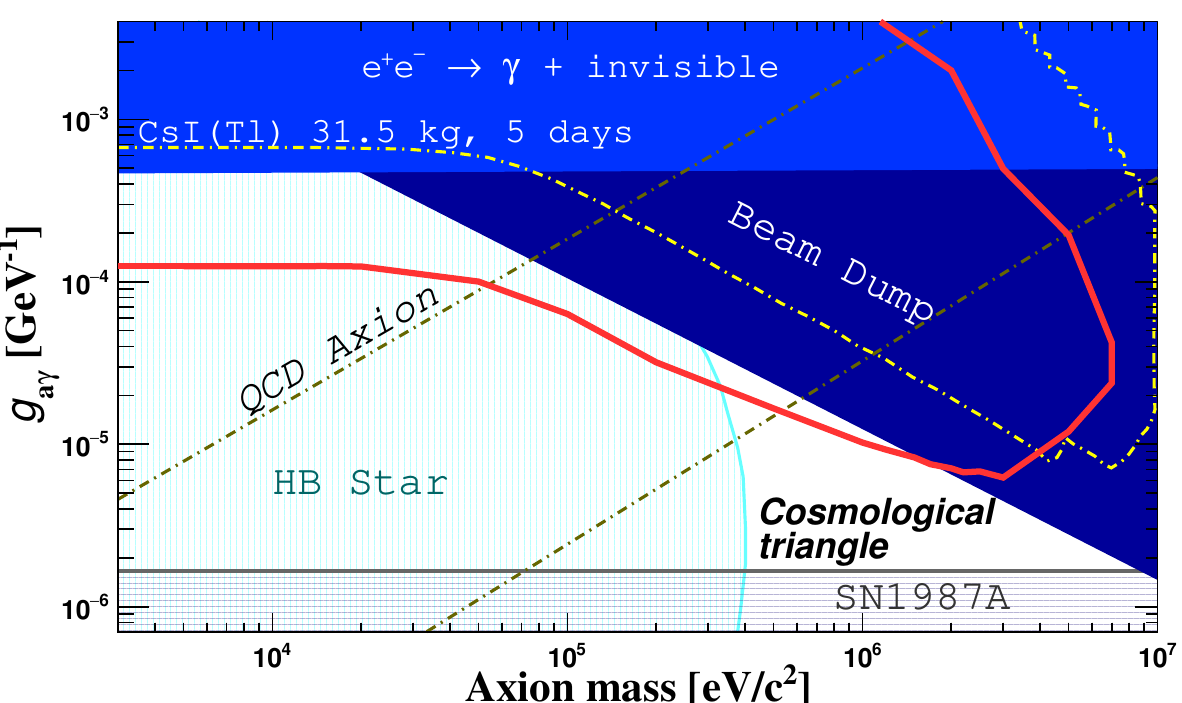}\\
			(a) LDM search & (b) ALP search (Axion-photon coupling) \\
    	\end{tabular}
			\caption{ (a) The observed 90\% C.L. exclusion limit (black solid line) on the LDM-electron scattering cross-section, derived from NEON data, under the assumption $m_{A'}=3m_{\chi}$, where $m_A{'}$ and $m_{\chi}$ are dark photon mass and DM mass, respectively. Figure adapted from Ref.~\cite{NEON:2024bpw}. (b) The observed 95\% C.L. exclusion limit (red solid line) on the ALP-photon coupling ($g_{a\gamma}$), derived from NEON data. Figure adapted from Ref.~\cite{NEON:2024kwv}.
    }
  \label{figure:NEON_result}
	\end{center}
\end{figure*}

The Neutrino Elastic Scattering Observation with NaI (NEON) experiment is designed to detect CE$\nu$NS from reactor antineutrinos~\cite{NEON:2022hbk}. The detector has been operating in the tendon gallery of Hanbit Nuclear Power Plant Unit 6 in Yeong-gwang, Korea, since December 2020. Full physics operations began in April 2022 following an upgrade to the detector encapsulation~\cite{NEON:2024bsc}. Situated just 23.7 meters from a 2.8 GW$_\text{th}$ reactor core, NEON benefits from an intense antineutrino flux, offering a favorable environment for rare event searches.

Physics data acquisition began in April 2022 and has remained stable through June 2024. Notably, the reactor operated at full power continuously—without interruption—between September 2022 and February 2023, enabling high-quality data collection. As of April 2024, NEON has accumulated 523 days of reactor-on and 144 days of reactor-off data. 

While originally designed to observe CE$\nu$NS, NEON also demonstrates that reactor neutrino detectors can serve as sensitive precision probes for bosonic dark sector particles, leveraging the intense photon flux from the reactor core. 
In the dark photon search, the experiment considers the scenario where the dark photon ($A'$) decays invisibly into a pair of DM particles ($\chi$), assuming the mass relation $m_{A'} = 3 m_{\chi}$. The resulting DM particles are then searched for via electron recoils in the NaI target. No significant excess was observed, leading to world-leading constraints on LDM with masses as low as 1 keV~\cite{NEON:2024bpw} as shown in Fig.~\ref{figure:NEON_result} (a).
In the ALP search, both ALP-photon and ALP-electron coupling channels were investigated. ALPs produced through Primakoff and Compton-like processes were considered, with detection relying on either their scattering or decay signatures in the detector. The analysis sets new exclusion limits on ALP couplings in the MeV mass range, probing regions of parameter space overlapping the ``cosmological triangle''---a region largely inaccessible to previous laboratory experiments~\cite{NEON:2024kwv} as shown in Fig.~\ref{figure:NEON_result} (b).
The results highlight the complementarity of low-energy, high-intensity reactor-based experiments in the search for dark sector particles.

\subsection{KamLAND/KamLAND-Zen}

The KamLAND detector, located 2700 m.w.e. underground in the Kamioka mine in Gifu prefecture in Japan, was originally designed to be a multi-purpose neutrino detector sensitive to anti-neutrinos, such as reactor neutrinos and geo-neutrinos.
KamLAND-Zen is a modification of the KamLAND detector to search for the neutrinoless double beta decay (0$\nu\beta\beta$) of $^{136}$Xe. KamLAND-Zen 800 performed the first search assuming inverted mass ordering and completed data-taking in January 2024~\cite{KamLAND-Zen:2024eml}.

KamLAND-Zen has also explored new Nambu-Goldstone bosons such as the Majoron, as well as charged excited states of DM.
The massless Majoron was searched for in KamLAND-Zen 400 Phase-I with an exposure of 112.3 days with 125 kg of $^{136}$Xe ~\cite{KamLAND-Zen:2012maj}.
With larger statistics and a tenfold reduction in internal background levels, KamLAND-Zen 800 is expected to enable searches with significantly improved sensitivity.
If the mass difference between the WIMP ($\chi^{0}$) and a negatively charged excited state ($\chi^{-}$) is less than 20 MeV, motivated by the charginos degenerate with neutralinos in supersymmetric dark matter theories, the $\chi^{-}$ can form a stable bound state with nuclei~\cite{An:2012bs}. 
The bound-state lifetimes accessible to the KamLAND-Zen experiment extend up to approximately 0.1 s; however, significantly longer lifetimes are theoretically achievable for the smaller mass gap and can be probed by other experiments, such as EXO-200.
KamLAND-Zen has set the most stringent upper limits on the recombination cross-section times velocity ($\langle\sigma v\rangle$), and the decay-width of $\chi^{-}$ ~\cite{KamLAND-Zen:2024dm}.

In addition, there are phenomenological studies on the possible contributions of  neutrino~\cite{Jana:2024xmc}, NSI~\cite{Dekens:2021qch}, and HNL~\cite{Abdullahi:2022jlv,Lisi:2023amm} to the $0\nu \beta \beta$ data in KamLAND-Zen. 
This opens up extra new opportunities for $0\nu \beta \beta$ experiments such as KamLAND-Zen, which is worth conducting dedicated studies by the experimental collaborations.

\section{Laboratory-Based Next-Generation Neutrino Experiments and Their Prospects}
\label{sec:futurelab}

The past decades have witnessed remarkable successes in neutrino experiments, with many previously elusive parameters measured for the first time. As we enter the next era, neutrino experiments are transitioning into a phase of precision measurements, aiming to probe the neutrino sector with unparalleled accuracy.
To meet this ambitious goal, next-generation neutrino experiments are being designed with unprecedented scale and resolution. 
Due to space limitations, this section presents only a selected list of future laboratory-based experiments and their prospects for searching for BSM signals, as outlined in Sec.~\ref{sec:landscape}, which is by no means exhaustive.

\subsection{Accelerator-based searches}

\subsubsection{DAMSA}

DArk Messenger Searches at an Accelerator (DAMSA)~\cite{Jang:2022tsp,DAMSA:2026fiw,Bhattarai:2026wxm} is a very short baseline, table-top scale beam dump experiment, aimed to probe the parameter space inaccessible in previous beam dump experiments.
Given the proximity to the beam dump, the primary background comes from the large number of beam-related neutrons (BRN) from the beam interactions in the dump.
In order to overcome these backgrounds, DAMSA aims to detect two-photon or $e^{+}e^{-}$ final states, resulting from the decays of dark sector particles. 

\begin{figure}[t]
    \centering
    \includegraphics[width=\textwidth,height=0.3\textwidth]{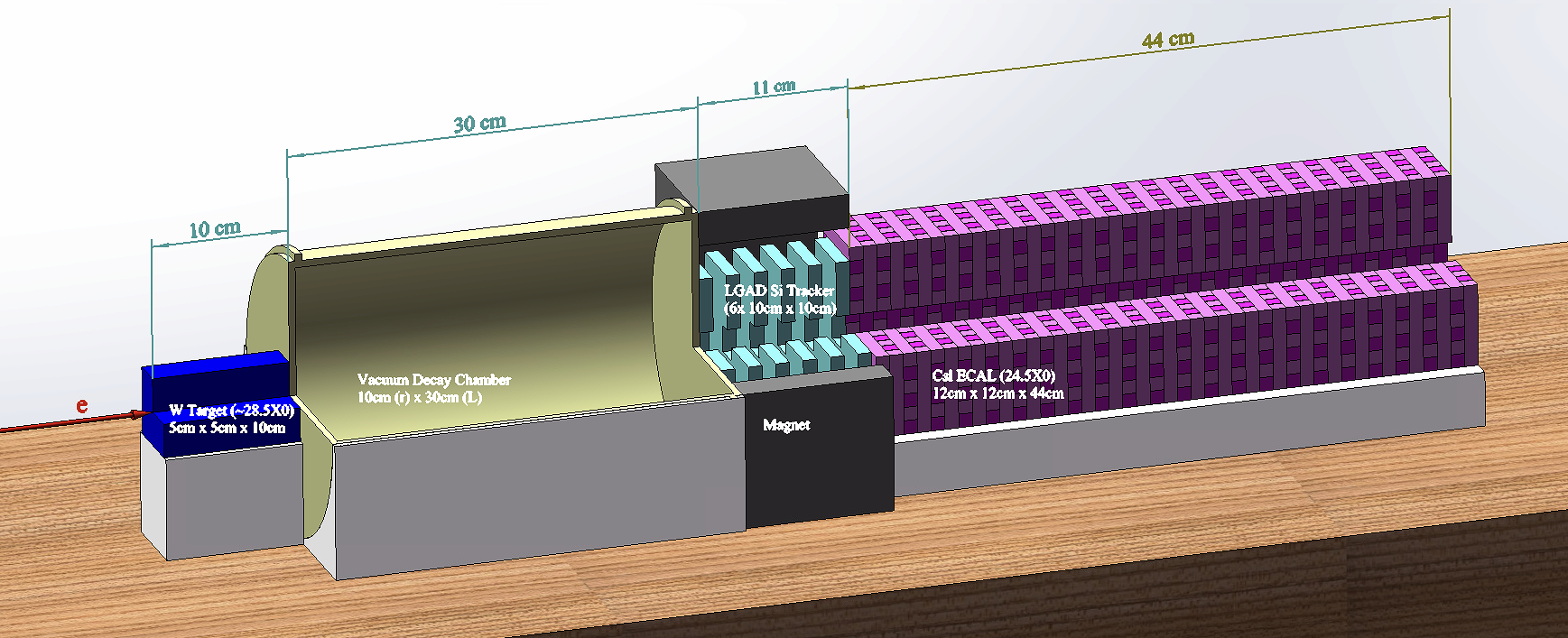}
    \caption{A 3D model of a table top-scale DAMSA detector at the SLAC LESA 8~GeV electron beams}
    \label{fig:little-damsa-det}
\end{figure}

Based on detailed GEANT4~\cite{GEANT4:2002zbu} studies, in order to mitigate the backgrounds from accidental overlaps of the photons from BRN spallation interactions by 10 orders of magnitude, the following detector capabilities are essential:
\begin{itemize}
\item Fine granular calorimeter with high precision shower position resolution to identify the decay vertex better than 1~cm 
\item Excellent timing resolution of the detector for sub-ns arrival time differences
\item As fine a mass resolution as possible, better than few MeV level
\item  Precision tracking system under a magnetic field to distinguish charged particles from photons, identify the sign of the charged particle, and measure momenta
\end{itemize}
Based on the above detector requirements and the subsequent detailed studies on the limitations in Ref.~\cite{Kim:2024vxg}, the DAMSA collaboration is planning on constructing a compact detector as shown in Fig.~\ref{fig:little-damsa-det}, named the Little DAMSA Pilot Project (LDPP).
The collaboration has established the following strategic staged approach:
\begin{itemize}
    \item {\bf Stage 0: Electron beam background validation} $\rightarrow$ This is to measure and validate GEANT4 simulation results of the neutron and photon background flux at the electron beams.
    \item {\bf Stage 1: The $a\rightarrow \gamma\gamma$ demonstrator at an electron beam} $\rightarrow$ This is to build a demonstrator experiment with only the tungsten target, vacuum decay chamber, and an ECAL.  While CsI ECAL would be ideal, this experiment could use an ECAL made of plastic scintillation counters.  A charged particle veto counter needs to be employed.  The likely candidate facility is the LESA facility at SLAC with the 8~GeV electron beams.
    \item {\bf Stage 2: The $a\rightarrow \gamma\gamma/ e^{+}e^{-}$ demonstrator at an electron beam} $\rightarrow$ The detector in Stage 1 demonstrator will be enhanced with the Si tracker (and magnet if needed) to demonstrate the expanded signal capture to charged particle final states.
    \item {\bf Stage 3: Proton beam background validation} $\rightarrow$ This is to measure and validate GEANT4 simulation results of the neutron and photon background fluxes at proton beams.
    \item {\bf Stage 4: Full scale DAMSA experiment at the proton beams} $\rightarrow$ The full scale DAMSA experiment with proper beam dumps at a proton beam facility, such as that the CERN's beam dump facility in which the SHiP experiment plans to commission in 2032.  The collaboration has submitted a white paper~\cite{damsa-ship} to the recently completed European Particle Physics Strategy update.
\end{itemize}
Based on both experimental results and simulation studies, the validation effort using the 2 GeV mixed beam (approximately 80\% electrons and 20\% charged pions) at the Fermilab Test Beam Facility showed that no significant neutron production was observed for either the 4 mm or 16 mm tungsten target configurations.

The original DAMSA proposal~\cite{Jang:2022tsp} considered the search for ALPs interacting with SM photons as a benchmark physics case. 
They envisioned the situation where a high-intensity 600~MeV proton beam at the RAON facility in Korea ($1.4\times 10^{23}$ protons on target) impinges on a 1-meter-long tungsten target, creating a number of photons via $\pi^0$ decays, electromagnetic cascade showers of secondary electrons, etc., inside the target. 
A photon then may convert to an ALP through the Primakoff process and decay into a couple of photons inside the 10-meter-long vacuum decay chamber. 
The electromagnetic calorimeter placed downstream of the decay chamber is responsible for detecting the two photons. 

\begin{figure}[t]
    \centering
    \includegraphics[width=0.48\textwidth]{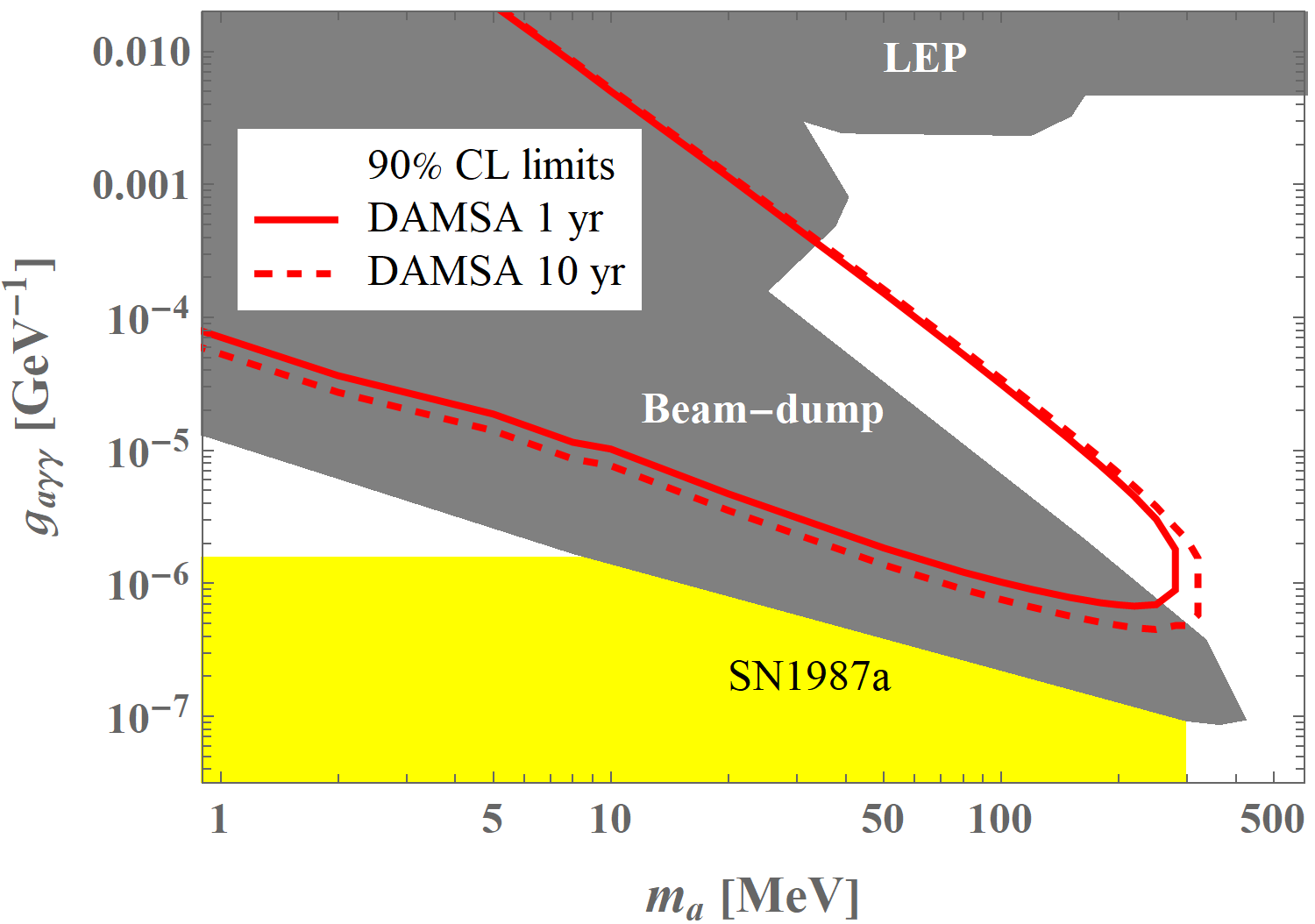}
    \includegraphics[width=0.48\textwidth]{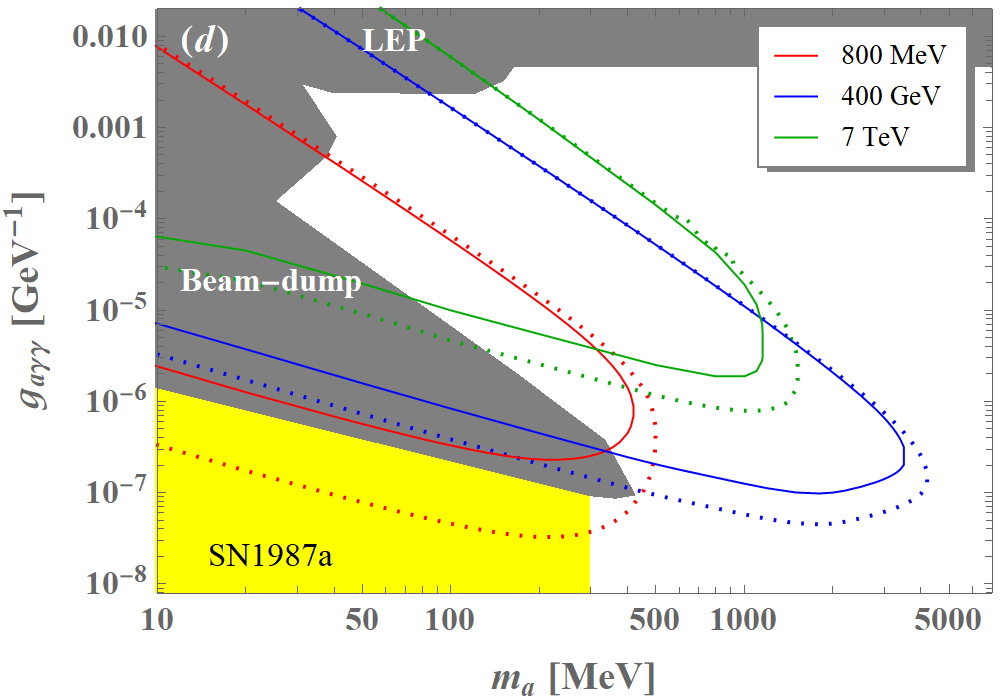}
    \caption{Left: 90\% C.L. expected sensitivity reaches for the ALP-photon coupling parameter $g_{a\gamma\gamma}$ in ALP mass $m_a$ at the RAON facility. Plots taken from Ref.~\cite{Jang:2022tsp}. Right: 90\% C.L. expected sensitivity reaches at PIP-II at Fermilab and SPS, and LHC-dump at CERN. Plots taken from Ref.~\cite{Kim:2024vxg}. }
    \label{fig:DAMSAsensitivity}
\end{figure}

The dominant background to this ALP signal is identified as the accidental di-photon signal made by the BRNs.  
A set of cuts~\cite{Jang:2022tsp}, including invariant mass window cuts, is applied to suppress backgrounds sufficiently while keeping the ALP signal events nearly intact. 
The 90\% C.L. expected sensitivity reaches for the ALP-photon coupling parameter are shown in the left panel of Fig.~\ref{fig:DAMSAsensitivity}. 
They found that with 1-year exposure, new regions of parameter space beyond the existing (beam-dump) limits can be explored, especially toward the ``prompt-decay'' region where most of the produced ALPs are likely to decay inside the target. 

The DAMSA program is further extended to other accelerator facilities with higher-energy particle beams, including PIP-II at Fermilab (800 MeV proton beam), SLAC LSEA (8~GeV electron beams), the SPS at CERN (400 GeV proton beam), and the LHC dump at CERN (7 TeV proton beam). A recent study~\cite{Kim:2024vxg} indicates that exploring the prompt-decay region does not require high beam intensity or wide angular detector coverage. As a result, a large-scale detector is not necessary; a ``tabletop-sized'' detector is sufficient to meet the experimental objectives.
Moreover, if the instantaneous beam intensity is low and the beam separation is greater than $10~\mu~sec$ the aforementioned reducible BRN-initiated backgrounds can be significantly suppressed to a manageable level. The right panel of Fig.~\ref{fig:DAMSAsensitivity} shows the 90\% C.L. expected sensitivity reaches at proton beam facilities, PIP-II, SPS, and the LHC-dump, assuming a meter-scale baseline and a 0.05 rad angular coverage. Notably, comparing the solid lines (3-month exposure) with the dotted lines (1-year exposure), the results demonstrate that even a short beam exposure is sufficient to probe the prompt-decay region nearly to its fullest extent.

\begin{figure}[t]
    \centering
    \includegraphics[width=0.48\linewidth]{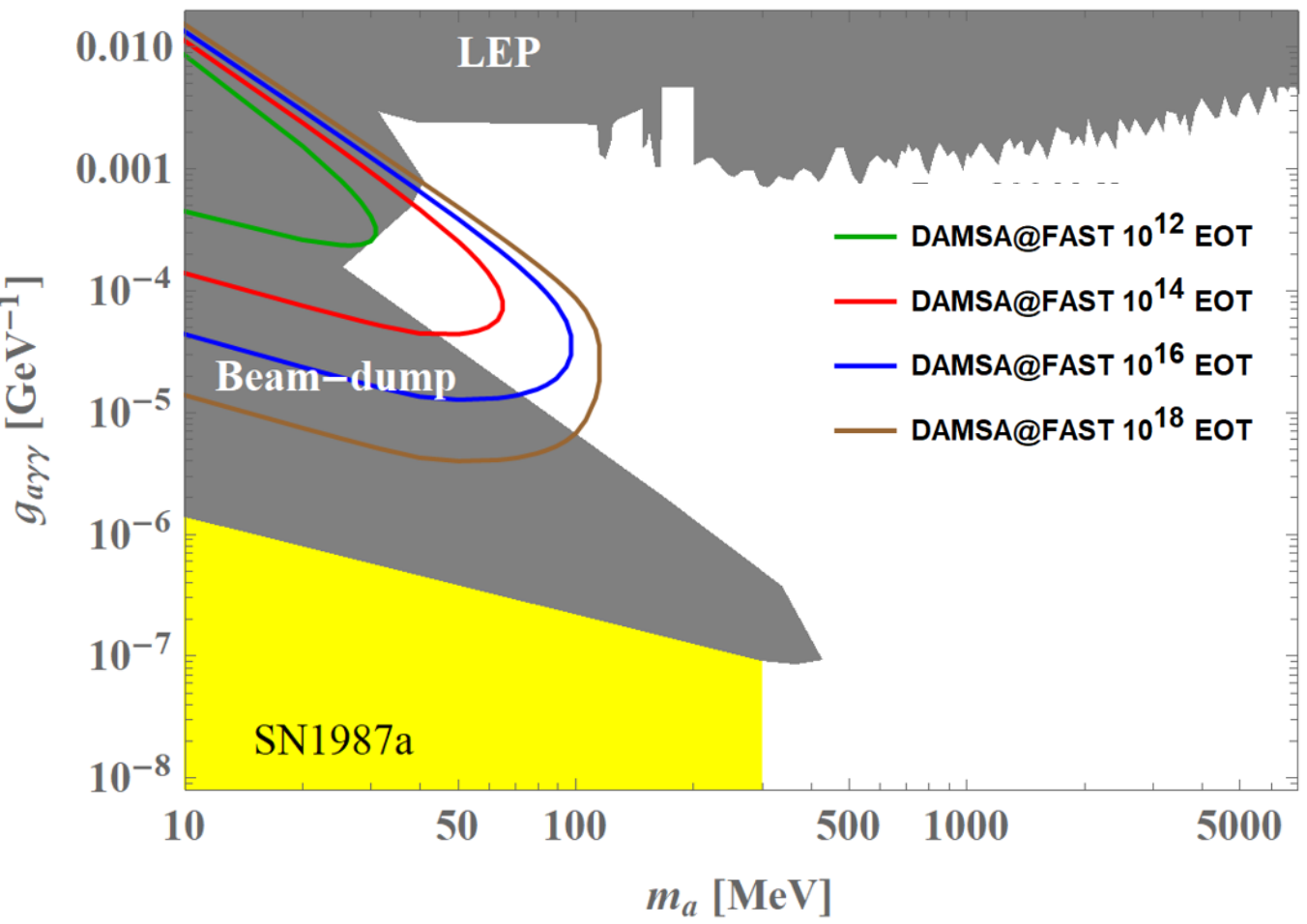}
    \includegraphics[width=0.48\linewidth]{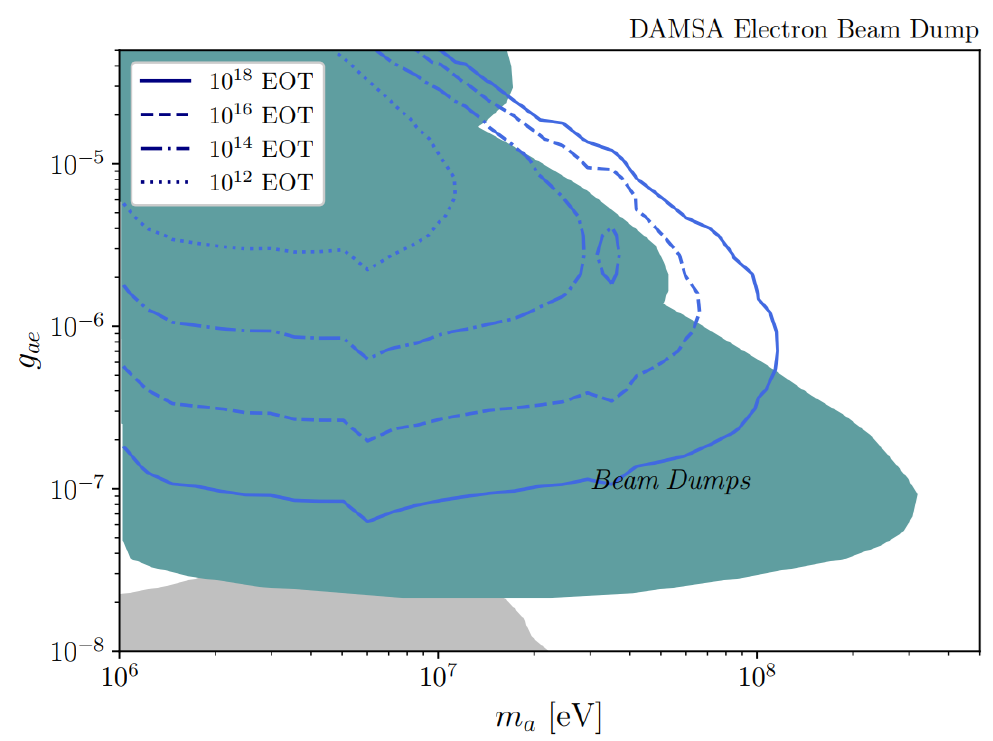} \\
    \includegraphics[width=0.48\linewidth]{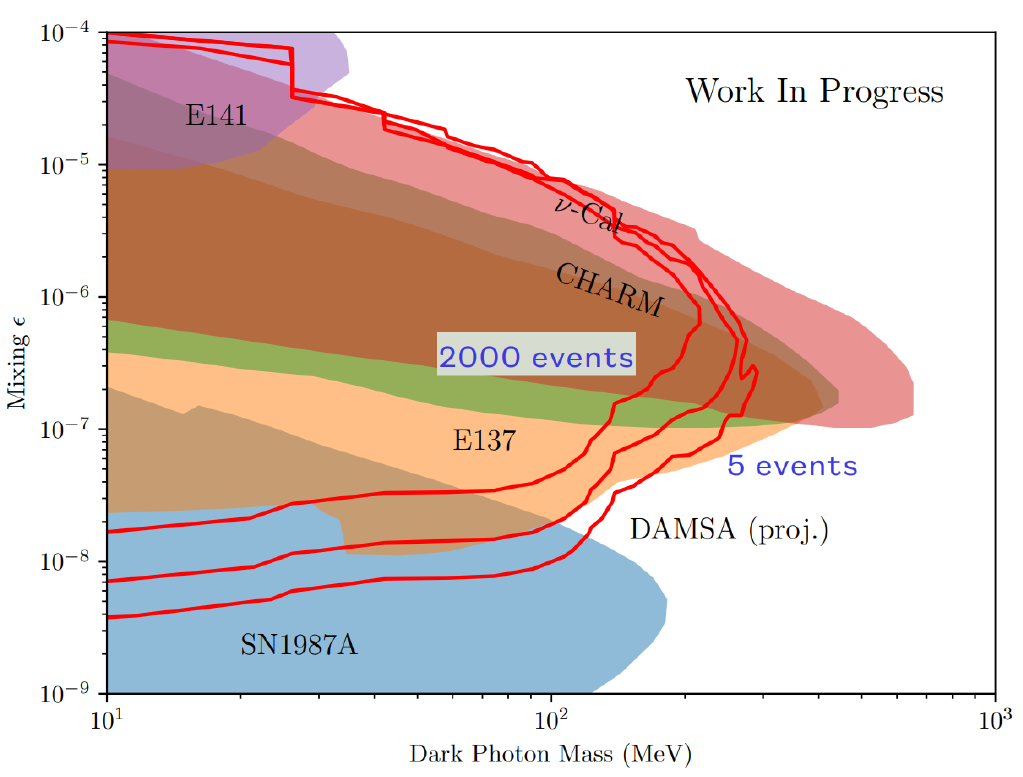}
    \caption{Top: 90\% C.L. expected sensitivity estimates with the 300 MeV FAST electron beam at Fermilab for the ALPs interacting with SM photons (left) and electrons (right). Bottom: 90\% C.L. expected sensitivity estimates at PIP-II for dark photons.}
    \label{fig:protodamsa-sense}
\end{figure}

Since DAMSA experiments operate with a short baseline, their success depends critically on understanding BRN-induced backgrounds. To address this, the ProtoDAMSA project has been launched, utilizing a 300~MeV FAST electron beam at Fermilab. Due to the beam's leptonic nature and relatively low energy, BRNs are expected to be more manageable, providing a controlled environment for background studies. Preliminary sensitivity estimates, assuming the previously described tabletop-scale detector, are shown in the left panel of Fig.~\ref{fig:protodamsa-sense}.

In addition to ALPs interacting with photons, DAMSA is also expected to be sensitive to ALPs coupling to electrons. In this case, ALPs can be produced in the target by photons and electrons through various processes, including Compton-like scattering and subsequent decays into electron-positron pairs. Preliminary sensitivity estimates with the 300~MeV FAST electron beam are shown in the top-right panel of Fig.~\ref{fig:protodamsa-sense}. Similarly, dark photons ($A'$) produced inside the beam target can be probed via their decay into $e^+e^-$ pairs. Preliminary sensitivity projections for dark photons at PIP-II are shown in the bottom panel of Fig.~\ref{fig:protodamsa-sense}.
At the time of writing this report, a serious discussion is ongoing to utilize the flexible 8~GeV electron beams at SLAC's LESA facility for DAMSA.
LESA is expected to start operation in late calendar year 2027 and is to accommodate the LDMX experiment which is in preparation.

\subsubsection{DUNE-ND}

The Deep Underground Neutrino Experiment (DUNE)~\cite{DUNE:2020lwj} is a next-generation long-baseline neutrino experiment that utilizes high flux neutrino beams resulting from the high intensity proton beams with 1.2~MW initial power ultimately reaching 2.4~MW. 
DUNE detector consists of a powerful near detector (ND) complex located 547~m downstream of the neutrino production target and the far detector (FD) 1,300~km away in a newly excavated cavern 1,500~m underground at the Sanford Underground Research Facility (SURF)~\cite{SURF}.
DUNE is expected to use the standard low energy (LE) tuned flux (having a peak around $2-3$ GeV and sharply falling at energies $E \gtrsim 4$ GeV) with a total runtime of 13 years distributed equally between the $\nu$ and $\bar{\nu}$ modes (6.5 years + 6.5 years) (see the DUNE Technical Design Report or TDR~\cite{DUNE:2020ypp, DUNE:2021cuw} for more details).

DUNE-ND complex consists of a Liquid Argon Time Projection Chamber detector (LArTPC), identical technology as the FD to control the systematic uncertainty from the neutrino flux, followed by a The Muon Spectrometer (TMS) to tag the muons resulting from the neutrino interactions in the LArTPC, in Phase-I.  The TMS is potentially to be replaced with a high-pressure (10~atm) gas argon TPC augmented with a precision electromagnetic calorimeter utilizing a 0.5~T magnetic field in Phase-II for precision tracking, EM particle energy, and muon momentum measurements.
The two upstream detectors are movable from 0~m on axis to a maximum 28.5~m off axis (53~mrad or $3^{\circ}$ off axis), acting as a prism to provide additional information on the neutrino flux.
The two upstream detectors are followed by the System for on-Axis Neutrino Detection (SAND), which consists of a straw-tube detector, a crystal EM calorimeter under a 0.6~T magnetic field.
SAND is responsible for a constant on-axis neutrino beam monitoring and capable for physics.

The neutrino beam with 1.2~MW starting beam power in the Long Baseline Neutrino Facility (LBNF) is expected to be completed by 2031. 
The construction of PIP-II Linac, an essential element for the DUNE neutrino beams is progressing quickly, in preparation for the Fermilab shutdown for the LBNF beamline switchover in 2027.

DUNE-ND not only serves to measure the neutrino flux at the source and neutrino cross sections, but also provides a unique opportunity to study new physics, such as NSI~\cite{Bakhti:2016gic},  SNI~\cite{Bakhti:2018avv}, heavy neutral leptons~\cite{Abdullahi:2023gdj}, eV scale sterile neutrino~\cite{Gandhi:2015xza, Dutta:2016glq}, LDM~\cite{DeRomeri:2019kic}, and dark photon decaying to the SM particles~\cite{Berryman:2019dme} thanks to the incredible level of precisions.
The expected sensitivity on CC NSI at DUNE-ND is expected to be an order of magnitude stronger than current limits. 
For SNI, the sensitivity at DUNE-ND is expected to be comparable to the current constraints from the kaon decay-in-flight experiment NA62.
However, unlike NA62, DUNE-ND has the advantage of flavor sensitivity, allowing it to explore neutrino interactions in a more detailed manner.

\begin{figure}
\centering
\includegraphics[width=0.7\linewidth]{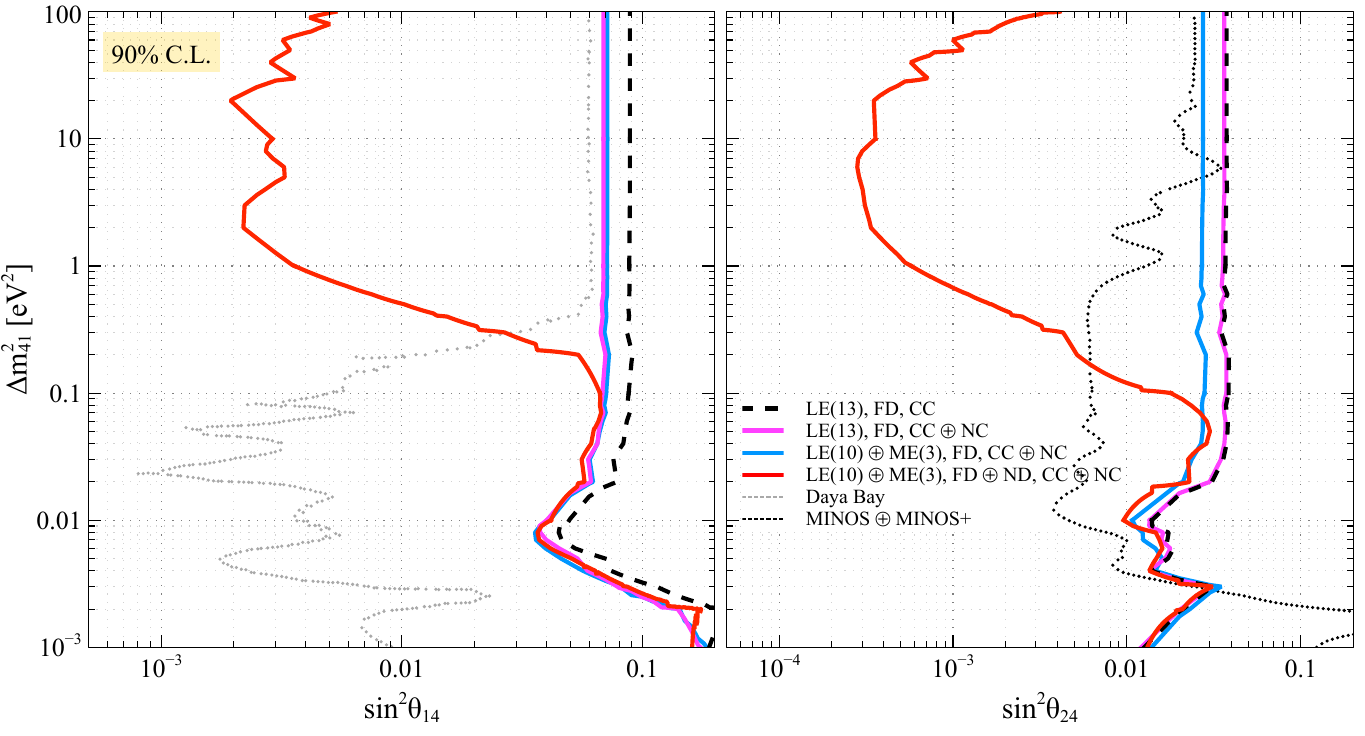}
\caption{\footnotesize{
The contours of $\Delta \chi^{2}$ at $90\%$ C.L. in the $\sin^{2}\theta_{i4}-\Delta m^{2}_{41}$  ($i=1,2$ in the two panels respectively) for various analysis configurations are shown, taken from Ref.~\cite{Parveen:2024bcc}.
The configurations considered are: LE
beam at the FD using the CC interaction of neutrinos (black dashed); LE beam at the FD using the CC and NC interactions (magenta dotted);  
a combination LE beam and ME beam at the FD using CC and NC interactions (blue solid); a combination of LE beam and ME beam at the FD augmented with the corresponding to ND analysis by using both CC and NC interactions of neutrino (red solid) The numbers in parentheses beside each legend indicates the corresponding total runtime (in years) shared 
equally in $\nu$ and $\bar{\nu}$ modes. For the comparison Daya Bay's $90\%$ C.L.\ \cite{DayaBay:2016lkk} in a plane of $\sin^{2}\theta_{14}-\Delta m^{2}_{41}$  (dotted gray in the left panel of figure) and MINOS $\&$ MINOS+ at $90\%$ C.L.\ \cite{MINOS:2017cae} in a plane of $\sin^{2}\theta_{24}-\Delta m^{2}_{41}$  (dotted black in the right panel of figure).
}}
\label{fig:chisq_dm41_thi4}
\end{figure}

For the eV-scale sterile neutrino searches, a combined fit for FD and ND at DUNE using GLoBES~\cite{Huber:2004ka, Huber:2007ji} can be performed.  
Considering the alternative beam tune (low energy or LE, and medium energy or ME beam) and runtime combinations, an improved sensitivity for a wide range of sterile neutrino mass splitting can be obtained~\cite{Parveen:2024bcc}, as displayed in Fig.~\ref{fig:chisq_dm41_thi4}.
Note that, using a shared runtime of LE (10 yrs) and ME beam (3 yrs), rather than using the LE beam alone (13 yrs), can improve the sensitivity, especially to $\sin^{2}\theta_{24}$. 
On the other hand, inclusion of neutral current 
improves the sensitivity to $\sin^{2}\theta_{14}$. 
More crucially, using a combined analysis with both FD and ND can dramatically increase the sensitivities to both the active-sterile mixing angles, especially for $\Delta m^{2}_{41} \gtrsim 0.1 \text{ eV}^{2}$.
Since the oscillation effect due to $\Delta m^{2}_{41} \sim \mathcal{O}(1 \text{ eV}^{2})$ is most pronounced at baselines $\lesssim 1 $ km, the ND can probe the sterile parameters very efficiently. 
In the combined analysis of FD and ND, the sensitivity is expected to reach the parameter region:
\begin{align}
\sin^{2}\theta_{14} &\gtrsim 2 \times 10^{-3}\, (90 \%\, \text{C.L.}) 
\text{ at } \Delta m^{2}_{41} \simeq 11 \text{ eV}^{2}\,, \nonumber \\
{\sin^{2}\theta_{24}} &\gtrsim 3 \times 10^{-4}\, (90 \%\, \text{C.L.})
\text{ at } \Delta m^{2}_{41} \simeq 7 \text{ eV}^{2}.\nonumber
\end{align}

\subsubsection{Forward neutrino experiments at LHC/HL-LHC}

The SND@LHC~\cite{SHiP:2020sos,SNDLHC:2022ihg} and FASER/FASER$\nu$~\cite{FASER:2018bac, FASER:2019aik, FASER:2020gpr} are currently operating at the LHC to measure the process $pp \rightarrow \nu X$ and to search for feebly interacting particles (FIPs) in the forward region. These experiments are located at a distance of 480 m from the ATLAS interaction point on either side, probing the pseudorapidity region $7.2 < \eta < 8.4$ and $\eta \gtrsim 8.5 $, respectively.
The location of SND@LHC is in the TI18 unused LEP transfer tunnel and shielded from collision debris by around 100 m of rock and concrete.

The emulsion-based detectors of the SND@LHC and FASER$\nu$ can identify three species of neutrinos and measure the energy spectrum, which will enable muon and electron neutrino contributions to be disentangled and a first-ever estimate to be made for heavy quark production in the forward region. This measurement will allow the study of quantum chromodynamics effects in an unexplored domain, which has significant implications for the simulation of heavy quarks produced in atmospheric swarms that are initiated by cosmic rays. 
The detector of SND@LHC consists of 830 kg target made of tungsten plates with nuclear emulsions and electronic trackers, followed by a hadronic calorimeter and a muon system. Each unit of  Emulsion Cloud Chamber (ECC) is a sequence of 57 nuclear emulsion films, 19.2 $\times$ 19.2 cm$^2$ and approximately 300 $\mu$m thick, interleaved with 56 tungsten plates, each 1 mm thick.
The high luminosity pp collision at LHC leads to a large neutrino flux, and its high energies (100 GeV to a few TeV) imply relatively large neutrino cross-sections, leading to a significant physics potential with a relatively moderate size. Increasing the distance from the interaction point and from the beam line reduces the backgrounds efficiently.

In addition to the ongoing experiments, 
there are proposals to extend the physics potentials during the high luminosity of the LHC (HL-LHC) stage by building Forward Physics Facility (FPF)~\cite{Anchordoqui:2021ghd, Feng:2022inv,FPF:2025bor} in charge of hosting a variety of upgraded forward experiments: FASER2/FASER$\nu$2 which is a 20-ton emulsion-tungsten detector, Forward Liquid Argon Experiment (FLArE) which is a LArTPC with an active volume of 10 tons, and FORMOSA for MCP search.
In addition, SND@HL-LHC~\cite{SNDLHC:2026why} which has a 5-ton fiducial volume, is proposed as an updated project of SND@LHC. SND@HL-LHC is supposed to have silicon trackers instead of nuclear emulsion in SND@LHC, with the addition of a new magnetic spectrometer to allow separate identification of neutrino and antineutrino interactions for both muon and tau neutrinos. The addition of a magnetic spectrometer would allow the first experimental direct observation and the study of tau antineutrinos, while extending the range of flavor conservation tests. It will also extend the reach for the possible discovery of new exotic phenomena.
With the increase of both luminosity and detector size, the proposed experiments in FPF are able to detect neutrinos with significantly higher statistics for all three flavors, thereby improving our understanding of neutrino properties and increasing the potential for various new physics discoveries. In the following, we introduce a few examples of the potential for BSM search that can be explored using neutrino detectors of the forward experiments.\footnote{See also Ref.~\cite{Anchordoqui:2021ghd} for the BSM scenarios not discussed here.}

\begin{figure}[t]
    \centering
    \includegraphics[width=0.7\textwidth]{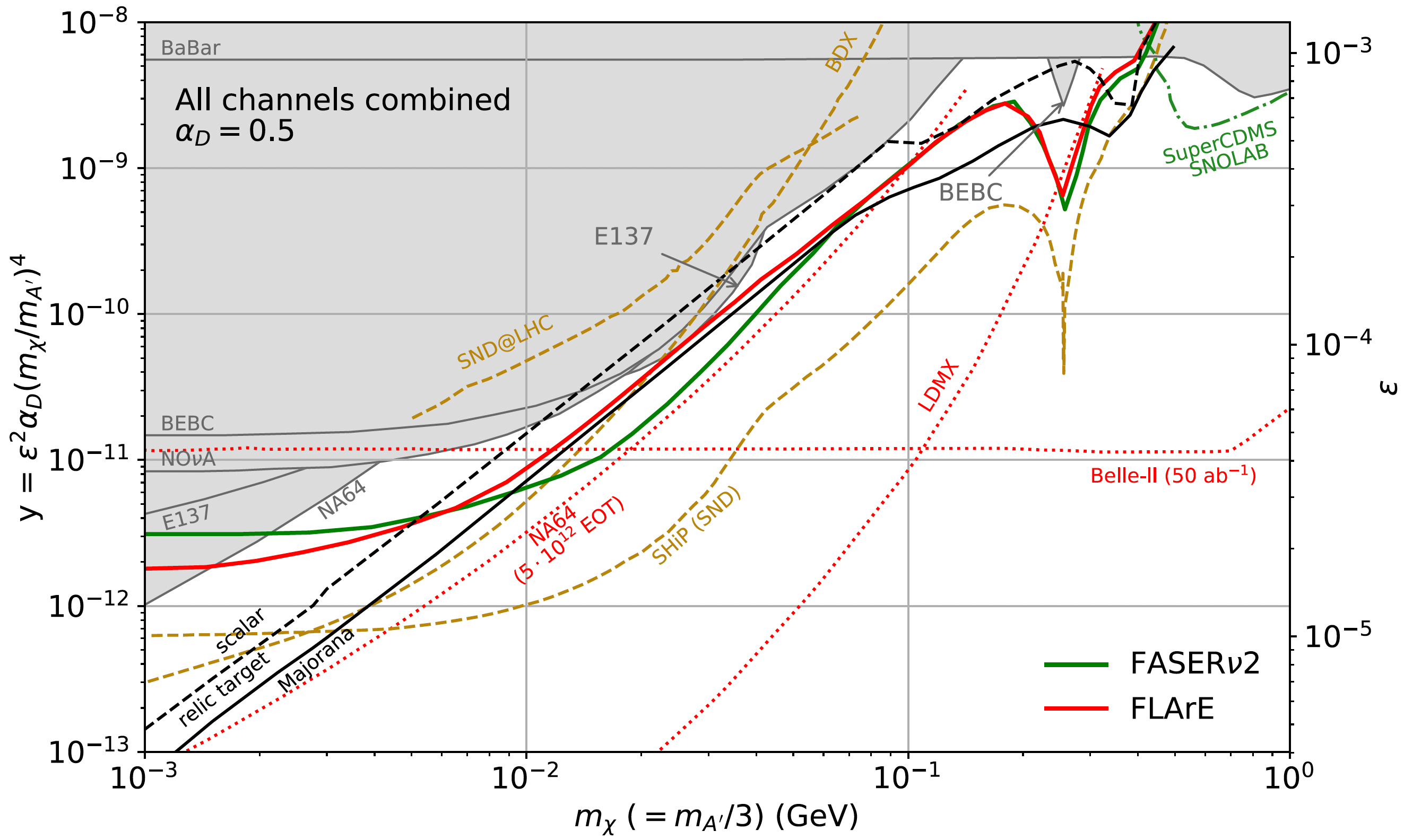}
    \caption{The projected exclusion limits on the DM model mediated by dark photon for two neutrino detectors at the proposed FPF, FASER$\nu$2 (red) and FLArE (green) for HL-LHC era. Figure is taken from Ref.~\cite{Feng:2022inv}}
    \label{fig:LDF_FPF}
\end{figure}

\paragraph{\bf Light dark matter (LDM) search}

produced at the LHC, LDM is predominantly directed to the forward region and can be detected at the neutrino detectors, SND@LHC and FASER$\nu$ in Run-3, as well as those in FPF in the HL-LHC stage.
Potential signatures come from their electron or nuclear scattering with matter of the detectors, similar to neutral current scattering of neutrinos. 
Figure~\ref{fig:LDF_FPF} presents the projected sensitivity reach for the two detectors at the FPF as well as the SND@LHC for a typical DM model mediated by dark photons, kinematically mixed with ordinary photons \cite{Feng:2022inv, Batell:2021aja, Batell:2021blf}. 
The gray region is the parameter space excluded by various experiments \cite{BaBar:2017tiz, BEBCWA66:1986err, Wang:2017tmk, Batell:2014mga, Banerjee:2019pds, deNiverville:2011it, MiniBooNEDM:2018cxm}, and the expected constraints from other future experiments \cite{Belle-II:2018jsg, LDMX:2018cma, Gninenko:2019qiv, SHiP:2020sos, SHiP:2020noy, BDX:2016akw} are shown with dashed curves. 
One can see that the neutrino experiments at the FPF will be able to probe unexplored parameter space, and more interestingly, can cover the region where the thermal relic density meets the observed DM abundance, shown in black lines. 
There are more studies on the LDM search at forward experiments for the different kinds of mediator models \cite{Batell:2021snh, Boyarsky:2021moj,Feng:2022inv}.  

\paragraph{\bf sterile neutrino oscillation}
The LHC produces a vast number of neutrinos in the forward direction over a wide energy range. 
The neutrino detectors of the forward experiments at the LHC can detect those neutrinos mostly in the energies from 100 GeV to $\mathcal{O}(1)$ TeV, and distinguish their flavors, which would make it possible to investigate neutrino oscillations and search for sterile neutrinos. 
Considering the neutrino energies mentioned above and the baseline of 500-600 m, oscillations between active neutrinos are negligible, hence any oscillation signals appearing on the event spectrum would be due to the existence of a sterile neutrino with $\Delta m^2_{41} \sim 1000 {\, \rm eV}^2$ \cite{Bai:2020ukz,Anchordoqui:2021ghd,Feng:2022inv}. 
\begin{figure}[t]
    \centering
    \includegraphics[width=0.45\textwidth]{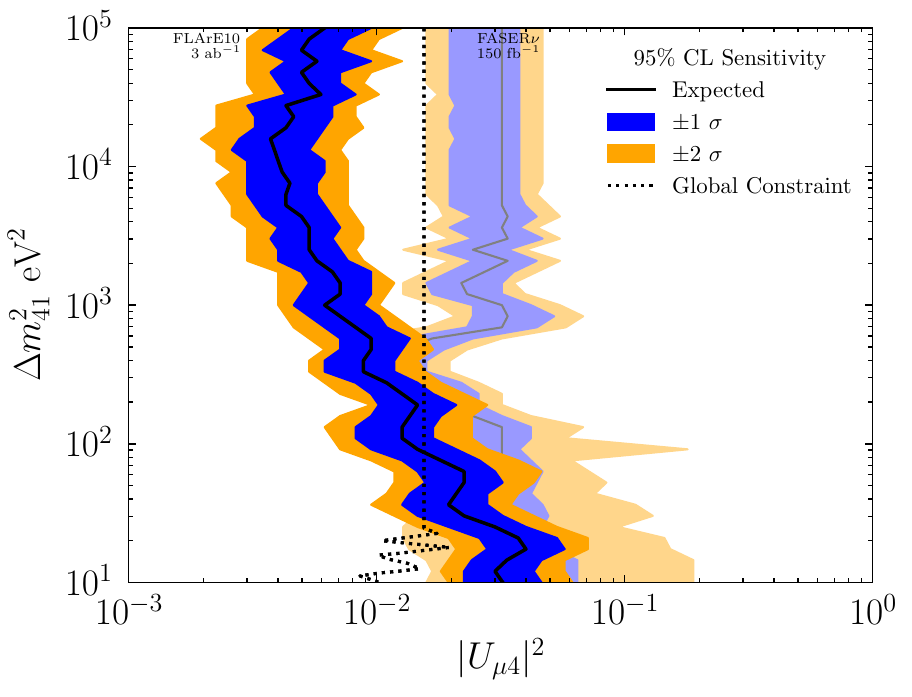}
    \includegraphics[width=0.48\textwidth]{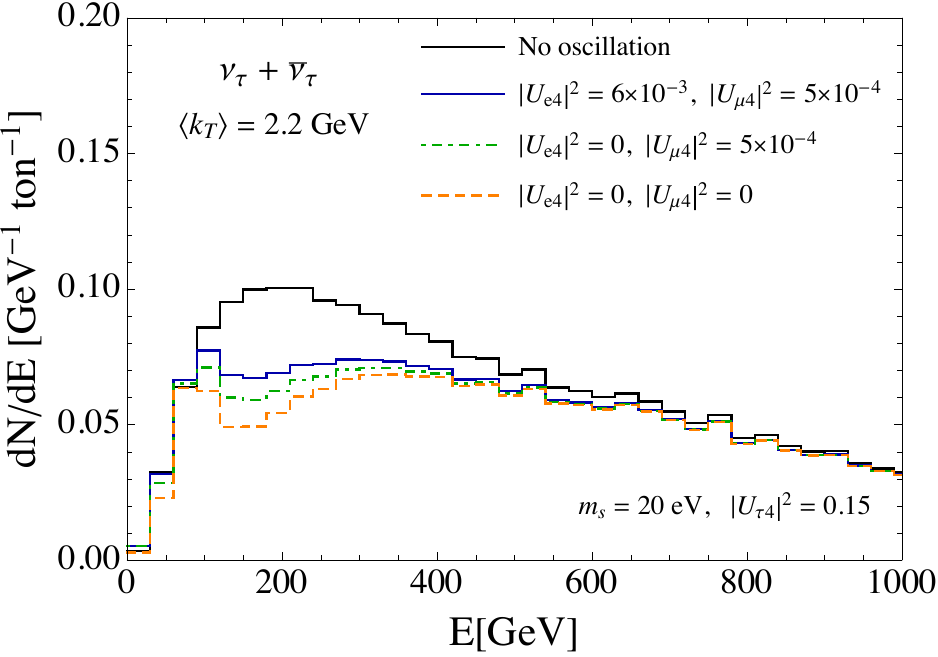}
    \caption{Left: Expected constraints on mixing parameters for muon neutrinos in the presence of a sterile neutrino for FASER$\nu$ in the Run-3 and FLArE in the HL-LHC. Right: Energy distribution of $\nu_\tau$ events with/without sterile neutrino oscillations. Figures are taken from Refs.~\cite{Feng:2022inv, Bai:2020ukz}}
    \label{fig:NuOsc_forward}
\end{figure}
Figure~\ref{fig:NuOsc_forward} (left) shows the expected sensitivity to the sterile neutrino parameter space estimated with $\nu_\mu$ disappearance channel for FASER$\nu$ in the Run-3 and FLArE in the HL-LHC stage. 
As shown in the figure, the FLArE at the FPF can provide better constraints than the global constraints for a large mass region above $\Delta m_{41}^2 \sim \mathcal{O}(10^2)~{\rm eV^2}$.
Figure~\ref{fig:NuOsc_forward} (right) shows the $\nu_\tau + \bar{\nu}_\tau$ event spectrum as a function of energy in the standard model and its distortion in the presence of a sterile neutrino, evaluated with a set of allowed parameters for demonstration. 
While sterile neutrino oscillations can be studied with muon neutrinos and electron neutrinos at the SND@LHC and FASER$\nu$, the study with tau neutrinos will only be possible at the FPF, where a sufficient number of tau neutrino events—on the order of thousands—can be obtained.

\paragraph{\bf neutrino-philic new particles}
In addition to the LDM and sterile neutrino, it is possible to search for a neutrino-philic new particle that is produced from the exotic three-body decays of charged mesons in forward detectors.
A promising example is the new gauge boson $Z'$ mediating the SNI~\cite{Bakhti:2023mvo}. 
In particular, due to the short baseline, unexpected $\nu_\tau$ appearance by the new particle decaying to tau neutrino can increase the sensitivity.

\subsubsection{SHiP}
The Search for Hidden Particles (SHiP) experiment~\cite{Albanese:2878604} is a proposed fixed-target facility at the CERN SPS designed to search for long-lived particles with feeble couplings to the SM particles, while also advancing neutrino physics.
A high-intensity $400\,\mathrm{GeV}$ proton beam at the SPS collides with a proton target, producing both active neutrinos of all flavors and potential hidden-sector states.
Downstream of the proton target, a hadron absorber stops most hadrons and electromagnetic radiation, and an active muon shield based on magnetic deflection suppresses residual muons to negligible levels.
Over a 15-year run accumulating $6\times10^{20}$ protons on target (POT), SHiP will collect the necessary statistics for both neutrino physics and hidden-sector portals from 2031~\cite{SHiP:2025ows}.

The Scattering and Neutrino Detector (SND), located right after the muon shield, will detect on the order of $10^{7}$ neutrino interactions over the full run, assuming a 3-ton target. 
SND will perform precision measurements of $\nu_\tau$ and $\bar\nu_\tau$ deep-inelastic scattering, and extract the structure functions $F_{4}$ and $F_{5}$~\cite{Alekhin:2015byh}.
Composed of an ECC followed by a muon spectrometer, it offers precise vertex reconstruction and real-time readout.
It will also collect over $10^{5}$ neutrino-induced charmed hadrons, and will search for NSIs, such as $Z'$ inducing SNI from $\nu_\tau$ appearance~\cite{Bakhti:2023mvo}, and neutrino electromagnetic properties. 
By considering SND as a short-baseline neutrino detector, possible $\nu_\tau$ oscillation into sterile neutrino can be probed by observing the $\nu_\tau$ disappearance, which is sensitive to the parameter region $10^2\,{\mathrm eV}^2 \lesssim \Delta m_{41}^2 \lesssim 10^4\,\mathrm{eV}^2$ from the $2 \times 10^{20}$ POT for the 5 years of operation~\cite{Choi:2024ips,Choi:2025vby}.
The mixing parameter $|U_{\tau 4}|^2 \sim 0.08$ (90\% CL) can be probed for $\Delta m_{41}^2 \sim 500\,{\rm eV}^2$ assuming the 10\% signal-to-background ratio.
SND can also search for LDM scattering off electrons and MCPs~\cite{SHiP:2020noy,Magill:2018tbb}.
Unlike LDM, MCPs are expected to leave low-ionization tracks and a multi-scattering signature.
Over the 15-year run, SHiP is expected to reach sensitivity to $10^{-3}e$ millicharges at $m_{\mathrm{MCP}} \lesssim 1\,\mathrm{GeV}$~\cite{Magill:2018tbb}.

Any long-lived hidden particles produced in the proton target of SHiP will travel through the SND and may decay inside a $50\,\mathrm{m}$ helium-filled decay volume. 
Visible decay products are captured by the Hidden Sector Decay Spectrometer (HSDS), whose high-precision magnetic tracker, electromagnetic calorimeter, and muon stations reconstruct vertices and momenta. 
A surrounding veto system and sub-nanosecond timing layers keep backgrounds nearly zero.
SHiP will probe heavy neutral leptons up to a few GeV with mixing angles as small as $10^{-9}$, dark photons with kinetic mixing $\epsilon^2\sim10^{-15}$, and scalar or axion-like portals across the MeV to GeV window~\cite{Albanese:2878604,Ovchynnikov:2023cry}.
By observing both dilepton and hadronic final states for each portal model, SHiP has the ability of measuring the branching fractions, lifetimes, and coupling structures of the long-lived particles in an ultra-low-background environment.

\subsubsection{Stopped pion neutrino experiments}

Stopped-pion neutrino experiments such as COHERENT, CCM, JSNS\textsuperscript{2}, and LSND provide a uniquely clean and well-characterized environment for the BSM search. These experiments utilize intense GeV-scale proton beams impinging on a fixed target, producing charged pions that rapidly come to rest and decay at rest, yielding neutrino fluxes with well-defined energy spectra and timing structures. The resulting neutrinos---particularly the prompt monoenergetic $\nu_\mu$ (from the $\pi^+$ decay) and delayed $\nu_e$ and $\bar{\nu}_\mu$ (from the $\mu^+$ decay)---enable precision measurements and time-correlated event tagging that are invaluable for BSM searches. Because of their low-energy neutrino spectra, stopped-pion experiments are ideally suited to probe light, weakly coupled particles such as sterile neutrinos~\cite{Formaggio:2011jt,Anderson:2012pn}, LDM~\cite{deNiverville:2015mwa,Dutta:2019nbn,Dutta:2020vop,CCM:2021leg}, dark-sector mediators~\cite{Dutta:2020vop}, heavy neutral lepton~\cite{Khan:2022bcl,Ema:2023buz}, and ALPs~\cite{CCM:2021jmk,COHERENT:2022nrm}. Furthermore, their sensitivity to CE$\nu$NS opens a window into NSIs~\cite{Liao:2017uzy,Giunti:2019xpr,Coloma:2019mbs,Flores:2020lji,Shoemaker:2021hvm}, neutrino electromagnetic properties~\cite{AtzoriCorona:2022qrf}, and potential signatures of light mediators. The pulsed nature of the beam and low background environment significantly enhance signal discrimination, making these facilities powerful tools in the broader BSM physics program.

The sterile neutrino search experiment JSNS$^2$ has a concrete upgrade plan, JSNS$^2$-II. The construction of an additional far detector~\cite{JSNS2:2020hmg} is progressing according to the original schedule and is anticipated to enhance the search sensitivity, particularly in the low $\Delta m^2$ region.
Separately, a dedicated detector for searching for MCPs, named SUB-Millicharge ExperimenT (SUBMET)~\cite{Kim:2021eix,Campagnari:2025icx}, has been installed at J-PARC.
SUBMET uses the 30 GeV proton beam (compared to the 3 GeV beam for JSNS$^2$) and aims to explore the parameter space of MCP with masses below 1.6 GeV and charges smaller than $10^{-3} e$.

High-intensity proton beams with energies of $\mathcal{O}({\rm GeV})$ or less striking a target create not only charged pions but also neutral pions that predominantly decay into two photons, one of which can kinematically mix to a dark photon and decay into a pair of DM particles. Some of these particles may travel unimpeded to a nearby detector and produce observable signatures through nuclear or electron recoils within the detector material. Some charged pions may undergo an absorption process in which a photon is emitted in the final state; in the presence of a dark-sector mediator, this photon can similarly be replaced with the mediator $V$, $\pi^-+p \to n+V$. Given that stopped-pion neutrino experiments operate with high-intensity proton beams delivering approximately $10^{22}-10^{23}$ protons-on-target per year, a substantial yield of DM particles is expected in scenarios involving dark-sector mediators. Furthermore, the close proximity of the detectors to the beam target---typically around $20-40$ meters---and their large mass, reaching up to the ton scale, provide these experiments with excellent sensitivity to potential DM signals. 

\begin{figure}[h]
    \centering
    \includegraphics[width=0.75\linewidth]{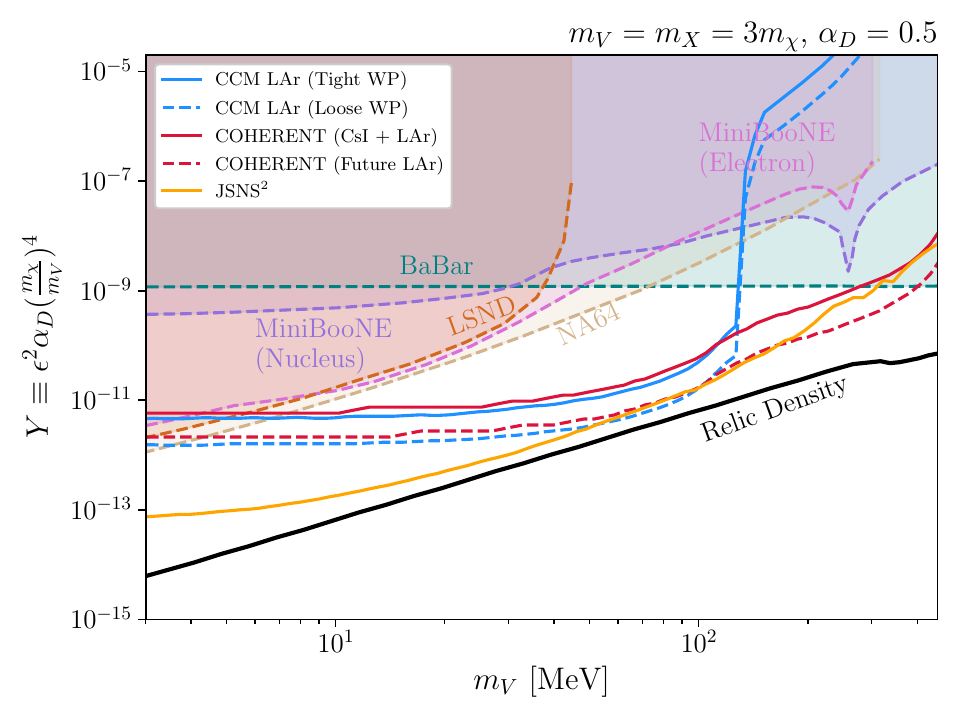}
    \caption{90\% C.L. projected experimental sensitivity to the vector-portal DM signals at COHERENT, CCM, and JSNS$^2$.
    The sensitivity estimates are shown for conventioanal $Y\equiv \epsilon^2 \alpha_D (\frac{m_\chi}{m_V})^4$ where $\alpha_D = g_D^2 / (4\pi)=0.5$ and $m_V/m_\chi=3$.
    The parameter sets that are consistent with the observed DM relic abundance are shown by the black solid line. Plot taken from Ref.~\cite{Dutta:2020vop}.}
    \label{fig:stoppedpion-sense}
\end{figure}

An earlier approach to background reduction involved applying an energy cut to suppress certain neutrino backgrounds whose energy deposit at the detector is kinematically bounded from above. Notable examples of this strategy include DM searches conducted at the LSND and MiniBooNE detectors~\cite{deNiverville:2011it}, as well as at the COHERENT and CENNS (now part of the COHERENT experiment) detectors~\cite{deNiverville:2015mwa,Ge:2017mcq}, and more recently at COHERENT and CCM~\cite{Berlin:2020uwy}. However, this method alone is insufficient to eliminate other neutrino backgrounds while preserving adequate signal statistics. It was later recognized that employing a timing cut offers a more effective way to reject neutrino events occurring in delayed timing bins, as demonstrated in COHERENT analyses~\cite{Dutta:2019nbn,COHERENT:2019kwz,Dutta:2020vop}. See Fig.~\ref{fig:stoppedpion-sense} for example sensitivity prospects of vector-portal DM signals at COHERENT, CCM, and JSNS$^2$. In particular, Ref.~\cite{COHERENT:2019kwz} utilizes delayed timing bins to extract the spectral shape of neutrino events in the prompt region through a side-band analysis, subject to statistical and systematic uncertainties. Nevertheless, this approach assumes the absence of NSIs, potentially introducing model dependencies. Moreover, the signal region may still contain a substantial number of neutrino-induced background events, which can ultimately limit sensitivity to dark-sector signals.

\begin{figure}[h]
    \centering
    \includegraphics[width=0.495\linewidth]{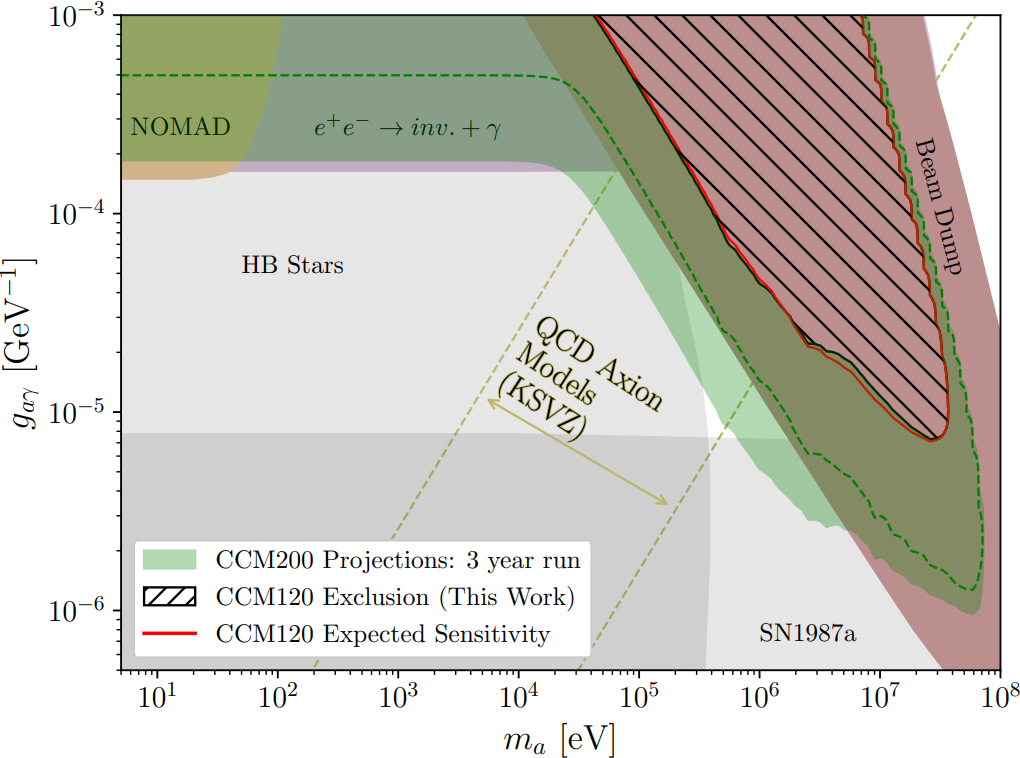}
    \includegraphics[width=0.49\linewidth]{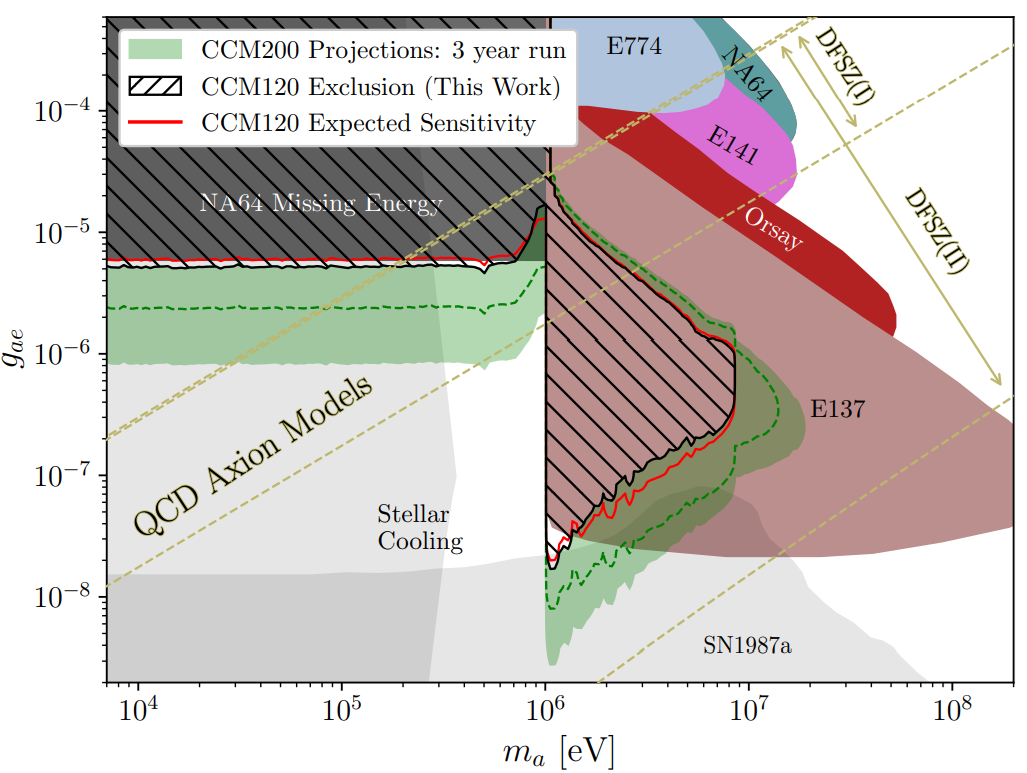}
    \caption{The expected and actual 90\% C.L. from
CCM120 for the ALP-photon coupling $g_{a\gamma}$ (left) and ALP-electron coupling $g_{ae}$ (right). Also included is the projection region for CCM200's three-year run using background taken from CCM120’s spectrum reduced by two orders of magnitude for various conservative improvements (dashed green line) and a background-free assumption (extent of shaded green region). Plots taken from Ref.~\cite{CCM:2021jmk}.}
    \label{fig:ccm-sense-est}
\end{figure}

While low-energy proton beams can generate an immense flux of charged and neutral pions, a copious number of photons and $e^\pm$ can also be produced within the beam target through electromagnetic shower processes. These secondary particles can give rise to ALPs through various mechanisms---including the Primakoff process, Compton-like scattering, resonant production, $e^\pm$-induced bremsstrahlung, and associated production---depending on the specifics of the underlying ALP model. As a result, stopped-pion neutrino experiments offer strong sensitivity to ALP signals, particularly those involving couplings to SM photons and electrons. Related search proposals targeting ALP decay and/or scattering signals include CCM~\cite{CCM:2021jmk}, PIP-II~\cite{Toups:2022yxs,Aguilar-Arevalo:2023dai}, and DAMSA implementations at the RAON beam facility~\cite{Jang:2022tsp} and PIP-II~\cite{Kim:2024vxg}. Example sensitivity prospects for the ALP-photon and ALP-electron couplings at CCM are shown in Fig.~\ref{fig:ccm-sense-est}.

Furthermore, pion decay-at-rest (DAR) experiments, along with $\mu$DAR, will be utilized to measure neutrino–nucleon cross sections, including those on $^{12}\mathrm{C}$ and $^{13}\mathrm{C}$. The latter is particularly important for improving the $^8\mathrm{B}$ neutrino flux measurement and addressing the so-called 5 MeV bump observed in reactor neutrino experiments with LS detectors~\cite{Bakhti:2024dcv}.

\subsubsection{T2HK}

The long-baseline aspect of the Hyper-Kamiokande, T2HK, will utilize several near detectors.
The T2K near detector, ND280, has recently completed the upgrade outlined in Ref.~\cite{T2K:2019bbb}. The previous $\pi^0$ detector is replaced by a combination of several novel detectors, namely the Time-of-Flight detector (TOF), the Super Fine-Grained Detector (SFGD), and two High Angle Time Projection Chambers (HA-TPC). 
TOF comprises 6 planes of 20 scintillation bars, providing an excellent timing resolution of about $0.14$ ns \cite{Korzenev:2021mny}. The HA-TPC has a new field cage design and resistive MicroMegas, leading to a larger tracking volume and better spatial resolution, respectively, compared to the existing TPCs in ND280. More specifically, its spatial and $\textrm{d}E/\textrm{d}x$  resolution could reach $500$ mm and $10\%$ respectively \cite{Attie:2019hua,Attie:2021yeh}.
SFGD is a 3D plastic scintillator made up of 2 million $1\textrm{cm}^3$ cubes. It is the new active target, weighing about 2 tons, and its granularity brings great tracking and particle identification capability, most importantly, isotropic acceptance. Its timing resolution can reach $1.14$ ns. More details of its performance can be found in beam tests in Refs.~\cite{Blondel:2020hml,Agarwal:2022kiv}.

In addition, J-PARC is upgrading its beam in preparation for the HK, leading to a much stronger flux of potential BSM particles passing or decaying in the near detectors.
It is highly promising that T2K could push the HNL limit even lower. Additionally, the larger tracking volumes and the stronger beam will make ND280 an ideal place for searching for exotic particles that could be produced in the beam, for example, MCPs, dark photons, LDM and many more.
In particular, the ND280's sensitivity to LDM models, which has provided potentially powerful sensitivity in different ranges for the mediator mass~\cite{deNiverville:2016rqh}, can be improved as the volume increases.
One can see that the upgraded ND280 will complement the dark photon searches by other future experiments such as FASER2, DarkQuest~\cite{Apyan:2022tsd}, and FACET~\cite{Araki:2023xgb}.
SuperFGD \cite{Artikov:2022mox} has the potential to detect double-hit signatures of hypothetical MCPs \cite{Gorbunov:2021jog}. With negligible background, its predicted sensitivity competes MilliQ@SLAC in the sub-GeV mass range and surpasses ArgoNeuT by an order of magnitude.

An additional near detector known as the Intermediate Water Cherenkov Detector (IWCD), a kiloton-scale water Cherenkov detector, is planned to be located around 1 km from the J-PARC beam, mainly to reduce the systematic error even further. The IWCD is designed to move vertically relative to the beam direction, changing the energies of neutrinos impinging on the detector. This prism concept allows IWCD to effectively measure the cross-sections at the neutrino energies relevant for the CP violation measurement, minimizing the systematic effects.
A varying ratio of DM versus neutrino flux is also expected to significantly enhance the sensitivity for beam-produced DM, by achieving improved signal-to-background ratios at off-axis positions and regulating systematic uncertainties~\cite{DeRomeri:2019kic,Breitbach:2021gvv}.

\subsubsection{Yemilab Experiments: $\nu$EYE with IsoDAR and $e$-beam dump}

Yemilab is a new foremost underground laboratory, inaugurated in October 2022, in Jeongseon, Gangwon Province, Korea, with a depth of 1,000 m. 
The tunnel is encased in limestone and accommodates 17 independent experimental spaces. Among
them, Yemilab includes a cylindrical pit with a volume of approximately 6,300 m$^3$, designed as a multipurpose laboratory for next-generation experiments involving neutrinos, DM, and related research such as COSINE-100U, COSINE-200, and AMoRE-II. 
Most importantly, an effort to develop liquid scintillators capable of separating
Cherenkov from scintillation light is planned to host about
2.26 ktons of neutrino detector, now called $\nu$EYE ($\nu$ Experiment at YEmilab), mainly to search for 
solar, reactor, geo-, supernova, and sterile neutrinos with low backgrounds.
A white paper discussing the physics potential~\cite{Seo:2023xku} and the conceptual design report~\cite{NuEYE:2026gyx} are now released.
In addition to the neutrino sector, $\nu$EYE is also able to search for other BSM particles such as dark photon, DM, and ALPs.
The primary method to realize this is to develop so-called ``slow LS'' by optimizing the fluor, solvent, and wavelength shifting material~\cite{Seo:2023xku,NuEYE:2026gyx}.
The capabilities of $\nu$EYE are further enhanced by potential integrations with an electron linear accelerator (linac), a proton cyclotron (IsoDAR source), or a radioactive source~\cite{Seo:2020dtx,Seo:2023xku}.

\paragraph{\bf IsoDAR}

The IsoDAR (Isotope Decay-At-Rest) presents a unique opportunity for a short-baseline neutrino physics program when paired with the proposed $\nu$EYE at Yemilab. This accelerator-driven facility produces a high-intensity, nearly isotropic flux of electron antineutrinos through the beta decay of $^8$Li, which is generated by bombarding a 
$^7$Li sleeve with neutrons from a proton beam incident on a beryllium target. The facility is designed to deliver about $1.15\times10^{23}$ $\bar{\nu}_e$ over four years of operation. The proximity (17 meters) of the IsoDAR target to the $\nu$EYE ensures a high-statistics dataset ideal for precision measurements. 

\begin{figure}[t]
    \centering
    \includegraphics[width=0.85\linewidth]{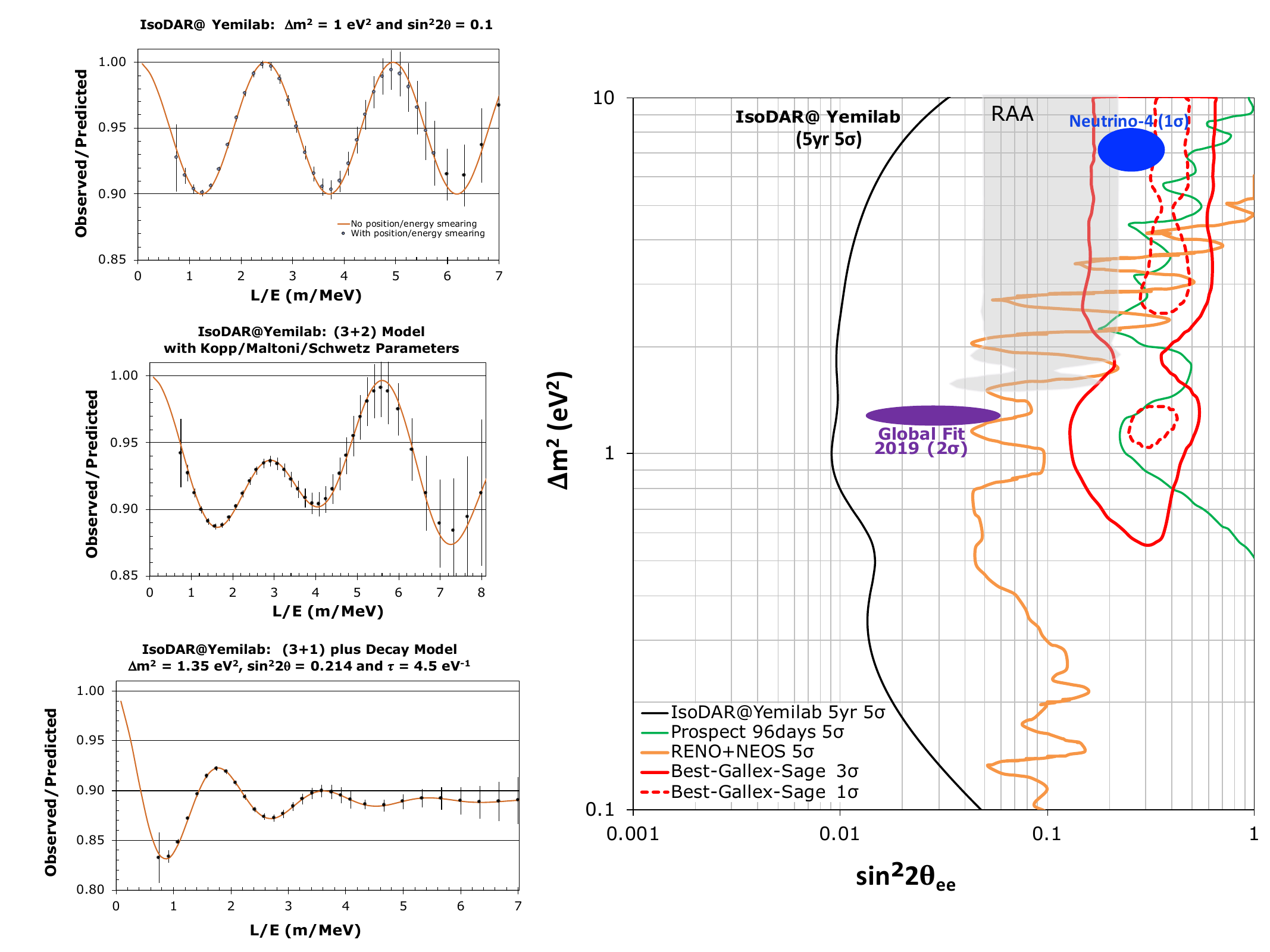}
    \caption{Left: Examples of three representative oscillation wave scenarios involving $\bar \nu_e$ disappearance that can be reconstructed at IsoDAR@Yemilab. Right: The state of the electron-flavor anomalies in February 2022, with the level of limit or allowed region indicated in the legend, along with the 5$\sigma$ sensitivity of IsoDAR@Yemilab for 4 years of run time. Plots taken from Ref.~\cite{Alonso:2021kyu}.}
    \label{fig:isodar-sense}
\end{figure}

A primary physics goal of IsoDAR@Yemilab is to search for evidence of sterile neutrino oscillations, particularly in the parameter space suggested by reactor and gallium anomalies. Using the IBD interaction, the experiment can detect spectral distortions indicative of active-sterile mixing. Because the distance and energy are precisely known, $\nu$EYE will be able to observe oscillation patterns in the L/E (baseline over neutrino energy) spectrum with excellent resolution. IsoDAR is expected to achieve sensitivity well beyond current and previous experiments like Daya Bay~\cite{DayaBay:2024nip}, NEOS~\cite{NEOS:2016wee}, and PROSPECT~\cite{PROSPECT:2024gps}, probing values of $\sin^2 2\theta_{ee}$ down to a few $\times 10^{-3}$ for $\Delta m^2 \sim 1~{\rm eV}^2$~\cite{Alonso:2021kyu}, as shown in Fig.~\ref{fig:isodar-sense}.

Recently, a new channel searching for reactor neutrinos by observing the $\nu \, {}^{13}{\rm C} \to \nu \, {}^{13}{\rm C}^\ast$ emitting a 3.685 MeV photon during the de-excitation has been proposed~\cite{Bakhti:2024dcv}. In order to enhance the sensitivity of the channel, it is important to precisely measure the cross section in solar or beam neutrino experiments such as IsoDAR@Yemilab. Its prospect in measuring the $\nu - {}^{13}{\rm C}$ cross section is shown in Ref.~\cite{Bakhti:2024dcv} in comparison with other experiments.
 
\begin{figure}[t]
    \centering
    \includegraphics[width=0.95\linewidth]{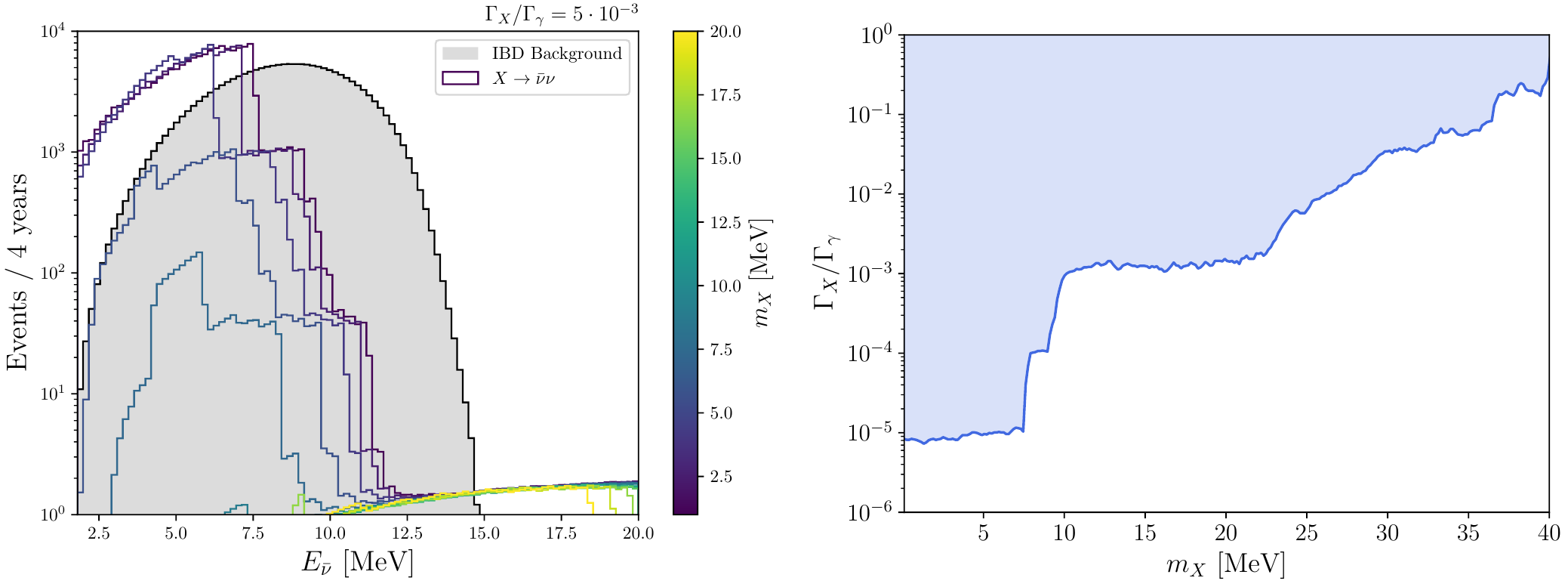}
    \includegraphics[width=0.95\linewidth]{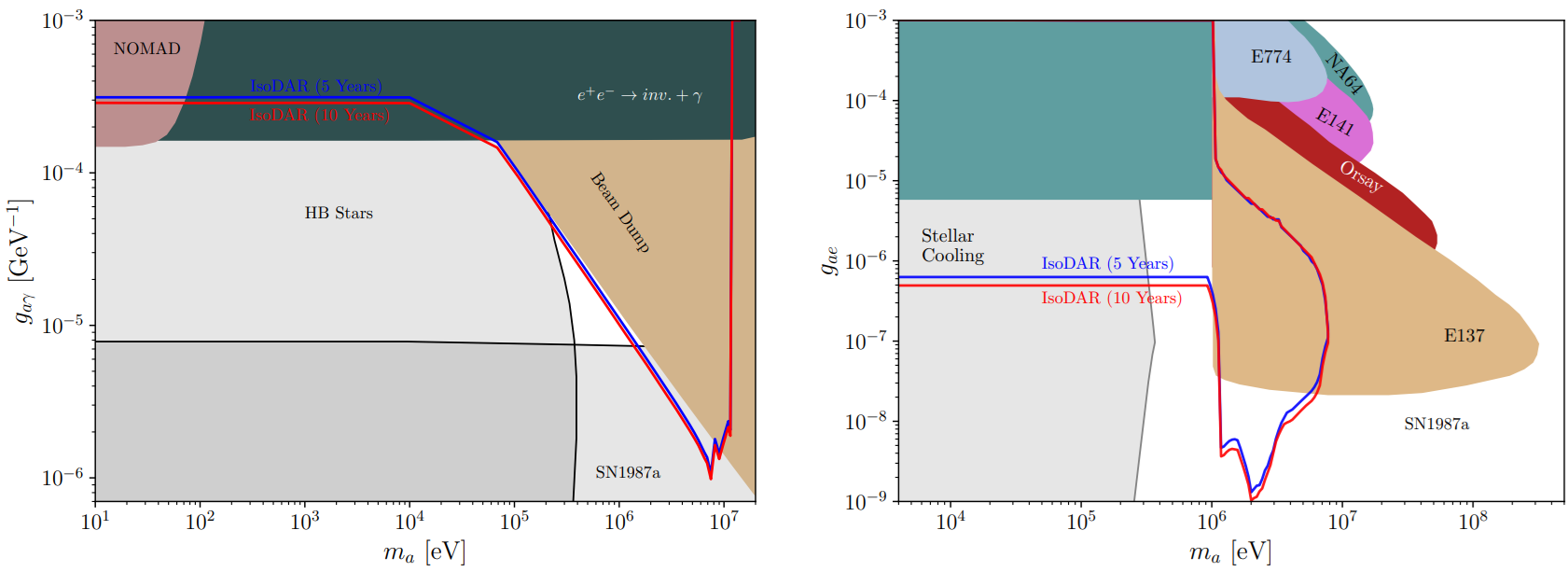}
    \caption{Top: Example IBD event peaks from $X \rightarrow \nu \bar \nu$, where the colors correspond to masses indicated by the side-bar and the IBD spectrum from $^8$Li decays is shown in gray (left). 90\% C.L. limit for the fraction of photon conversions to $X$ from observing the peak above the IBD background (right). Plots taken from Ref.~\cite{Seo:2023xku}.
    Bottom: Sensitivity contours of IsoDAR at 90\% C.L., for 5 (blue lines) and 10 (red lines) year exposures, using ALP couplings to photons (left) and ALP couplings to electrons (right). Plots taken from Ref.~\cite{Waites:2022tov}.}
    \label{fig:isodar-bumpalp}
\end{figure}

IsoDAR enables searches for new physics via ``bump hunting'' in the IBD spectrum, such as signals from light vector or scalar bosons (e.g., $X$ bosons) that decay into neutrino pairs~\cite{Seo:2023xku} (see the top panel of Fig.~\ref{fig:isodar-bumpalp}) and signals from ALPs that give rise to electromagnetic signatures including $\gamma$, $\gamma\gamma$, $\gamma e^-$, and $e^+e^-$~\cite{Waites:2022tov} (see the bottom panel of Fig.~\ref{fig:isodar-bumpalp}). Such searches are motivated by anomalies like the Atomki anomaly~\cite{Krasznahorkay:2015iga} and the unexplained 5 MeV reactor bump. The distinct high-statistics and low-background environment of $\nu$EYE allows for a sensitive scan of the antineutrino energy spectrum for localized excesses corresponding to new particle decays. Additionally, elastic scattering of $\bar{\nu}_e$ on electrons provides a clean probe for electroweak parameters like the weak mixing angle and potential NSI. This suite of physics opportunities makes IsoDAR@Yemilab a powerful, multi-purpose facility in the landscape of precision neutrino physics.

\paragraph{\bf Electron beam dump}

\begin{figure}[t]
    \centering
    \includegraphics[width=0.49\linewidth, height=6.6cm]{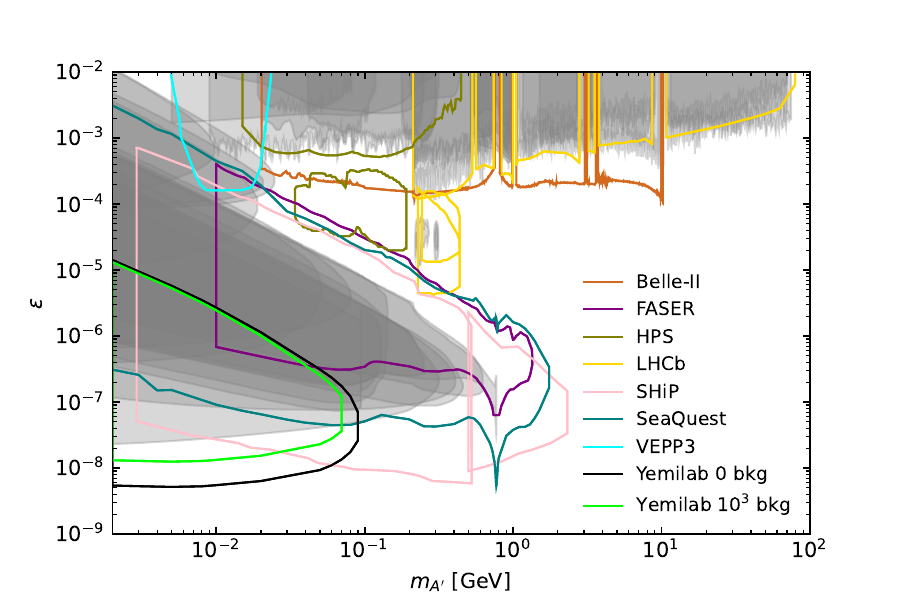} 
    \includegraphics[width=0.49\linewidth]{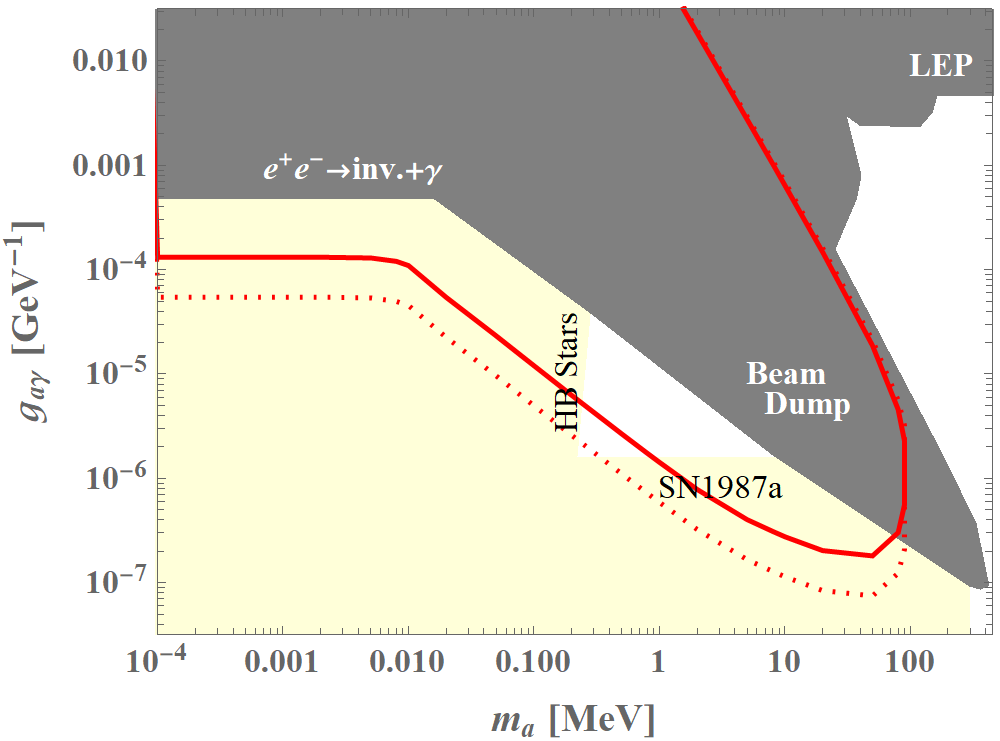} \\
    \includegraphics[width=0.49\linewidth]{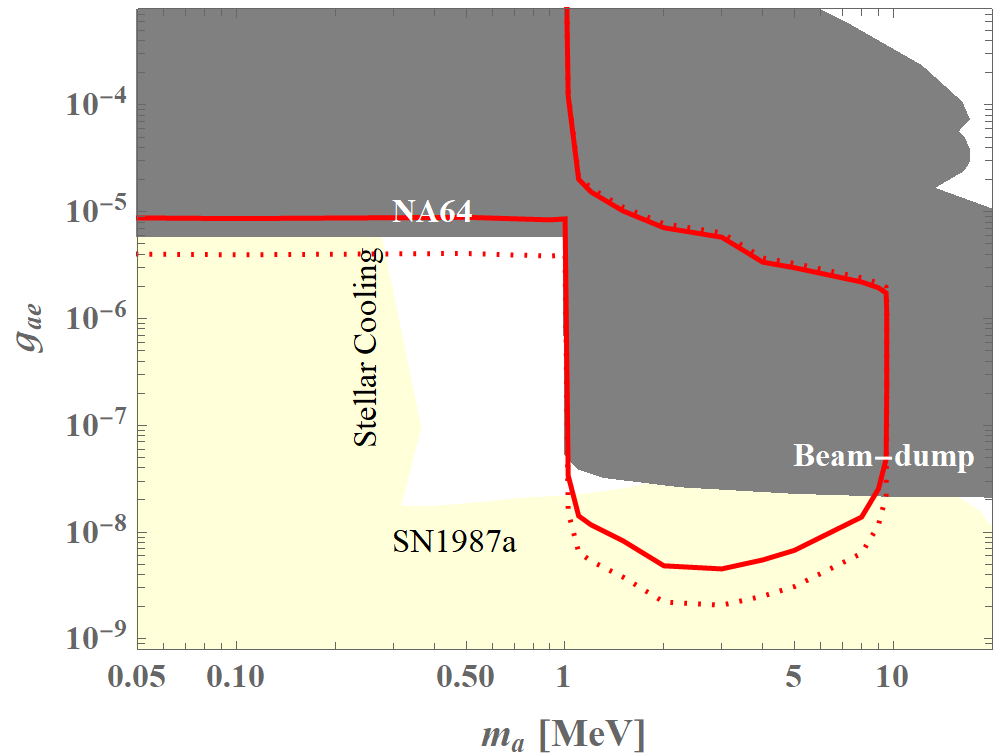}
    \includegraphics[width=0.49\linewidth]{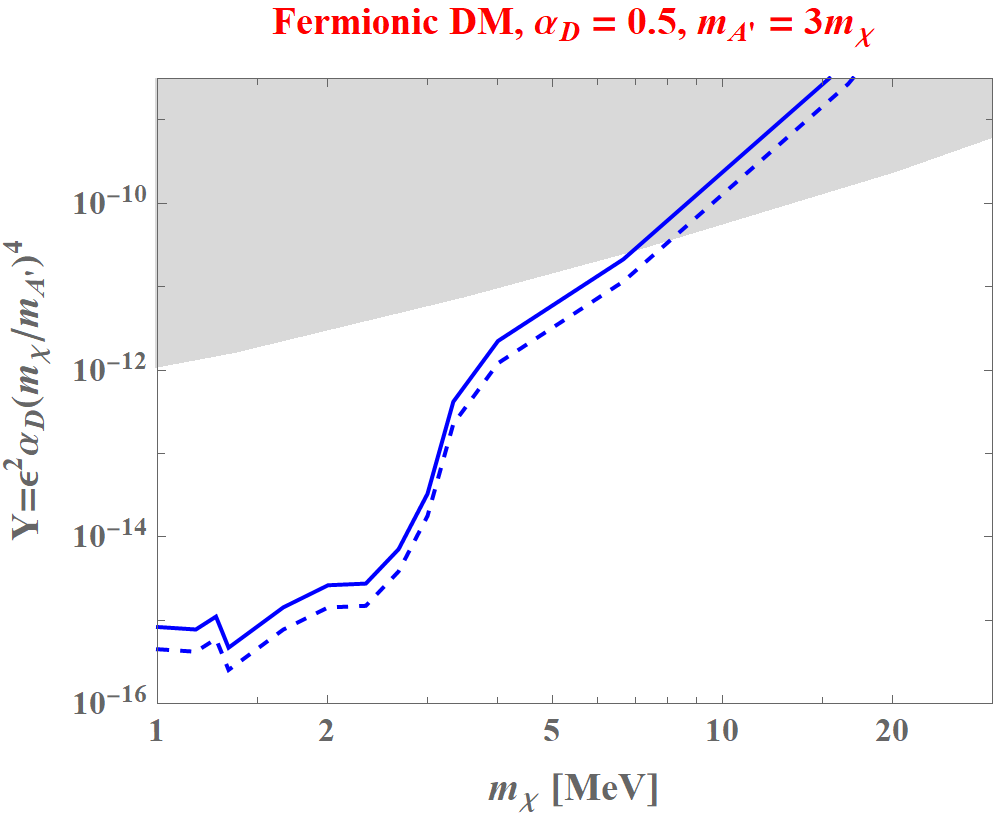}
    \caption{Sensitivity estimates of the electron beam-dump program for Yemilab to $A'\to e^+e^-$ signal (upper left), $a\to \gamma\gamma$ signal (upper right), $a\to e^+e^-$ signal (lower left), and LDM-$e^-$ scattering signal (lower right). For the sensitivity estimates in the upper-right, lower-left, and lower-right panels, a one-year exposure is taken into account under the assumptions of zero backgrounds (dashed lines) and $\sim 1000$ backgrounds (solid lines).
    The upper left panel is adapted from Ref.~\cite{Seo:2020dtx} and the others are from Ref.~\cite{Seo:2023xku}.}
    \label{fig:yemilab-bsm}
\end{figure}

The electron beam-dump program proposed for Yemilab aims to explore feebly interacting particles up to a few tens of MeV mass range using a high-intensity 100 MeV electron linac~\cite{Seo:2020dtx, Seo:2023xku}. When this beam strikes a dense target, it produces a cascade of secondary particles, including potential dark sector mediators such as dark photons, ALPs, and LDM. These particles could traverse shielding and interact in the adjacent $\nu$EYE, providing observable signatures such as electromagnetic showers or recoil events. The underground setting of Yemilab provides an ultra-low-background environment, enhancing the discovery potential of rare, weakly coupled particles that would be invisible in conventional collider experiments.

The beam-dump setup is particularly sensitive to dark photons ($A'$) that kinetically mix with the SM photon. Once produced in the dump, dark photons can decay into electron-positron pairs or scatter off electrons in the $\nu$EYE. Because of the short beam-target distance and the $\nu$EYE's large fiducial volume, Yemilab can explore regions of parameter space that are inaccessible to many other beam-dump experiments, as shown in the upper-left panel of Fig.~\ref{fig:yemilab-bsm}. Similarly, ALPs ($a$) interacting with the SM photon could be generated through the Primakoff process of photons inside the beam target and decay into photons within the detector. The expected sensitivity estimate is shown in the upper-right panel of Fig.~\ref{fig:yemilab-bsm}. For ALPs interacting with the electron, they could be produced by a Compton-like process of photons inside the beam target and decay into electron pairs. The lower-left panel of Fig.~\ref{fig:yemilab-bsm} shows the expected sensitivity reaches. 

Another major objective is the search for LDM particles, which could be produced via dark photon decays or directly in the beam-target interactions. These LDM particles would traverse shielding and scatter elastically with electrons in the $\nu$EYE. The experiment is designed to detect these low-energy recoils with high efficiency, thanks to the $\nu$EYE’s fine energy resolution and low threshold, achieving the sensitivity to new regions of LDM parameter space as shown in the lower-right panel of Fig.~\ref{fig:yemilab-bsm}.

Altogether, the electron beam-dump program at Yemilab represents a cost-effective and powerful platform to test a wide class of dark-sector models beyond the reach of past beam-dump-type experiments.

\paragraph{\bf Sterile neutrino search with radioactive sources}

This program involves deploying intense radioactive neutrino sources (e.g., $^{51}$Cr, $^{37}$Ar for electron neutrinos, or $^{144}$Ce for electron anti-neutrinos) within or around $\nu$EYE. 
By searching for spatial oscillation patterns, $\nu$EYE can test for short-distance active-to-sterile neutrino transitions. 
We expect the $\nu$EYE detector to have a low radioactive background and good position reconstruction resolution, enabling us to exploit the disappearance of the neutrino flux and the appearance as a function of distance.
Better than 1\% level calibration of the neutrino and anti-neutrino flux is expected to be achievable with careful calorimetric measurement and the establishment of a precise correspondence between heat and the neutrino flux.

The measurement of short-distance neutrino oscillations is important in probing the BSM contributions to the weak mixing angle at very low energy and neutrino magnetic dipole moments. This complements reactor and IsoDAR-based sterile neutrino searches and can cover different regions of parameter space, enhancing discovery potential or setting strong exclusion limits.

\subsection{Non-accelerator based searches}

\subsubsection{NEON Upgrade}
Although NEON has collected physics data for nearly four years, the recent searches for BSM physics, such as LDM~\cite{NEON:2024bpw} and ALPs~\cite{NEON:2024kwv}, utilized only the initial 1.6 years of data. Even with this limited dataset, NEON has already surpassed existing experimental limits. Ongoing efforts aim to extend these searches by using the full dataset and refining analysis techniques, which are expected to further enhance sensitivity to dark sector particles. Additional detection channels, including alternative modes of dark photon interaction, are also under consideration.

To support these extended searches, the NEON collaboration is planning a significant detector upgrade. This includes increasing the target mass, installing a muon veto system, and increasing the lead shielding thickness. These upgrades are expected to substantially reduce background rates, thereby enabling the exploration of new regions of parameter space that remain inaccessible to current experiments.

\subsubsection{RENE}

The Reactor Experiment for Neutrinos and Exotics (RENE) is a newly proposed reactor neutrino experiment that aims to explore the parameter space around $\Delta m^2_{41} \sim 2\,{\rm eV}^2$, as suggested by the combined analysis of NEOS and RENO~\cite{RENO:2020hva} and as outlined in Ref.~\cite{RENETDR}.
The prototype detector consists of a cylindrical target filled with Gd-loaded LS and a box-shaped gamma catcher filled with LS.
The detector will be located in the tendon gallery of a reactor in the Hanbit Nuclear Power Plant Complex, Yeong-gwang, Korea, with a baseline of 24\,m~\cite{RENETDR}.
The detector commissioning is now ongoing at the site of Chonnam National University (CNU) and the collaboration is waiting for the official approval of the installation in the tendon gallery.

Compared to NEOS, the RENE detector will feature a gamma catcher to reduce $\gamma$-ray background and improve energy resolution.
In NEOS, which lacks a gamma catcher, the energy spectrum of monoenergetic positrons exhibits a secondary peak below the main peak. 
This feature arises from partial energy loss when $\gamma$-rays—typically produced by positron annihilation or subsequent interactions—escape the target volume without depositing their full energy in the detector.
As a result, the reconstructed energy is underestimated, leading to the formation of the secondary peak.
To address this issue, the RENE detector incorporates a gamma catcher filled with LS surrounding the target region.
By absorbing the escaping $\gamma$-rays, the gamma catcher recovers the lost energy, suppresses the secondary peak, and enables more accurate energy reconstruction, thereby enhancing the detector’s overall energy resolution.

The inner structure of the detector, located inside a box-shaped stainless steel gamma catcher filled with LS, 
consists of a cylindrical acrylic target filled with 0.5\% Gd-LS, two 20-inch PMTs mounted on opposite sides, 
and a PMT holder with mu-metal shielding.
A Teflon reflector cone is installed around the rim of each PMT to improve light collection.
In addition, the inner walls of the gamma catcher are lined with Tyvek to enhance reflectivity. 
These design features are optimized to maximize PMT light collection efficiency, thereby contributing to improved energy resolution.

The shielding system consists of nine borated polyethylene (PE) panels, each 100 mm thick, for the sides and top, and high-density PE panels of the same thickness for the bottom section.
The outer veto detector comprises 15 plastic scintillator panels, positioned outside the lead bricks, along with 30 2-inch PMTs.

In order to identify the 5 MeV bump, the authors in Ref.~\cite{Bakhti:2024dcv} proposed to add moderate neutron shielding such as water, polyethylene, borated polyethylene, and cadmium in addition to the current one, as well as an extra muon veto system.
This may increase the potential of RENE in observing the alternative channel $\bar \nu_e {}^{13}{\rm C} \to \bar \nu_e {}^{13}{\rm C}^\ast$ emitting a 3.685 MeV photon, which opens up its scope beyond the search for sterile neutrino.

\subsubsection{$\nu$EYE at Yemilab: Hanul nuclear power plant}

The Hanul Nuclear Power Plant at Uljin, Gyeongsangbuk-do, located 65 km away from Yemilab, can be used as a source for a long-baseline reactor neutrino program. 
$1,488\pm 46$ IBD and $690\pm 25$ $\bar\nu_e$-$e$ elastic scattering (ES) events are expected from these nuclear reactor cores, assuming full reactor power~\cite{Seo:2023xku}. 
Combined analysis of the reactor data and solar data at $\nu$EYE in Yemilab would help investigate the oscillation parameters of solar neutrinos ($\theta_{12}$ and $\Delta m^2_{21}$) precisely~\cite{Bakhti:2023vzn}.
$\nu$EYE may also be able to monitor reactor activities by determining the distance and direction to reactor complexes via oscillation patterns and displacement vectors.

\subsubsection{JUNO \& JUNO-TAO}

The Jiangmen Underground Neutrino Observatory (JUNO) is a multipurpose neutrino experiment located in Jiangmen, Kaiping, China~\cite{JUNO:2015zny,JUNO:2021vlw}. It is designed to study neutrino properties with unprecedented precision, observe both astrophysical and terrestrial neutrinos, and search for new physics beyond the Standard Model. The JUNO detector is equipped with a 20-kiloton liquid scintillator target located 700 meters underground, monitored by 17,612 20-inch photomultiplier tubes (PMTs) and 25,600 3-inch PMTs to achieve an excellent energy resolution of 3\% at 1 MeV. Construction of the JUNO detector was completed in 2024, and the liquid filling process began in December 2024, with completion expected in the summer of 2025.\\
The primary goal of JUNO is to determine the neutrino mass ordering by detecting the reactor antineutrinos emitted from eight nuclear reactor cores at the Taishan and Yangjiang nuclear power plants. This measurement is planned by studying the fine interference pattern caused by quasi-vacuum oscillations in the oscillated antineutrino spectrum at a baseline of 52.5 km~\cite{JUNO:2015zny,JUNO:2021vlw}. With 6.7 years of data, the neutrino mass ordering can be determined at a 3$\sigma$ significance~\cite{JUNO:2024jaw} and the neutrino oscillation parameters $\sin^2\theta_{12}$, $\Delta m_{21}^2$, $\Delta m_{31}^2$, can be measured to a precision of 0.6\% or better~\cite{JUNO:2022mxj}.
JUNO has announced its first results of reactor neutrino oscillations from the 59.1 days of data collection and reported the simultaneous high precision determination of the solar oscillation parameters $\sin^2\theta_{12} = 0.3092 \pm 0.0087$ and $\Delta m_{21}^2 = (7.50 \pm 0.12) \times 10^{-5}\,{\rm eV}^2$ for the normal mass ordering, which is improved by a factor of 1.6 compared to the combined previous results~\cite{JUNO:2025gmd}.

The Taishan Antineutrino Observatory (TAO)~\cite{JUNO:2020ijm}, also referred to as JUNO-TAO, is a satellite experiment of JUNO~\cite{JUNO:2015zny}. A ton-level liquid scintillator detector will be located 30 - 44 meters away from the core of the Taishan Nuclear Power Plant. This experiment aims to measure the reactor antineutrino spectrum with an energy resolution better than 2\%, providing a reference spectrum for future reactor neutrino experiments and serving as a benchmark measurement to validate nuclear databases. It will intensively investigate the reactor rate and spectrum anomalies, as well as provide a model-independent approach to test the sterile neutrino hypothesis.

The TAO detector features a spherical acrylic vessel containing 2.8 tons of gadolinium-doped liquid scintillator, which will be illuminated by silicon photomultipliers (SiPM) with a total area of 10 m² and a photon detection efficiency of more than 50\%, ensuring nearly complete coverage. The expected photoelectron yield is approximately 4500 per MeV, which is significantly higher than that of any existing large-scale liquid scintillator detectors. The detector is designed to operate at -50°C to reduce the dark noise of the SiPMs to an acceptable level. Furthermore, it will be equipped with effective shielding from cosmogenic backgrounds and ambient radioactivity, aiming for a background-to-signal ratio of about 10\%. The experiment is expected to start operations in 2025.

\subsubsection{KamLAND2-Zen}

The KamLAND2-Zen is the next generation of the KamLAND-Zen experiment. The research and deconstruction work for KamLAND is currently underway~\cite{Nakamura:2020szx}.
The goal of KamLAND2-Zen experiment is the first search to cover the 3$\sigma$ band of the Majorana nature of neutrinos in the inverted mass ordering (IO) region.
In order to achieved this goal, we need to reduce the main problem for neutrinoless double beta decay which are the BG from $^{214}$Bi, 2$\nu\beta\beta$ and xenon spallation products.

\subsubsection{AMoRE-II}

AMoRE-II aims to search for neutrinoless double beta decay of  100 kg of ${}^{100}$Mo nuclei using molybdate scintillating crystals operating at milli-Kelvin temperatures.
Currently, AMoRE-II is under construction and will start data-taking at Yemilab in 2027.
The muon veto system is now installed and its functionality is being tested~\cite{Kim:2024ioo}.
The level of background is expected to be $\sim 10^{-4}$ counts/keV/kg/year in the region
of interest (ROI), thanks to the low background detector materials, an optimized shielding structure
at Yemilab with around 1000 m overburden.

Interestingly, the signal of $2\nu\beta\beta$ can also lead extra BSM searches in AMoRE-II.
Such examples include the searches for Majoron, sterile neutrino, Lorentz violation, Pauli exclusion violation, right-handed current, and neutrino self-interactions~\cite{Bossio:2023wpj}.
Those possibilities will be tested in AMoRE-II.

\section{Next Generation Neutrino Observatories and Their Prospects}\label{ss:ng-bsm}
\label{sec:futurecosmo}

Despite this workshop focused on the capabilities and potential of accelerator-based neutrino experiments, we also discussed the future prospect of cosmogenic neutrinos and BSM signals which can be probed in neutrino telescopes. Next-generation large neutrino observatories such as DUNE-FD, HK, IceCube, and smaller detectors such as JUNO and $\nu$EYE at Yemilab, open unique avenues for testing BSM scenarios, including heavy neutral leptons, NSIs, sterile neutrinos, DM interactions, and signatures from exotic high-energy phenomena.

\subsection{DUNE-FD}

DUNE FD is a precision 3D LArTPC~\cite{Nygren:1976fe,ICARUSbible} in four independent large scale cryostats of which two will be constructed for the Phase-I operation with two additional ones to follow as resources become available.
Each FD module has total LAr mass of 17~kt, $\sim10$~kt active mass.
Thus, the DUNE FD in its final configuration is expected to provide a total of 40~kt active mass.
The excavation of the new DUNE FD cavern has been completed as of January 2024, and the construction of the infrastructure for the detector system is expected to complete by July, 2026, as of writing this whitepaper.
All the cryostat steel components have been delivered to the storage facility in Rapid City, South Dakota.
The SURF crew has began transporting the steel beams down to the cavern 1500~m underground early May.
The delivery of the steel beams to the cavern for the first FD (FD1) is expected to be completed in late 2026 followed by the beginning of the cryostat construction, which is expected to be completed by August 2027.
The detector assembly and installation are expected to begin late 2027, aiming for completion in the middle of 2028.

While the FD1 construction is ongoing at SURF, the detector parts for the second FD (FD2) will be prepared with the goal of starting the installation in 2028, completing in about one year. 
It is expected to take about a year to fill FD1 with LAr, during which the FD2 assembly and installation will occur.
The cosmic data taking for FD1 is expected to be ready in 2029 after filling FD1 with LAr, while FD2 will be ready for cosmic data taking in 2030.

In addition to neutrinos from its accelerator-based beam, DUNE will also detect naturally occurring solar and atmospheric neutrinos, providing a valuable opportunity to test BSM physics.  
The precise measurement of solar neutrinos at experiments like DUNE and HK, with their large datasets, will not only enhance our understanding of the Sun’s internal structure and Earth’s composition through Day-Night asymmetry~\cite{Bakhti:2020tcj}, but also place constraints on NSI~\cite{Bakhti:2020hbz} and probe the mixing of super-light sterile neutrinos at low energies~\cite{deHolanda:2003tx,deHolanda:2010am,Bakhti:2013ora}.
Similarly, the detection of atmospheric neutrinos at DUNE will provide constraints on NSI~\cite{Bakhti:2022axo} and offer insights into SNI~\cite{Bakhti:2023mvo}.
Note that for low-energy neutrinos such as solar neutrinos and supernova neutrinos with $E_\nu \lesssim$ 10 MeV, the neutral current scattering $\nu$-${}^{40}$Ar$\to \nu$-${}^{40}$Ar$^\ast$ producing photons in 5 MeV - 10 MeV photon can be also considered as a complementing channel~\cite{Tornow:2022kmo,Meighen-Berger:2024xbx}.

DUNE-FD has been considered as one of the ideal neutrino experiments capable of directly detecting highly boosted DM signals. 
Pioneering early work on the theoretical proposals of BDM analyzed the prospects of DUNE in directly searching for the DM scattering signals with energies at least $\mathcal O$(MeV), which are unprecedented in neutrino experiments~\cite{Agashe:2014yua,Kim:2016zjx,Giudice:2017zke,Kim:2019had,Berger:2019ttc,Kim:2020ipj,DeRoeck:2020ntj}.
Compared to larger size neutrino experiments such as HK and IceCUBE, DUNE-FD has unique advantage in probing nuclear recoil BDM signals or possible secondary signature expected in a class of DM scenarios such as inelastic BDM (iBDM)~\cite{Kim:2016zjx} due to lower $E_{\rm th}$ and better detector position/angular resolutions~\cite{Kim:2020ipj}.

\begin{figure}[h]
\includegraphics[width=0.4\linewidth]{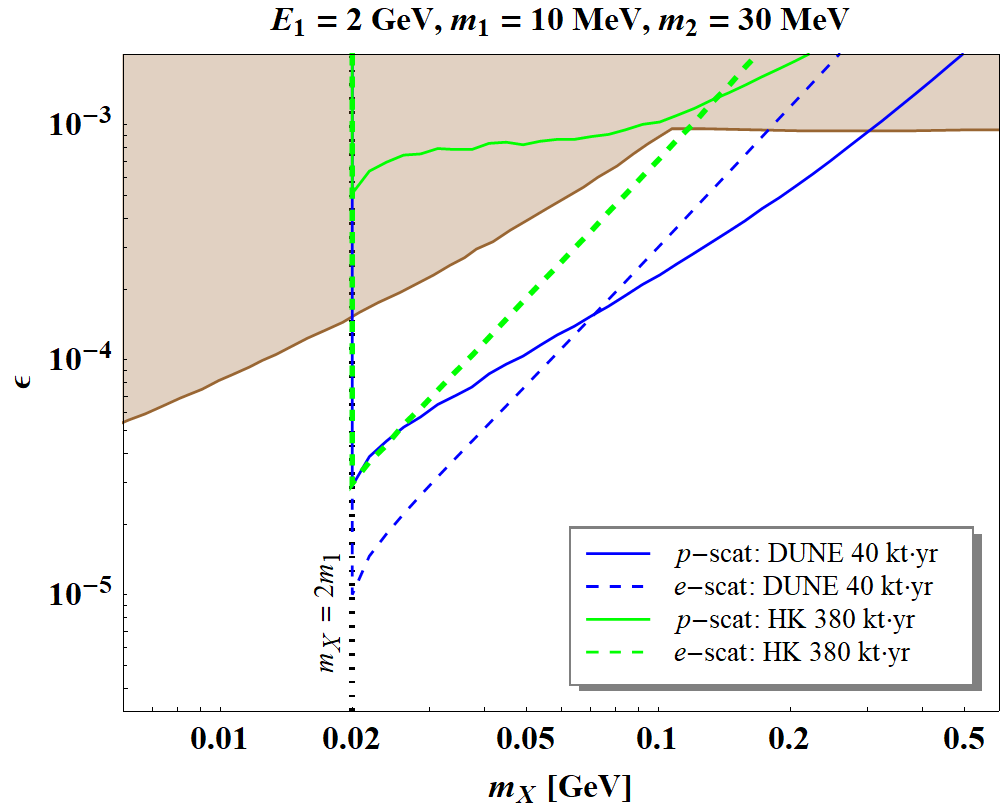} 
\includegraphics[width=0.4\linewidth]{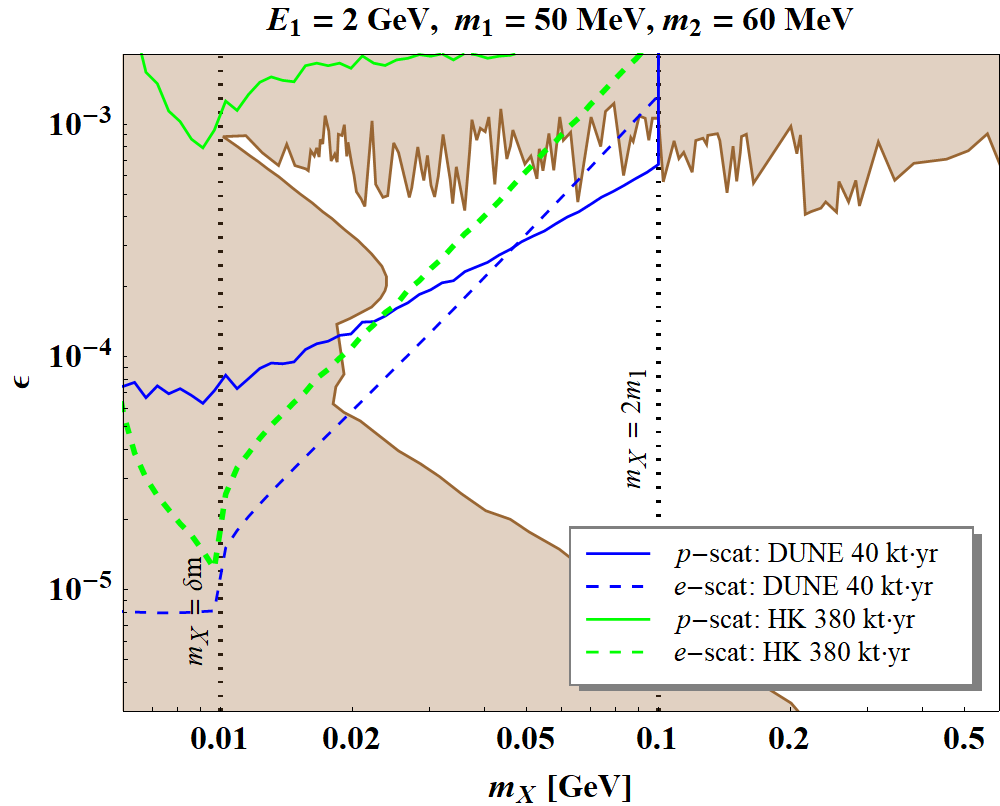} \\
\caption{
The bottom panels of Fig.~8 of Ref.~\cite{Kim:2020ipj} comparing the sensitivities of DUNE and HK in searching for the iBDM signal in terms of the mass of the mediator $m_X$ and the kinetic mixing parameter $\epsilon$. The left panel is for $m_X > 2 m_{\chi_1}$ and the right is for $m_X < 2 m_{\chi_1}$ where $\chi_1$ is the light BDM.
}
\label{fig:iBDMDUNEHK}
\end{figure}

Figure~\ref{fig:iBDMDUNEHK}, originally from Fig.~8 of Ref.~\cite{Kim:2020ipj}, shows the experimental sensitivity of DUNE for iBDM signals. In this scenario, a dark sector particle $\chi_2$ is produced via inelastic scattering of the BDM $\chi_1$ with targets and then decay back into $\chi_1$ associated with the SM particles (here $e^+ e^-$) from the decay of the mediator $X$ (dark photon as a reference), providing a secondary signature which enables subtracting most of the background events.
Notably, the sensitivities of DUNE-FD are dominant over those of HK even with about 10 times larger exposure, thanks to the good particle identification and excellent energy and angular resolutions. 
More dedicated analysis considering DUNE-FD's expected abilities of particle identification and $dE/dx$ measurements is presented in Ref.~\cite{DeRoeck:2020ntj}.

Interestingly, it is possible to test those search proposals for iBDM as well as the minimal BDM signals in the proto-type detector now being operated in CERN, ProtoDUNE~\cite{Chatterjee:2018mej,Kim:2018veo}.
The ProtoDUNE collaboration is now in the process of data analysis to search for these BDM signals.
Note that although the discussion so far is limited to the multi-component BDM scenarios where the BDM $\chi_1$ is produced with energy $E_1$ from the annihilation or decay~\cite{Bhattacharya:2014yha,Kopp:2015bfa,Heurtier:2019rkz} of heavier DM component whose mass determines $E_1$ and the flux of $\chi_1$, the same analysis strategy and method would be applied to other classes of BDM, e.g., DM boosted by scattering with high energy cosmic-rays or neutrinos.

In addition, DUNE-FD has a potential in searching for other classes of BSM such as new physics in proton decays, large extra dimensions, heavy neutral lepton, ALP~\cite{Brdar:2025hqi}, MCP~\cite{Plestid:2020kdm}, magnetic monopoles~\cite{Candela:2025gwp}, and WIMP-like DM annihilation in the Sun.
Some of the details are well summarized in Ref.~\cite{DUNE:2020fgq}.

\subsection{Hyper-Kamiokande}

The successor of the SK, HK, is currently under construction in Japan~\cite{Hyper-Kamiokande:2018ofw,Hyper-Kamiokande:2025fci}. 
The HK far detector is situated beneath the peak of Mt. Nijyugo, 295 km from the production target. It is designed as a cylindrical tank with a diameter of 68~m and a height of 71~m, containing a total water mass of 258~kton. 
The inner detector (ID) region has a volume of approximately 217~kton and will be instrumented with approximately 20,000 50~cm diameter PMTs and 1,000 multi-PMT photosensor modules \cite{Hyper-Kamiokande:2025fci}. 
While the photocathode coverage in the ID is about 20\%—half of the 40\% coverage in SK-IV, the light collection efficiency of the new 50~cm PMTs is doubled. Consequently, the overall photon detection efficiency of HK is expected to be approximately equal to that of SK~\cite{Hyper-Kamiokande:2025fci}.

The atmosphere acts as a natural collider, providing a robust and universal source of flux for BSM searches. Cosmic-ray cascades that give rise to atmospheric neutrino oscillation signals also continuously produce exotic particles in the atmosphere. This allows HK to set stringent, cosmology-independent limits on their flux. Quantitative phenomenological studies have demonstrated the feasibility of this approach, showing that HK can probe a wide region of parameter space for MCPs~\cite{Plestid:2020kdm,Du:2022hms,Du:2023hsv}, ALPs~\cite{Cui:2022owf}, magnetic monopoles~\cite{Candela:2025gwp}, and HNLs~\cite{Coloma:2019htx,Atkinson:2021rnp}. 
Hyper-Kamiokande is expected to play a leading role in indirect searches for low-mass WIMPs, lowering the limits on the spin-dependent DM–nucleon scattering cross-section by a factor of 2–3 relative to current SK results~\cite{Bell:2021esh}.
Neutrinos produced from thermal relic DM annihilation in the Galactic halo are particularly important for probing models in which DM predominantly couples to neutrinos, such as neutrino portal scenarios~\cite{Primulando:2017kxf,Olivares-DelCampo:2017feq,Olivares-DelCampo:2018pdl,Blennow:2019fhy}. The potential of the HK to discover or exclude thermal DM with s-wave annihilation in the mass range of 15–30 MeV has been highlighted by numerous theoretical studies~\cite{Palomares-Ruiz:2007trf} and has been further substantiated through detailed simulation studies~\cite{Bell:2020rkw}.
It is also worth noting that the sensitivity of the direct DM searches discussed in Sec.~\ref{sk} would be further improved by utilizing the HK data in place of that from the SK.

\subsection{IceCube-Upgrade \& IceCube-Gen2}

The IceCube Upgrade~\cite{Ishihara:2019aao}, scheduled to be constructed in the polar season of 2025-2026, represents a significant improvement in the capabilities of the current IceCube Neutrino Observatory. By deploying seven new strings with more compact spacing between sensors (3~m vertically and 20~m horizontally), featuring multi-PMT digital optical modules and other calibration sensors, the IceCube Upgrade will provide a diverse platform in neutrino detection, enabling more precise measurements in energy and direction. With a deeper understanding of the detector medium, sensitivity is expected to improve both in low and high energy neutrino physics analyses ranging from GeV to PeV.
In particular, with expected improvements for low energy neutrino interactions, the IceCube Upgrade will also enhance sensitivity for DM indirect detections below a few tens of GeV masses. In this energy region, the IceCube Upgrade is expected to increase $\nu_\mu + \bar\nu_\mu$ effective areas and therefore DM annihilation cross section sensitivity more than 10 times larger than the existing array. 

IceCube-Gen2 is a major expansion to the IceCube Neutrino Observatory, aimed at enhancing its ability to detect the highest-energy neutrinos~\cite{IceCube-Gen2:2020qha}. The IceCube-Gen2 optical array will deploy around 10,000 PMTs, more than doubling the current number in IceCube. Most of these sensors will be placed in adjacent ice, extending the detector's volume to roughly 8 times larger than that of the existing IceCube array. The expanded array is expected to increase the number of high-energy neutrinos and its highest energy reach, significantly improving the angular resolution and muon effective areas with increasing neutrino energies.
The IceCube-Gen2 surface array, an air shower array consisting of scintillation panels and radio antennas that sits on top of the footprint of the optical array, and the IceCube-Gen2 radio array consisting of surface and deep antenna stations covering an area of $\sim$ 500 km$^2$, are also under discussion.
Besides enhancing neutrino astronomy and providing high discovery potential for cosmic-ray astrophysics, IceCube-Gen2 will also probe fundamental properties of the neutrino itself. The neutrino interaction cross-section can be probed indirectly through absorption in the Earth, which will allow us to test Standard Model cross-section predictions and constrain BSM physics~\cite{IceCube:2017roe,Valera:2023boc}. Detecting energy-dependent variations in the flavor composition of the cosmic neutrino flux will enable us to differentiate between acceleration scenarios and source environments, as well as the structure of space-time itself~\cite{IceCube-Gen2:2023rds}.

IceCube-Gen2 will also improve the search for the annihilation or decay of heavy DM, extending to the DM lifetimes beyond $10^{29}$~s scale~\cite{IceCube-Gen2:2020qha}.
Additional BSM scenarios such as sterile neutrinos~\cite{IceCube-Gen2:2020qha}, NSI~\cite{IceCube:2022ubv}, HNL~\cite{IceCube:2025kve}, SNI~\cite{Ng:2014pca}, and  MCP~\cite{ArguellesDelgado:2021lek} are expected to be probed.

\subsection{JUNO}

JUNO is expected not only to achieve high-precision measurements of neutrino properties but also to open new avenues for studying both Standard Model processes and potential signatures of BSM physics.
The diffuse neutrinos from all past core-collapse supernovae could be observed at a significance of better than 5$\sigma$ for ten years of JUNO running~\cite{JUNO:2022lpc}. 
Meanwhile, a typical core-collapse supernova located at a distance of 10 kpc within our Galaxy would result in approximately 5,000 inverse beta decay events, around 2,000 all-flavor neutrino-proton elastic scattering events, and about 300 electron elastic scattering events at the JUNO detector. Additionally, there would be several hundred events from charged-current and neutral-current interactions on carbon. This would open unique windows for multi-flavor supernova neutrino observations and have significant implications for both particle physics and astrophysics~\cite{JUNO:2023dnp}.

For terrestrial neutrinos produced from radioactive decays within the Earth, approximately 400 inverse beta decay events can be recorded each year. 
Depending on the levels of radioactivity control~\cite{JUNO:2021kxb}, JUNO could achieve solar neutrino spectroscopy across a range of neutrino types, from $^8$B neutrinos~\cite{JUNO:2020hqc,JUNO:2022jkf} to $^7$Be, hep and CNO species of solar neutrinos~\cite{JUNO:2023zty}. This capability would provide new insights into the solar metallicity problem and enable the examination of the vacuum-matter transition region. 

JUNO also has a high potential in discovering new physics beyond the Standard Model.
Several representative new physics models, including light sterile neutrinos, non-standard interactions, lepton flavor unitarity violation, and Lorentz and CPT violations, can be explored using reactor, solar, and atmospheric neutrinos~\cite{JUNO:2015zny,JUNO:2021vlw}. Moreover, as a cutting-edge detector with unprecedented high resolution and low background, the JUNO detector could effectively search for novel hypothetical particles, including WIMPs~\cite{JUNO:2015zny,JUNO:2021vlw,Guo:2015hsy}, axions and ALPs~\cite{Lucente:2022esm,AristizabalSierra:2025myf}, heavy neutral leptons~\cite{vanRemortel:2024wcf}, and dark photons~\cite{Smirnov:2021wgi}. Finally, searching for baryon and lepton number-violating processes is also an important goal of JUNO. This includes studying proton decay or nucleon decay in general~\cite{JUNO:2022qgr,JUNO:2024pur}, and neutrinoless double beta decay~\cite{JUNO:2021vlw,Zhao:2016brs}. As a liquid scintillator detector with a relatively low threshold and low background, JUNO is competitive in the decay mode of $p\to\bar{\nu}+K^{+}$, where it can achieve a lower limit for the proton lifetime of $9.6\times10^{33}$ years (90\%C.L.) with ten years of data taking~\cite{JUNO:2022qgr}. 

Future upgrades~\cite{JUNO:2021vlw,Zhao:2016brs} involving the incorporation of double beta decay isotopes into the scintillator would position JUNO as one of the leading efforts globally in this area, enabling the exploration of the normal neutrino mass ordering region~\cite{Cao:2019hli}. Research and development on loading these isotopes, developing purification methods, and advancing analysis techniques for background suppression are currently underway~\cite{Ding:2023oxd}.

\subsection{$\nu$EYE at Yemilab}

The multi-purpose LS detector $\nu$EYE at Yemilab is expected to probe BSM physics through observations of solar, geo, supernova, and diffuse supernova neutrino background (DSNB)~\cite{Seo:2023xku,NuEYE:2026gyx}. We also briefly highlight other interesting SM physics topics accessible with this new detector.

\paragraph{\bf Solar neutrinos} 

\begin{table}[t]
\centering
\scalebox{1.0}{}
\begin{tabular}{|c|c|c|c|}
\hline\hline
\multirow{2}{*}{Solar $\nu$ type} & Rate & Flux & Flux-SSM prediction  \\
           & (counts/day/100 ton) & (${\rm cm}^{-1}{\rm s}^{-1}$) & (${\rm cm}^{-1}{\rm s}^{-1}$) \\
\hline
\multirow{2}{*}{$pp$}       & \multirow{2}{*}{$134\pm2^{+6}_{-10}$}  & \multirow{2}{*}{$(6.1\pm0.08^{+0.3}_{-0.5})\times10^{10}$} & $5.98(1.0\pm0.006)\times10^{10}$ (HZ)  \\
  & & & $6.03(1.0\pm0.005)\times10^{10}$ (LZ) \\
\hline
\multirow{2}{*}{$^7$Be}      & \multirow{2}{*}{$48.3\pm0.2^{+0.4}_{-0.7}$}   & \multirow{2}{*}{$(4.99\pm0.02^{+0.06}_{-0.08})\times10^{9}$} & $4.93(1.0\pm0.006)\times10^{9}$ (HZ) \\
& & & $4.50(1.0\pm0.006)\times 10^{9}$ (LZ)\\
\hline
\multirow{2}{*}{$pep$ (HZ)} & \multirow{2}{*}{$2.43\pm0.06^{+0.15}_{-0.22}$}  & \multirow{2}{*}{$(1.27\pm0.03^{+0.08}_{-0.12})\times10^{8}$} & $1.44(1.0\pm0.01)\times10^{8}$ (HZ)  \\
& & & $1.46(1.0\pm0.0009)\times10^{8}$ (LZ)\\
\multirow{2}{*}{$pep$ (LZ)}   & \multirow{2}{*}{$2.65\pm0.06^{+0.15}_{-0.22}$} & \multirow{2}{*}{$(1.39\pm0.03^{+0.08}_{-0.13})\times10^{8}$} & $1.44(1.0\pm0.01)\times10^{8}$ (HZ) \\
& & & $1.46(1.0\pm0.009)\times10^{8}$ (LZ)\\
\hline
\multirow{2}{*}{$^8$B}      & \multirow{2}{*}{$0.223^{+0.002}_{-0.003}\pm0.006$}        & \multirow{2}{*}{($5.68\pm0.06\pm{0.03})\times10^{6}$} & $5.46(1.0\pm0.12)\times10^{6}$ (HZ) \\
& & & $4.50(1.0\pm0.12)\times10^{6}$ (LZ)\\
\hline
\multirow{2}{*}{CNO}    &        &   & $4.88(1.0\pm0.11)\times10^{8}$ (HZ)  \\
& & & $3.51(1.0\pm0.10)\times10^{8}$ (LZ) \\
\hline
\multirow{2}{*}{$hep$}     & \multirow{2}{*}{any}      & \multirow{2}{*}{$<2.2\times10^5$ (90\% C.L.)}   & $7.98(1.0\pm0.30)\times10^{3}$ (HZ)  \\
& & & $8.25(1.0\pm0.12)\times10^{3}$ (LZ)\\
\hline\hline
\end{tabular}
\caption{
Expected solar neutrino measurements for 5 years of operation at $\nu$EYE in Yemilab assuming the same central values and systematic uncertainties as those of Borexino~\cite{BOREXINO:2018ohr}. HZ and LZ stand for the high and low metallicity models, respectively. Table taken from Ref.~\cite{Seo:2023xku}.
}
\label{tab:solarNu_fluxes}
\end{table}

The $\nu$EYE will be able to observe all solar neutrino species, thanks to its high target mass and low background. These observations include precision measurements of the solar neutrino flux (especially $^7$Be, pep, and $^8$B) as shown in Table~\ref{tab:solarNu_fluxes}, the study of matter-enhanced flavor conversion via the MSW-LMA solution including solar neutrino parameter measurements (e.g., $\theta_{12}$~\cite{Bakhti:2023vzn}), and exploration of possible new physics, such as NSIs and sterile neutrinos.
The detector’s sensitivity will also allow precise testing of solar metallicity models. Borexino’s recent measurement of CNO neutrinos~\cite{BOREXINO:2022abl} has motivated further exploration into the solar core's chemical composition. By comparing flux predictions from high- and low-metallicity models, $\nu$EYE is expected to significantly improve constraints on solar models.

\paragraph{\bf Geo neutrinos}

Using tools from reactors.geoneutrinos.org, the expected IBD (inverse beta decay) and ES (elastic scattering) event rates in $\nu$EYE have been estimated. For example, the annual geo-neutrino IBD and ES rates are $60.6\pm 13.6$ and $819\pm 174$ events, respectively~\cite{Seo:2023xku}. By detecting these events, $\nu$EYE will contribute to geophysical studies by constraining Earth's heat production models.

\paragraph{\bf Supernova and pre-supernova neutrinos}

$\nu$EYE is expected to detect hundreds of neutrino events from a galactic supernova at 10 kpc, primarily via IBD interactions. The number of expected events ranges from $\sim430$ to 820, depending on progenitor mass and neutrino mass ordering~\cite{Seo:2023xku}.
$\nu$EYE is also suited to detect low-energy pre-supernova neutrinos due to its large target mass and low backgrounds.
With sensitivity reaching out to $\sim1$~kpc, $\nu$EYE can act as an early warning system for supernovae such as SNEWS 2.0~\cite{SNEWS:2020tbu}, complementing other experiments like KamLAND and JUNO.

$\nu$EYE's clean environment with long operational time ($\sim10$ years) may make the detection of diffuse supernova neutrino background (DSNB) feasible. Assuming performance comparable to the JUNO detector technology~\cite{JUNO:2022lpc}, a preliminary estimate suggests that $\nu$EYE could measure a few DSNB events per year~\cite{Seo:2023xku}. 

\paragraph{\bf Cosmogenic BSM}

\begin{figure}[!ht]
    \centering
    \includegraphics[width=0.4\linewidth]{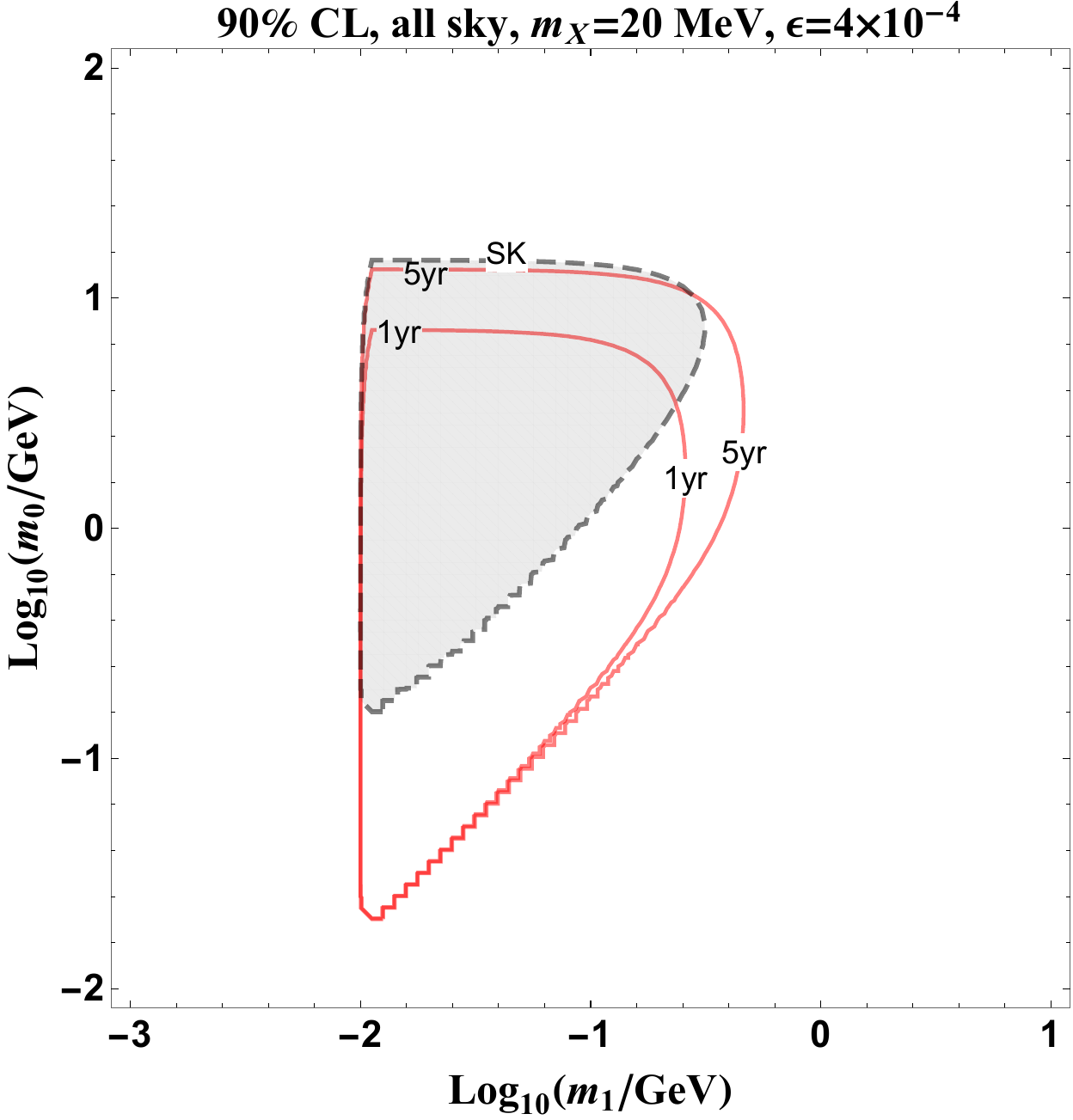} \\
    \includegraphics[width=0.45\linewidth]{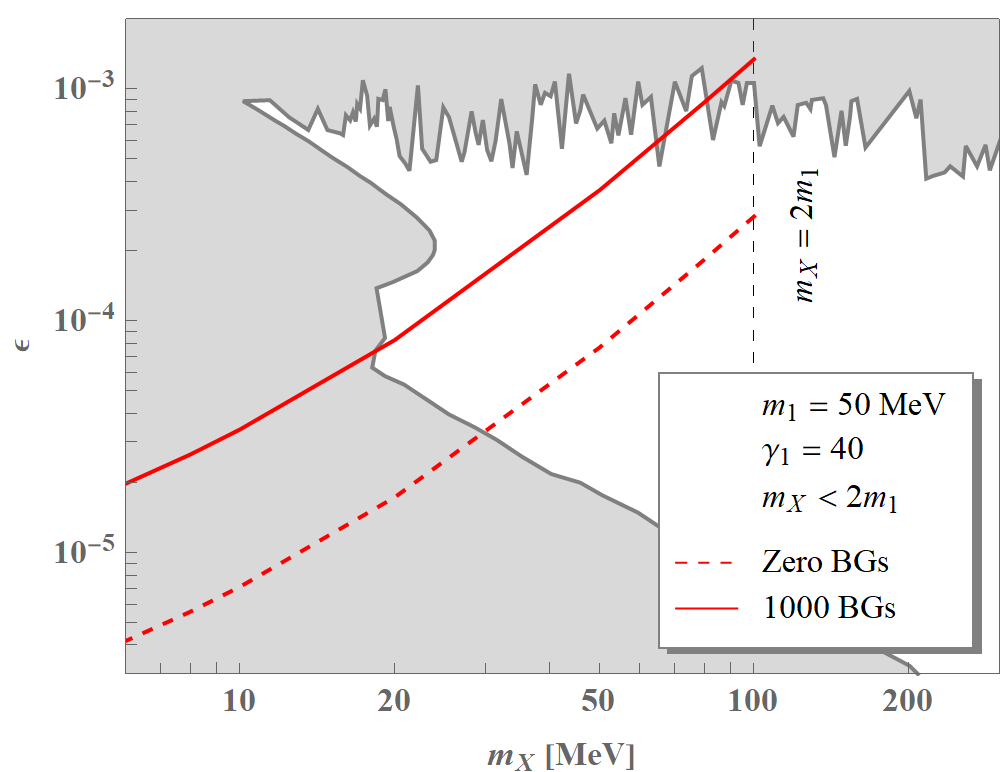}
    \includegraphics[width=0.45\linewidth]{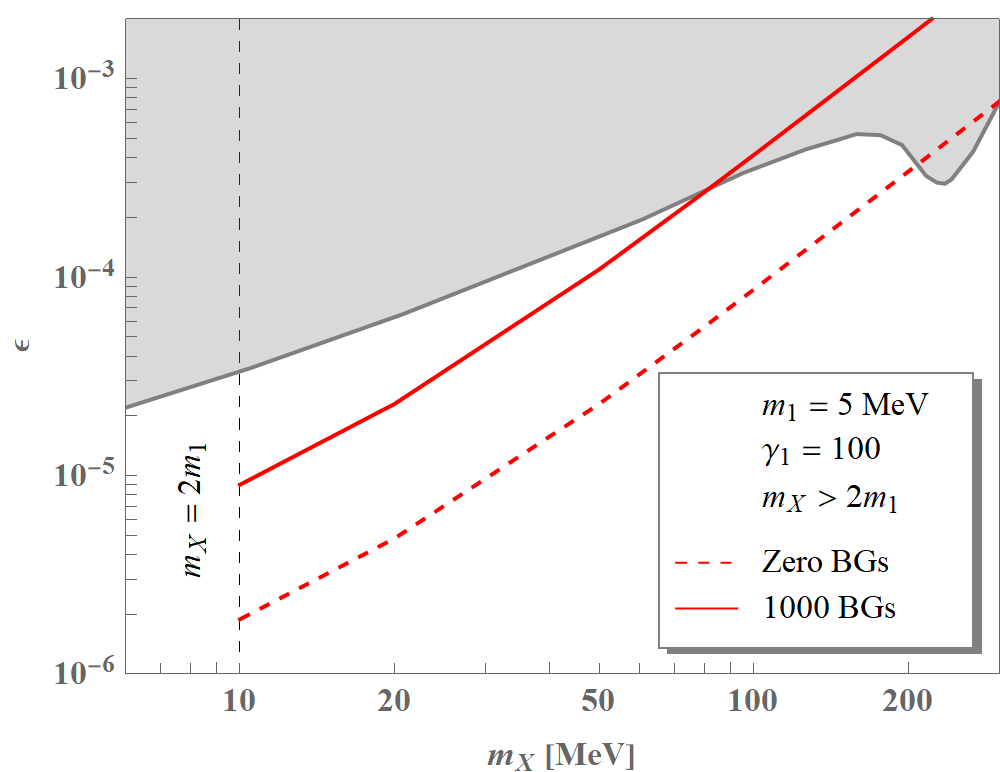}
    \caption{Top: The expected $90\%$ C.L. sensitivities from 1-year and 5-year running of $\nu$EYE with $E_{\rm th}=3~{\rm MeV}$ for the BDM flux from all sky, assuming $\sim1000/ {\rm year}$ backgrounds. Bottom: Experimental sensitivities of $\nu$EYE in the dark photon model parameters $m_X-\epsilon$ for the cases of $m_X<2m_1$ (left) and $m_X>2m_1$ (right). 
    A one-year exposure is considered with zero backgrounds (red dashed lines) and $\sim1000$ backgrounds (red solid lines). Plots taken from Ref.~\cite{Seo:2023xku}.}
    \label{fig:LSC-BDM}
\end{figure}

The deep-underground location and large volume of $\nu$EYE enable searches for exotic, high-energy particles that may originate from cosmogenic sources. Examples include BDM and atmospheric collider signals, both of which leverage the unique combination of cosmic particle flux and low-background detection environments. 
The $\nu$EYE is particularly well-suited for identifying BDM through elastic scattering off electrons or nuclei.
The large target mass and excellent background rejection of the $\nu$EYE allow it to probe BDM parameter space far beyond current experimental bounds, as shown in Fig.~\ref{fig:LSC-BDM}.  
The $\nu$EYE could also act as a ``cosmic beam-dump'' experiment, sensitive to a broad class of BSM scenarios, including heavy neutral leptons, scalar mediators, and other exotic LLPs.

\section{Tasks to Support Accomplishing the Prospective Results}
\label{sec:timeline}

To ensure accessibility to the BSM physics discussed in the previous section, there are several tasks that must be prepared.
Since the BSM signatures are at the tail end of the Standard Model phenomena, any fluctuations or uncertainties in SM background interactions could easily mask the extremely rare BSM features.
This section lists the detector and simulation tasks that have to be accomplished to make it possible to detect various BSM signatures.

\subsection{Necessary Detector Capabilities for BSM Signals }

The discovery potential for BSM physics at neutrino facilities depends on the ability of detectors to capture rare or subtle deviations from standard neutrino interactions. While many BSM signals may closely resemble known processes, they can manifest as small distortions in kinematic distributions, rare event topologies, or unexpected flavor transitions. 
Therefore, a successful BSM physics program demands detector capabilities tailored to enhance sensitivity to rare, exotic, or misidentified events while maintaining strong background rejection and high-precision measurements. 

First, detectors must achieve excellent spatial and vertex resolutions to identify possible displaced vertices from long-lived particle decays, such as ALPs or heavy neutral leptons. In order to accomplish this, detectors must achieve excellent momentum resolution to precisely reconstruct the trajectories of charged particles. This enables accurate determination of the invariant mass of the parent particle and improves the discrimination between new physics signals and SM backgrounds. Furthermore, precise momentum measurements contribute to refining the position of the displaced vertex by constraining the back-extrapolated track trajectories. Detectors like DAMSA and SHiP employ magnetic spectrometers and precision trackers for this purpose.

Second, a low energy threshold and high light yield are necessary to detect sub-MeV to few MeV signals, which are especially important in the search for coherent elastic neutrino-nucleus scattering (CE$\nu$NS), LDM, or dark photon-electron scattering. Experiments such as NEON, COHERENT, and JUNO-TAO exploit this capability to probe light mediators and other weakly coupled new particles.

Third, fine timing resolution—at the sub-nanosecond level in some cases—plays a key role in separating prompt and delayed signals, rejecting accidental backgrounds, and probing BSM particles with delayed decays. This feature is inherent in many pulsed-beam facilities, such as DUNE, JSNS$^{2}$, COHERENT, and CCM, where the structured beam timing naturally provides a low-background window.

Fourth, particle identification or flavor identification capabilities are essential for many BSM studies. For example, sterile neutrinos or NSIs often lead to flavor-dependent disappearance or appearance patterns, such as unexpected $\nu_{e}\rightarrow\nu_{\tau}$ transitions, which cannot be interpreted without clear identification of the final-state lepton flavor. The ability to distinguish $\nu_{e}, \nu_{\mu}$ and $\nu_{\tau}$ enhances sensitivity to such effects, particularly in long-baseline experiments like DUNE. In addition to flavor tagging, the ability to distinguish between electron- and photon-induced showers is important for identifying BSM signals involving LDM or ALPs. Both LDM scattering off electrons and ALP decays into two photons can produce electromagnetic signatures that resemble standard neutrino interactions, particularly neutral current events with $\pi^{0}\rightarrow\gamma\gamma$ decays. However, while photon-induced showers typically exhibit a conversion gap and are characterized by two-track topologies or broader transverse profiles, electron-induced showers originate directly at the interaction vertex and tend to be more collimated. High-resolution calorimetry and tracking systems are therefore essential to resolve the shower initiation point, measure opening angles, and reconstruct the event topology with precision.

Fifth, background suppression remains a central challenge in the detection of rare BSM processes. Shielding from cosmic rays and environmental radioactivity, use of passive and active veto systems, and the employment of side-band or reactor-off data are widely used to model and subtract backgrounds. In particular, surface or shallow-depth detectors like NEON or DAMSA require stringent background modeling to extract small signals from large rates of reactor- or beam-related neutrons and photons.

Finally, it is advantageous for detectors to have flexibility in configuration, data acquisition, and runtime operation. For example, the ability to move detectors off-axis, as in the IWCD and the DUNE-PRISM program, enables sampling of different neutrino flux spectra and compositions, which is essential for controlling systematic uncertainties in standard oscillation analyses. For BSM searches, this flexibility can also be beneficial, as moving off-axis reduces the overall neutrino background, thereby improving the signal-to-background ratio for rare or non-standard processes. Modular or movable detector setups further allow experiments to optimize their sensitivity to specific BSM scenarios or to perform systematic cross-checks under varying beam exposures and configurations.

These capabilities, taken together, will allow the next generation of neutrino detectors to comprehensively probe new physics models across a wide range of masses, couplings, and production mechanisms. The convergence of high beam intensities, precision detector design, and dedicated BSM analysis frameworks places neutrino experiments at the frontier of beyond Standard Model searches in the coming decade. For clarity, Table~\ref{tab:capabilitysummary} summarizes the detector capabilities discussed above, together with indicative target performance benchmarks and representative examples of the present state of the art.

\begin{table}[t]
\centering
\small
\begin{tabular}{|p{0.2\columnwidth}|p{0.36\columnwidth}|p{0.4\columnwidth}|}
\hline \hline
\textbf{Detector capability} & \textbf{Target performance} & \textbf{State of the art} \\
\hline
Vertex / tracking &
cm or sub-cm vertexing; precision momentum and charge reconstruction &
LArTPCs, emulsion detectors, magnetic spectrometers; DAMSA/SHiP-style LLP searches \\
\hline
Invariant-mass reconstruction &
MeV-scale resolution for $\gamma\gamma$ and $e^+e^-$ resonances &
Compact ECAL + tracker systems; DAMSA targets sub-MeV--MeV-scale mass resolution \\
\hline
Low threshold / high light yield &
sub-MeV--few-MeV recoils or photons; high photoelectron yield &
CE$\nu$NS detectors, NaI/CsI/LAr, JUNO, JUNO--TAO, NEON \\
\hline
Timing &
sub-ns--ns timing for prompt/delayed separation and accidental rejection &
Pulsed-beam experiments; fast scintillator and photodetector systems \\
\hline
PID / flavor ID &
$e/\gamma$, $\mu$, hadron, neutron, and flavor separation where applicable &
LArTPC $dE/dx$, WC ring imaging, LS timing/PSD, emulsion vertexing \\
\hline
Background control &
beam-off/reactor-off subtraction; veto, shielding, side-band constraints &
COHERENT, CCM, JSNS$^2$, NEON, reactor and beam-dump programs \\
\hline
Configurability / systematics &
off-axis or movable operation; percent-level flux and detector-response control &
DUNE-PRISM, T2HK IWCD, near-detector complexes \\
\hline
Large exposure / stability &
multi-year stable operation; large fiducial mass or high-intensity source &
DUNE, HK, JUNO, IceCube/Gen2, $\nu$EYE; compact high-rate beam dumps \\
\hline \hline
\end{tabular}
\caption{
Compact summary of representative detector capabilities relevant for BSM searches at neutrino facilities.
The listed performance goals are indicative and depend on the signal topology and detector technology.
}
\label{tab:capabilitysummary}
\end{table}

\subsection{Signal and Background Simulation Tools}
Over the past few years, many BSM physics signatures have been proposed, and the theoretical calculations have become available for the searches.
These theoretical tools, however, must be integrated into some Monte Carlo generators for experiments to be able to evaluate the feasibility of searching for such signatures in a realistic manner.
A handful of theoretical tools have been integrated into neutrino interaction generators, such as GENIE~\cite{Andreopoulos:2009rq,GENIE:2021npt}.

The interactions of DM have been integrated into GENIE via a GENIE plugin named GENIE-BDM~\cite{Berger:2018urf,Berger:2019ttc}. 
The plugin simulates interactions of nuclei and DM, that produces recoiled electrons and nucleons. For example, in the LDM search at DUNE, dark photons are produced from the two-photon decay of neutral mesons produced in the beam and target (graphite) interaction process, such as $\pi^{0}$ and $\eta$. The generated dark photons then decay into two DM particles, considering the DM mass. The resulting DM flux can be injected into the DUNE Near Detector using the GENIE-BDM plugin to produce signal events of recoiled electrons. The generated events contain information such as the position and four-momentum of knocked-off electrons or nucleons. 

HNLs from decays of light mesons such as pions and kaons have also been integrated into GENIE, via the BeamHNL module \cite{PhysRevD.107.055003}.
The module generates HNL decays from a list of possibilities in the context of the effective field theory presented in \cite{Ballett:2019bgd, Coloma:2020lgy} inside an arbitrarily complicated detector volume, generating information on the delay of the final state with respect to a Standard Model neutrino, as well as some basic information about the angular distribution of the final state, and the momenta and energies of each final-state particle.
Input fluxes are handled as direct ROOT tuples with each entry being one neutrino-producing particle, produced in the beamline.
This allows beamline simulations such as GEANT4 to hook into the module almost directly, with a simple flattening script translating simulation output to BeamHNL input.
Finally, the module performs its own calculation of the expected number of protons on target event-by-event, by taking into account detector size and position effects.
The BeamHNL module is available in GENIE and can be used readily for beam-produced HNL searches. 
The logic contained in the module can readily be extended to other long-lived unstable particle types, to accommodate model-independent searches.
Moreover, the hook with a detailed simulation of neutrino-producing particles can be augmented to serve the needs of non-beam experiments, such as atmospheric searches for new physics.

Following the generation of BSM signals, the resulting detectable particles are propagated through the detector medium. For example, edep-sim, an Energy Deposition Simulation, is a tool  based on the GEANT4 to calculate how particles deposit energy as they traverse materials such as liquid argon \cite{edepsim}. 
The results are subsequently processed by detector-specific simulation tools to generate realistic signals --- an example tool developed for LArTPCs such as DUNE is larnd-sim~\cite{DUNE:2022gxa}, which models physics processes including charge quenching, electron drifting under an electric field, current accumulation, and the final electronics response.
For Water Cherenkov Detectors, the WCSim~\cite{wcsim} tool is utilized to simulate Cherenkov light production and detection. These tools can be accessed by anyone and be used to produce sensitivity results.

A transformative effort is currently underway to convert these simulators into differentiable ones using automatic differentiation libraries such as PyTorch, EagerPy, or JAX. By rewriting tools like larnd-sim to be differentiable~\cite{Gasiorowski_2024}, researchers can calculate exact gradients for detector parameters, enabling simultaneous high-dimensional calibration.
Similar advancements are being made in the WCD community, notably with the CIDeR-ML project~\cite{Alterkait:2026ocv}, which focuses on differentiable photon propagation.
These machine learning enhancements are deeply relevant to BSM physics searches, in a way that can significantly reduce systematic discrepancies between simulation and data, thereby boosting sensitivity to rare BSM-induced phenomena.

While software packages like GENIE have successfully centralized BSM event generation, the lack of a common and robust background simulation package hinders a fair comparison and forecasting of sensitivities across different experimental setups. Expanding the collaborative approach to encapsulate universal background components and shared systematic error models would be a major breakthrough for the community. Utilizing such a unified framework lays the crucial technical groundwork for future joint analyses between different experiments, allowing cross-experimental validation and maximizing the ultimate statistical power of global BSM searches.

\section{Conclusions}
\label{sec:conclusions}

A wide range of neutrino programs---located near reactors, underground, underwater, in ice, or at accelerator facilities---are currently operating or under serious development worldwide to explore physics beyond the Standard Model through precision studies of the neutrino sector. Remarkably, neutrino experiments also offer novel opportunities to search for other BSM signals beyond neutrino oscillations. Numerous pioneering studies have proposed leveraging neutrino experiments to probe phenomena such as dark photons, DM, ALPs and neutrino-philic BSM particles.

In this paper, we have summarized the promising prospects for new physics searches in current and next-generation accelerator-produced neutrino experiments, including T2K, T2HK, SHiP, SND@LHC, SND@HL-LHC FASER/FASER$\nu$, FPF detectors (FASER2, FASER$\nu$2, FLArE, and FORMOSA), COHERENT, CCM, JSNS$^2$, DAMSA, and the electron beam dump and IsoDAR program at Yemilab. 
Accelerator-based experiments using low-energy but high-intensity beams offer advantages in precision measurements, while forward detectors at the LHC and HL-LHC are particularly well-suited for probing heavy new particles with relatively low background. Compared to large-volume neutrino detectors, these setups allow for the integration of advanced detector technologies and the development of innovative strategies---leveraging beam properties for precise signal reconstruction and effective background suppression.
As a result, accelerator-based neutrino experiments significantly broaden the landscape for exploring new physics.

\begin{figure}[t]
    \centering
    \includegraphics[width=1.05\textwidth]{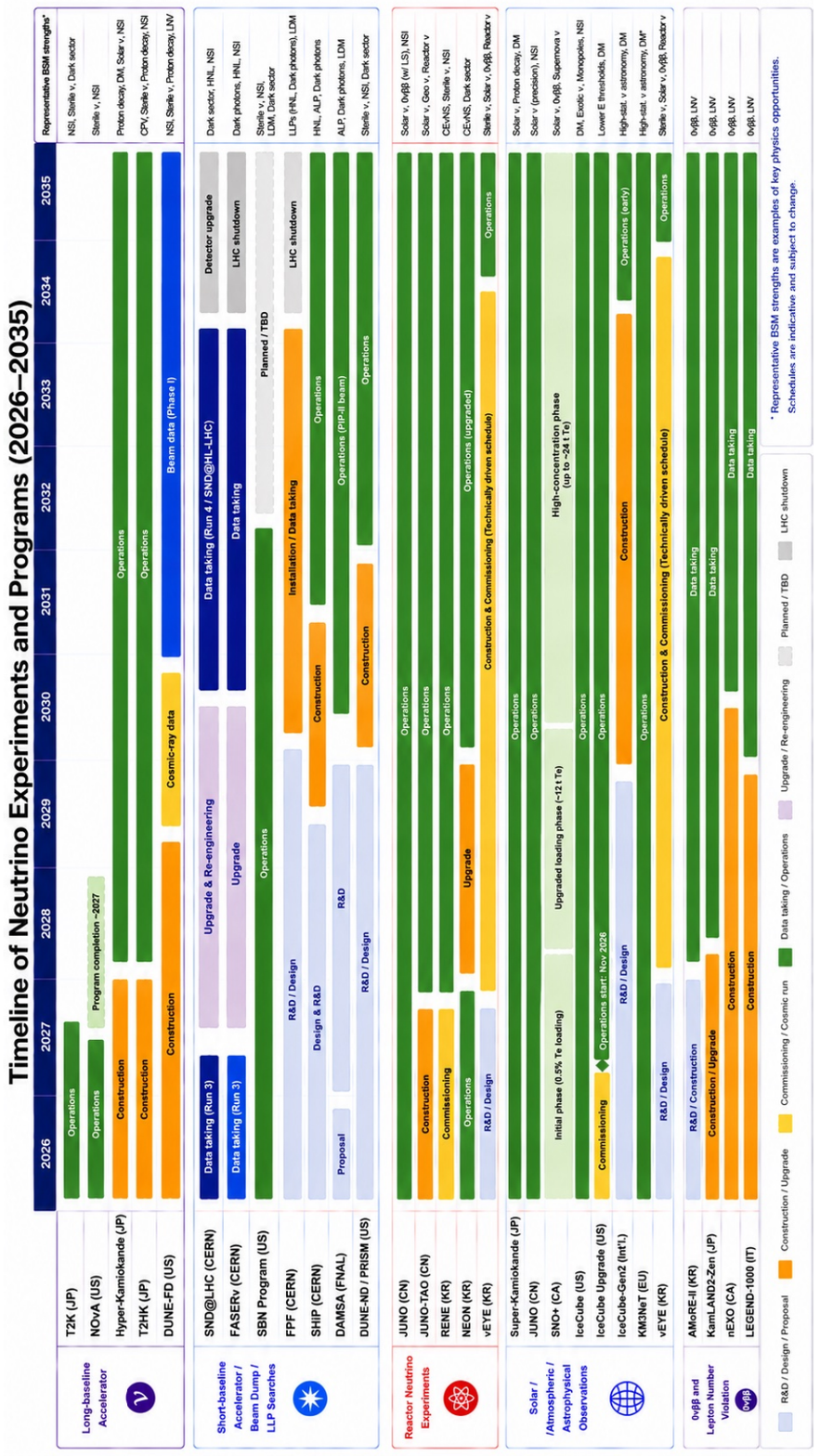}
    \caption{Schematic view of the timeline and the BSM search capabilities of global neutrino programs.}
    \label{fig:timeline}
\end{figure}

It is encouraging to see that a wide range of ambitious neutrino programs are currently operating, under construction, or in serious development across the East Asian region. 
In this paper, we have discussed the most up-to-date results in SK, T2K, KamLAND/KamLAND-Zen, JSNS$^2$, NEOS, and NEON.
Additionally, we have outlined future prospects for experiments such as HK, T2HK, JUNO, KamLAND2-Zen, RENE, and the experiments at Yemilab.
In particular, the new underground laboratory Yemilab in Korea plans to host a diverse suite of experiments, including $\nu$EYE and AMoRE-II.
Among them, the 2.26 kt liquid scintillator detector $\nu$EYE is designed as a multi-purpose instrument capable of probing neutrinos and BSM particles originating from reactors, accelerators, the Earth, and the cosmos. This neutrino complex is expected to synergistically enhance the region’s capabilities in fundamental physics. 
Moreover, many institutions in East Asia are actively participating in the worldwide neutrino programs such as DUNE, IceCUBE, DAMSA, SND@LHC, SHiP, and the experiments in FPF at the time of HL-LHC---further strengthening international collaboration and scientific impact.
In Fig.~\ref{fig:timeline}, we schematically show the global timeline of the neutrino programs, placing the experiments discussed in this paper within the wider international landscape and highlighting their complementarity with other major neutrino programs.

As a final remark, we strongly encourage deeper collaboration among experiments and researchers in the East Asian region, given the outstanding potential of their experimental facilities and phenomenological efforts highlighted in this paper. A coordinated effort to probe new physics through neutrino experiments will be essential to achieving breakthroughs in high-energy physics. We also hope that this paper will serve as a foundation for future collaborative work and help facilitate upcoming workshops and conferences focused on new physics opportunities at next-generation neutrino facilities.

\section*{Acknowledgments}

The authors appreciate the Institute for Basic Science for providing the venue and financial support for the 4th workshop on New Physics Opportunities at Neutrino Facilities.
YSJ thanks Felix Kling for the permission to use Figure \ref{fig:LDF_FPF} and the left panel of Figure \ref{fig:NuOsc_forward}. 
This work is supported in part by the National Research Foundation of Korea (NRF) grant funded by the Korea government through Ministry of Science and ICT Grant No. RS-2025-00555834 (YSJ, SMY). 
The work of JCP is supported by the National Research Foundation of Korea (NRF) grant funded by the Ministry of Science and ICT (RS-2024-00356960) and the Ministry of Education (RS-2025-25442707).
The work of PB, M.-G.P, MR, and SS is supported by the NRF with Grant No. RS-2025-00562917. 
MM acknowledges support from the grant NRF-2022R1A2C1009686. 
SS is partly supported by the IBS fund IBS-R018-D1. HSL is supported by the IBS under project code IBS-R016-A1 and KC is supported by the IBS under project code IBS-R016-D1.
JY is supported in part by the U.S. Department of Energy under Grant No. DE-SC0011686.
BSY is supported by the National Research Foundation of Korea (NRF) grants funded by the Ministry of Science and ICT (RS-2022-NR070836, RS-2025-23525600) and by the Ministry of Education (RS-2024-00442775).

\bibliography{refs}

\end{document}